\documentclass[preprint,11pt, 3p]{elsarticle}

\usepackage{amssymb}
\usepackage{amsmath}

\usepackage{xcolor} 
\usepackage{hyperref} 
\usepackage[textsize=small]{todonotes} 
\usepackage{tikz-feynman} 
\usepackage{tcolorbox} 
\tcbuselibrary{breakable, skins}
\tcbset{
    before skip = {20pt},
    after skip = {20pt},
    skin = enhanced
}

\usepackage{listings} 
\newcommand{\FeynCraft}{\lstinline!FeynCraft!}
\newcommand{\figref}[1]{\hbox{Fig. \ref{#1}}} 
\renewcommand{\eqref}[1]{\hbox{Eq. \ref{#1}}} 

\newcounter{bla}

\journal{Computer Physics Communications}

\begin{document}

\begin{frontmatter}



\title{ 
   \includegraphics[width=0.7\textwidth]{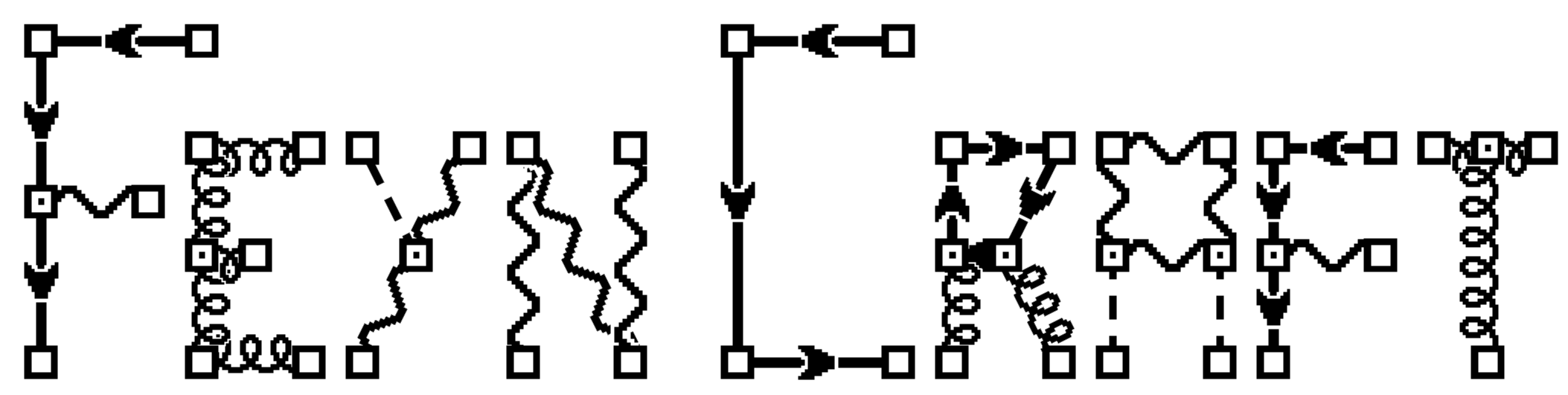}
   \\
   FeynCraft: A Game of Feynman Diagrams
 }


\author[a,b]{Jonathan R.~Gaunt\corref{author}}
\author[a,1]{Adam Owen}

\fntext[fn1]{\textit{Deceased.}}

\cortext[author] {Corresponding author.\\\textit{E-mail address:} gaunt.jonathan@ucy.ac.cy}
\address[a]{Department of Physics and Astronomy, University of Manchester, Manchester, M13 9PL, United
Kingdom}
\address[b]{Department of Physics, University of Cyprus, Nicosia 1678, Cyprus}






\begin{abstract}
\FeynCraft{} is a browser-based game that is designed to teach players the particle interactions of the Standard Model of particle physics, and how to link these interactions together to produce valid Feynman diagrams. It is primarily targeted at undergraduates and lecturers in introductory courses in particle physics, but we anticipate that it should also be useful for school pupils and teachers studying the basics of particle physics, and perhaps also current researchers. Users may draw particle lines and link them together to form vertices and complete diagrams, and \FeynCraft\ determines invalid vertices using a sequence of simple rules, showing users which vertices are invalid and why. Diagrams may be drawn that involve both fundamental Standard Model particles and hadrons (where hadrons are represented by their constituent quark content). Users can also be presented with a process for which they must draw valid Feynman diagrams -- \FeynCraft{} is able to generate such `problems' at random, but there is also the facility to create, share, import and solve curated sets of problems. Alternatively, one is able to specify the process, and ask \FeynCraft{} itself to generate the Feynman diagrams. Finally, we include several overlay options that give more information on a Feynman diagram (e.g. QCD colour flow, interaction strengths), and the option to export a drawn diagram as \LaTeX{} code.
\end{abstract}

\end{frontmatter}


\newpage

\begin{center} \includegraphics[width=0.3\textwidth]{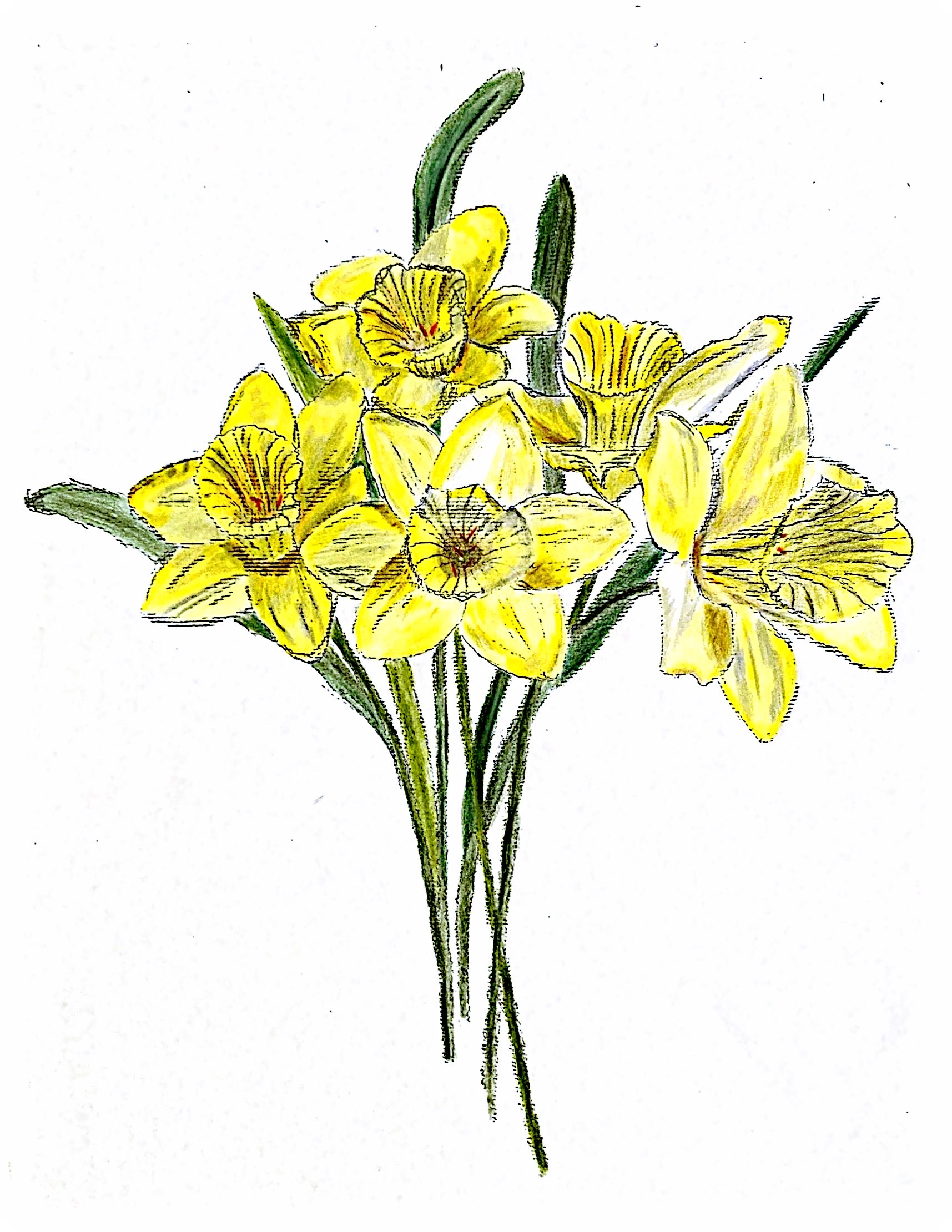} 
\\ \textit{Dedicated to the memory of Adam Owen MPhys (2000-2025), whose considerable talent, limitless enthusiasm, and remarkable creativity were the key driving forces behind this project.} \end{center}

\vspace{3cm}

\noindent{\bf PROGRAM SUMMARY/NEW VERSION PROGRAM SUMMARY}

\begin{small}
\noindent
{\em Program Title:} \FeynCraft                                          \\
{\em CPC Library link to program files:} (to be added by Technical Editor) \\
{\em Developer's repository link:} \href{https://github.com/JonathanGaunt/feyncraft-src}{https://github.com/JonathanGaunt/feyncraft-src} \\
{\em Code Ocean capsule:} (to be added by Technical Editor)\\
{\em Licensing provisions(please choose one):} MIT  \\
{\em Programming language:}  \lstinline{GDScript}       \\
{\em Operating systems:} Web \\
{\em Nature of problem:} An accessible way to draw, export, and practice the rules of, Feynman diagrams.\\
{\em Solution method:} A set of tools to draw and validate diagrams, which can be used to solve either user-created or randomly generated problems, or be exported into \LaTeX{}. \\
{\em Restrictions:} The addition of momentum lines and accents other than those available are required to be added manually after exporting into \LaTeX{}\ code. \\

\end{small}

\tableofcontents

\section{Introduction}
\label{sec:introduction}
Feynman diagrams are a key tool in modern particle physics computations. In order to compute the scattering amplitude for a given process (e.g. $e^+e^-$ annihilation to photons, $e^+e^- \to \gamma\gamma$), the typical first step is to draw all possible Feynman diagrams for that process (up to the given order in perturbation theory that one is interested in). These diagrams can be converted into the mathematical expression for the scattering amplitude, via `Feynman rules' that assign expressions to the elements of the diagrams, and the amplitude for the process subsequently computed.

As such, in an introductory course in particle physics one of the first things the students must learn to do is draw all the Feynman diagrams for a given process (initially at the lowest order in perturbation theory, in the Standard Model of particle physics). From our experience, this is something that students often seem to have trouble with; common errors include drawing vertices that do not exist (e.g. $e^+e^-\gamma\gamma$), forgetting one or more Feynman diagrams, and/or drawing diagrams that are topologically equivalent to ones that have already been drawn. The older author recalls this being true from his time as an undergraduate student and when teaching courses in particle physics as a PhD student, and judging from recent feedback from particle physics exams at the University of Manchester, the issue persists today. 

In our view, the best approach to tackling this issue was the development of a computer game (or app). The process of learning to draw Feynman diagrams is ultimately one of learning a set of rules and applying them to solve problems (or puzzles), which is the broad structure shared by a wide variety of puzzle games. Many bestselling puzzle games have rules that are ultimately relatively complicated, but players nonetheless clearly enjoy engaging with such rules if they are put within the context of a game -- see for example Braid \cite{Braid}, The Talos Principle \cite{Talos}, Baba is You \cite{Baba}, and Portal 1 \cite{Portal1} and 2 \cite{Portal2}, all of which have won multiple game awards. By developing a game about Feynman diagrams one may hope to tap into the same mechanisms where users quickly and enjoyably learn the rules through play. Further, the graphical nature of Feynman diagrams fits well with the computer game format. Finally, within the context of a game one can randomly generate an essentially endless set of problems to solve, allowing students to practice their skills as much as they wish.

In this paper we describe the game that we have developed along these lines, which we call \FeynCraft. The key features of the game, which have been developed with our primary target of advanced undergraduate students in mind, are as follows: 
\begin{itemize}
    \item One may draw Feynman diagrams by clicking and dragging particles of the Standard Model (SM), and connecting them at vertices. The game checks whether vertices are valid SM vertices, and informs users when and why vertices are incorrect. Vertices are not checked by comparing against a hard-coded list of SM vertices, but rather by using a simple set of rules (these are listed in section \ref{sec:validation}). We believe that these rules could be a useful way to teach and check valid vertices even outside the scope of \FeynCraft, so we highlight these in bold in the relevant section. One may draw diagrams involving both fundamental SM particles and also involving hadrons; we have included this feature since an important component of introductory particle physics courses (at least in the United Kingdom) is to draw (illustrative) Feynman diagrams for processes involving hadrons (such as neutron decay, $n \to p e^- \bar{\nu}_e$).
    \item \FeynCraft\ can generate problems at random, where a process is displayed and the user has to draw the Feynman diagrams for that process. It can generate problems that involve fundamental SM particles, or also involving hadrons, according to the preference of the user. The game only generates processes for which valid Feynman diagrams exist, and is able to show sample solutions if the user is stuck. The game has a dedicated `daily problem' feature where one random problem is generated for users to solve; the game stores the number of consecutive days for which they have managed to solve the problem (their `streak') which resets if they fail to solve the problem on a particular day.
    \item Users can also create tailored sets of problems, in which they can specify a list of processes for which other users should draw the Feynman diagrams. For each process they can also limit the available SM particles that can be used to solve the problem. Problems can easily be exported and imported via a text file, allowing easy sharing of problem sets between users. We envisage this feature as being useful for lecturers, who can convert the list of Feynman diagram problems in their examples sheets to \FeynCraft\ problem sets, and distribute these to their students via the export feature.
    \item If a user specifies the initial and final state of a particular process, then \FeynCraft\ can generate the full set of Feynman diagrams for that process at (a) particular perturbative order(s). This can be useful for a student to check that they have managed to draw all the diagrams for a process.
    \item There is the option to add various overlays to a Feynman diagram that convey additional information. For processes involving strongly interacting particles, one may view a possible QCD colour flow in the process. We include this as it is common in introductory course segments on QCD to encounter Feynman diagrams with illustrative colour flows drawn on (see for example Chapter 10 of \cite{Thomson:2013zua}), so students may wish to see the corresponding ways in which colour flows in \FeynCraft\ diagrams (at a more advanced level, the nature of the colour flows allowed in a diagram determines the size of its QCD colour factor \cite{Kilian:2012pz}). We also include an option to view an illustrative `electroweak colour flow' in the diagram for processes involving electroweak particles (for more details see section \ref{sec:vision_tab}). Finally, we have an option to visualise the coupling strengths associated with the vertices in a diagram -- this allows students to roughly gauge how important different diagrams are for a particular process by comparing the size of the coupling constants involved, which is a skill often taught in introductory particle physics courses.
    \item We have included a facility to export drawn Feynman diagrams to \LaTeX\ code, which students can use (for example) to generate figures for project reports.
\end{itemize}

Although we have primarily targeted \FeynCraft\ at advanced undergraduate students, we hope that the game will be of interest and use to a wider audience. It should also be of use to physics teachers and (advanced) school students who have a segment on particle physics as part of their curriculum. In the United Kingdom, many A-level courses have such a segment, which typically involves learning the particle content of the SM, the composition of hadrons according to the quark model, some basic particle physics processes, and/or drawing Feynman diagrams for these processes (see e.g.~\cite{EdExcel, OCR, AQA}). \FeynCraft\ could be used to assist with these goals, possibly via the development of a tailored problem set that is focussed on the processes and particles studied in that particular curriculum. We also hope that aspects of \FeynCraft\ will be of use to researchers in particle physics; for example the utility to draw all diagrams for a particular process, and the export feature (to draw diagrams for a publication). 

FeynCraft is accessed via a web browser at \href{https://jgaunt.itch.io/feyncraft}{https://jgaunt.itch.io/feyncraft}, such that there is no complex installation process for users to worry about. It can be used on Windows, Linux, Mac, iOS and Android, and switches to a touchscreen interface when a device with a touchscreen input mode is detected. The game is thus very portable and accessible, allowing it to be easily used (for example) both at home and at lectures and examples classes.

We note that there are other packages on the market with some overlap in features with \FeynCraft, but none of these share the full set of features described above, and/or have a different focus. In particular, there are other graphical tools for drawing Feynman diagrams that are more sophisticated than \FeynCraft; namely \lstinline{JaxoDraw}\cite{Binosi:2008ig} and \lstinline{FeynGame}\cite{Harlander:2020cyh, Harlander:2024qbn, Bundgen:2025utt}. The latter of these includes a game where the user must connect a given initial and final state via a Feynman diagram (and since this works via `model' files where one specifies the particle content and allowed interactions, one is not restricted to the SM but may also include hypothetical particles Beyond the Standard Model in this game). \lstinline{FeynGame} is also able to output the amplitude (in \LaTeX\ code) corresponding to a drawn Feynman diagram, which is a feature not currently present in \FeynCraft.

In this paper, we will describe in detail all the features of \FeynCraft\ outlined above, and explain how to use these features with the aid of examples. The examples are highlighted with grey boxes, and we encourage a first-time user to go through these examples in particular to familiarise themselves with the different functionalities of \FeynCraft{}. After a short section describing the main menu, section \ref{sec:feyncraft_mainmenu}, the first few sections cover general aspects of drawing and validating diagrams. Section \ref{sec:feyncraft_diagrams} describes the terminology used in this paper and in \FeynCraft\ to describe elements of a Feynman diagram, section \ref{sec:drawing} explains how to draw a diagram in \FeynCraft\, and section \ref{sec:validation} describes how \FeynCraft\ checks if a drawn diagram is valid in the SM. This section includes the set of rules to validate vertices in the SM discussed above. Section \ref{sec:sandbox} describes all of the features in the basic `Sandbox' game mode, including creating and solving a random problem, generating the Feynman diagrams for a given process, toggling the visual overlays, and exporting a diagram to \LaTeX\ code. Section \ref{sec:problems} covers the `Problem Set' game mode, describing how to create, solve, export and import problem sets. For interested readers, section \ref{sec:algorithms} looks `under the hood' of \FeynCraft\ and describes some of the key algorithms it uses; in particular, how it generates random problems, how it finds the Feynman diagrams for a user-specified process and the procedure used to explicitly draw these diagrams, how it constructs the colour and electroweak flow for a diagram, and how it checks if two diagrams are duplicates of one another (i.e. if they are topologically equivalent). We conclude in section \ref{sec: conclusions}.

\begin{figure}
    \centering
    \includegraphics[width=0.75\linewidth]{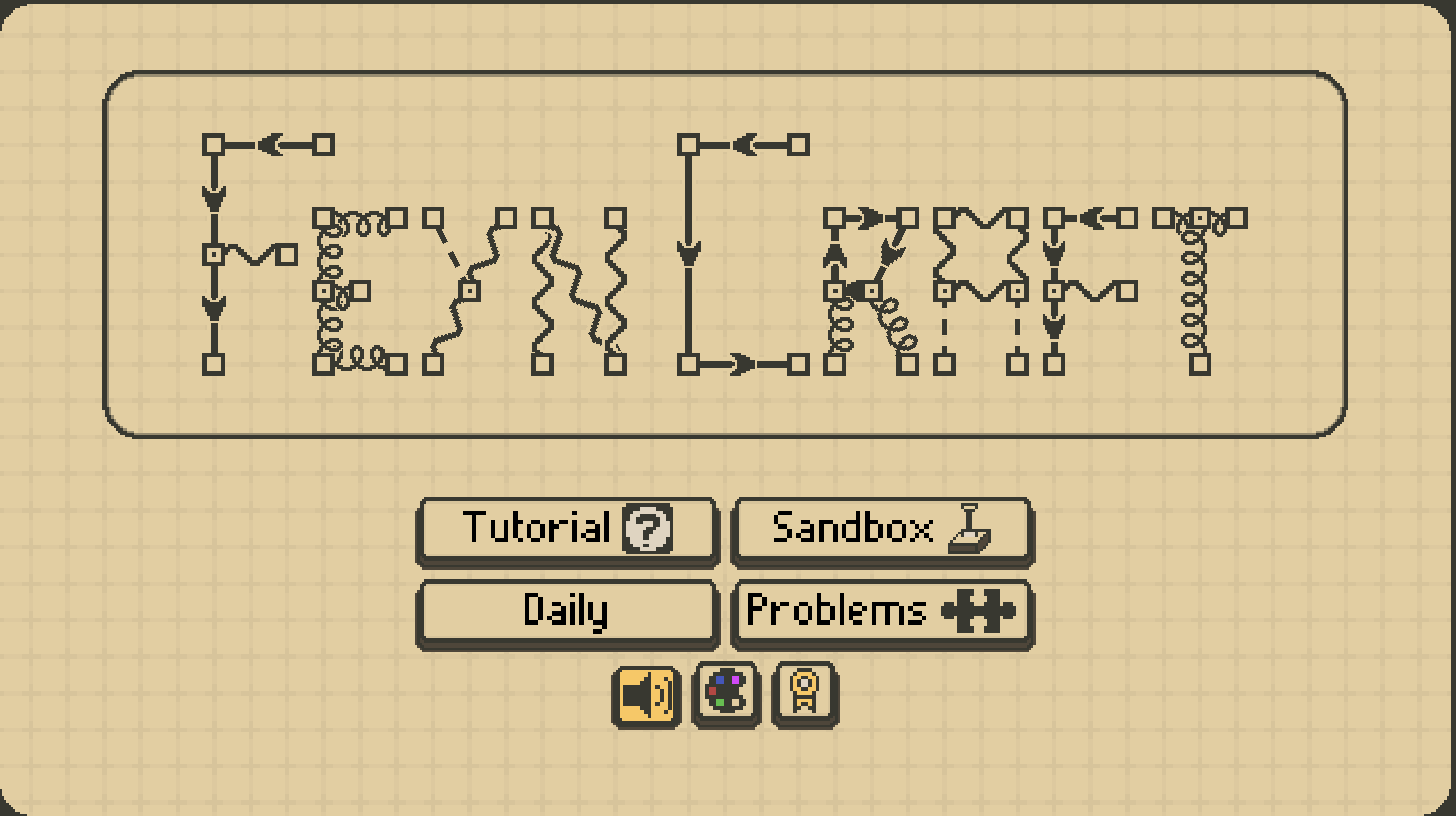}
    \caption{The main menu, the opening screen. From here users can play the tutorial, enter the sandbox, start solving problems, play the daily problem, and access palettes and credits.}
    \label{fig:main_menu}
\end{figure}

\section{\FeynCraft\ main menu}
\label{sec:feyncraft_mainmenu}

The main menu of \FeynCraft\ is shown in \figref{fig:main_menu}. The Sandbox mode allows access to most of the game features described in the previous section. `Daily' is the daily problem feature, and `Problems' allows access to the user-created problem set functionality. We have also included a brief Tutorial. The buttons along the bottom, from left to right, allow one to turn on/off sound effects, change the \FeynCraft{} colour palette, and view the credits. The colour palette button also allows users to create their own colour palettes, to share their created palettes, and import palettes from other users. In the various game modes (Sandbox, Problems, and Daily), the sound on/off and colour palette buttons are still accessible via an option menu in the bottom right (which also has buttons to quit to the main menu, and to show/hide the labels for particle lines).

\section{\FeynCraft\ diagrams}
\label{sec:feyncraft_diagrams}
Here we will demonstrate how Feynman diagrams are depicted in \FeynCraft, and introduce the terminology we will use in this paper and in \FeynCraft\ to refer to the anatomy of a Feynman diagram.

\begin{figure}
    \centering
    \includegraphics[width=0.75\linewidth]{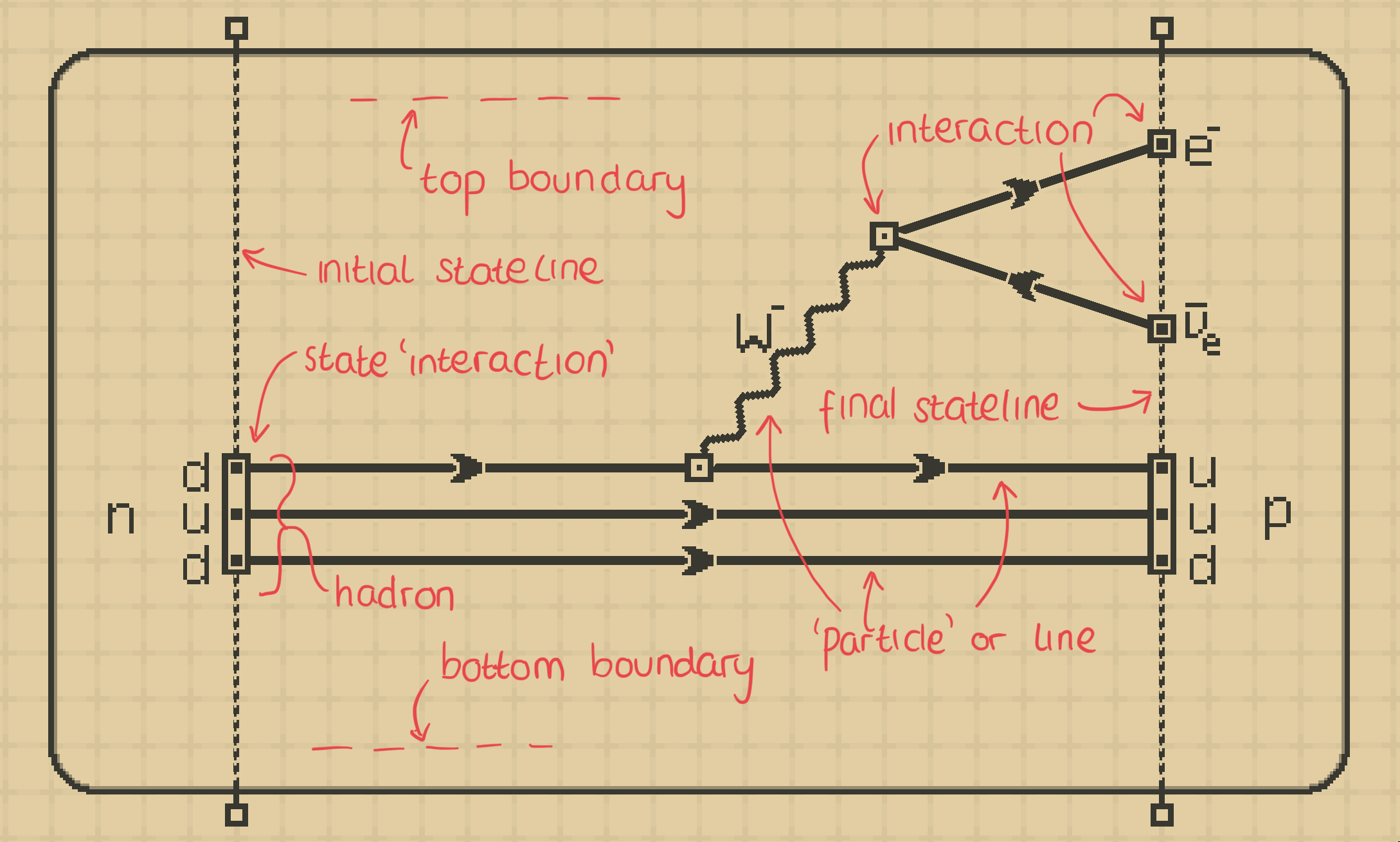}
    \caption{A labelled example of a FeynCraft diagram, showing neutron decay.}
    \label{fig:labelled_diagram}
\end{figure}

A labelled example diagram is shown in \figref{fig:labelled_diagram}. Lines or particle lines  are referred to as \textbf{particles} or particle lines and are drawn on a set of grid points (the `diagram grid') inside the `diagram box'\footnote{We have purposefully designed the drawing area to somewhat resemble a breadboard in electronics, with particle lines resembling wires connecting up holes in the breadboard. This is because there is a close analogy between Feynman diagrams and electrical circuits -- see footnote \ref{footnote: circuits} -- so it is helpful to gently nudge students towards thinking about Feynman diagrams in this way via the aesthetic of the game.}. This box is bounded on the left by the \textbf{initial state line} and on the right by the \textbf{final state line}, which are used to group the incoming and outgoing particles for a process. A particle line with an end on the initial state-line is part of the incoming particles (\textbf{initial state}), and a particle line with an end on the final state-line is part of the outgoing particles (\textbf{final state}). If one draws multiple quark lines with endpoints directly above each other on a state line, and the set of quarks corresponds to a hadron (according to the constituent quark model), then those quark lines are automatically grouped as that hadron. One sees this in \figref{fig:labelled_diagram} with the $dud$ quark combination being labeled as a neutron $n$, and the $udd$ as a proton $p$. For more details see section \ref{sec:drawing_particles}.

The ends of particle lines are referred to as \textbf{interactions}, and depicted as squares in \FeynCraft{} -- these interactions may only be placed on grid points. By placing the endpoints of $\ge 3$ particles on the same grid point, one may create a vertex interaction corresponding to a particle interaction. However, we may also have `interactions' where only one particle enters, in particular the ones on the initial and final state lines. We may also have `interactions' where two of the same particle are connected, which in fact simply corresponds to that particle continuing along without interacting. Even though such `interactions' are trivial in terms of the structure of the Feynman diagram, they are useful for aesthetic reasons and allow an emulation of curved lines within the grid system of \FeynCraft{} -- they are needed, for example, to draw legible bubble diagrams. Of course the one- and two-particle interactions do not actually correspond to particle interactions, but we still refer to them as such within \FeynCraft. In this paper, we will denote interactions in the text using square braces, for example $[W^+, W^-, \gamma]$.

Associated with each interaction is the \textbf{degree}, which effectively counts the number of coupling constants associated with that interaction. An interaction with three connected particles has degree 1, whilst an interaction with four coupling constants has degree 2, since SM vertices with three particles have one power of a coupling constant, whilst vertices with four particles have a product of two couplings. Interactions with fewer than three connected particles are of degree-0, as they do not actually correspond to a particle interaction. The degree of an interaction is displayed in the diagram by the number of dots inside the interaction square; the exception to this is the interactions on the state line, where every interaction has a larger dot, but still corresponds to degree 0. The \textbf{total degree of a diagram} is the sum of the degrees of all interactions in the diagram. This essentially corresponds to the perturbative order of the diagram, and gives some rough qualitative representation of the weight of the diagram (the higher the degree, the more coupling constants involved, and the smaller the contribution to the scattering amplitude from that diagram). One should bear in mind that this measure simply counts the number of coupling constants and does not take account of the (very) different coupling strengths of the different interactions, and also does not account for other factors in the vertex Feynman rules (e.g. factors of masses in the Higgs boson vertices) and in the propagators of the diagram. 

Note that in a Feynman diagram, even identical particles in the initial or final state are distinguished via a momentum assignment (so we can distinguish between an attachment to (say) final-state gluon `$1$' with momentum $k_1$ and one to final-state gluon `$2$' with momentum $k_2$). In \FeynCraft{} we do not write momentum labels explicitly, so identical particles on a state line are distinguished by their vertical position -- so the top gluon is always `gluon $1$', the next highest one is `gluon $2$', etc. We encourage users to fix the positions of the state-line interactions when drawing multiple diagrams for a given process, otherwise it may be difficult to deduce when one has drawn a diagram that is a duplicate of one drawn previously.

In code, and particularly in the solution generation section of this paper, section \ref{sec:sandbox-generating_solutions}, a particle and its anti-particle might not be distinguished. Instead, particle lines are drawn and described in terms of the `base' particle and its direction of flow. For example, rather than thinking of a positron going forward in time, left-to-right in the diagram, we may instead choose to think of this as an electron (the base particle) flowing backward in time, right-to-left (this is essentially the Feynman-St\"{u}ckelberg interpretation of antiparticles \cite{Feynman:1949hz}). This way of thinking is useful in the code (and more generally when thinking about Feynman diagrams!) as the validity of an interaction in a Feynman diagram does not depend on whether the attached lines are `before' or `after' the vertex, or the location/orientation of the vertex, only on whether the `flow' of the connected particles in and out of that vertex is correct. For example a valid $[e,e,\gamma]$ vertex always has a connected photon, an electron flowing into the vertex, and an electron flowing out of the vertex. The fundamental principle here is that a single Feynman diagram actually represents all possible time orderings of the internal vertices within the diagram at once. Thus, in a Feynman diagram what is relevant is not the positioning of the vertices and whether lines attached to vertices are `before'/`after' the vertex, but rather how the vertices are {\em connected} to each other and the {\em `flow'} of particles between the vertices\footnote{One may think of Feynman diagrams as being analogous to electrical circuits -- what matters is not where the components are located, but how they are connected to each other and the direction of the current flow. Moving the components around without changing the connections between them does not yield a different electric circuit, much as moving vertices around in a Feynman diagram without altering the connections does not yield a distinct Feynman diagram. \label{footnote: circuits}}. For the $W$ boson, the $W^-$ is taken to be the $W$ base particle, and for bosons that are their own anti-particles ($\gamma$, $g$, $Z$, $H$), the direction of flow has no effect, and there is only the base particle.

\section{Drawing Diagrams}
\label{sec:drawing}
In this section we will go over how users draw Feynman diagrams. We follow the example of how to draw the diagram shown in \figref{fig:labelled_diagram}, neutron decay:

\begin{equation}
\text{n} \rightarrow  \text{p} + e^- +\overline{\nu}_e,
\end{equation}
or, as we will be drawing the quark contents:
\begin{equation}
\{u, d, d\} \rightarrow \{u, u, d\} + e^- +\overline{\nu}_e.
\end{equation}

\subsection{Drawing Particles and Interactions}
\label{sec:drawing_particles}

\begin{figure}
    \centering
    \includegraphics[width=0.75\linewidth]{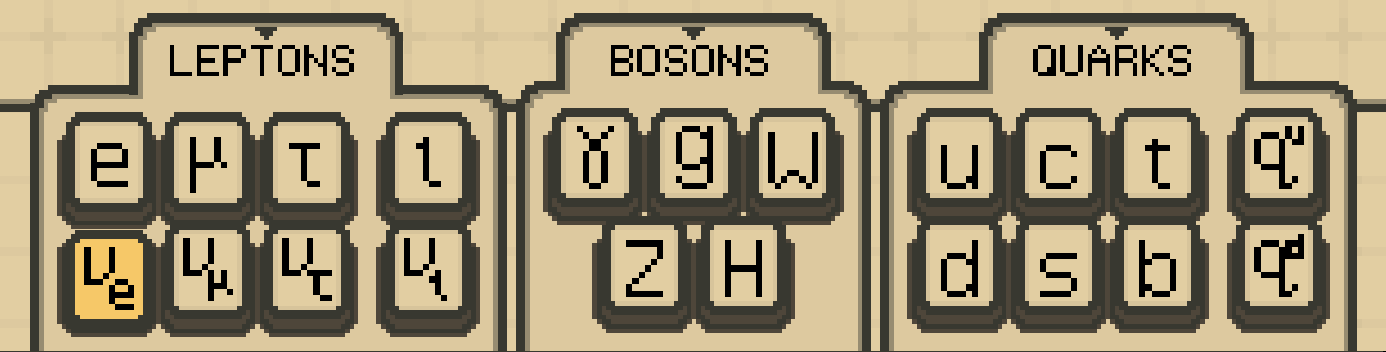}
    \caption{The particle tabs, containing the particles to be drawn. The electron neutrino is selected, making it the currently drawn particle.}
    \label{fig:particle_tabs}
\end{figure}

Before drawing a particle, a particle must first be selected. These particles can be found inside the three `particle tabs' at the bottom of the screen, shown in \figref{fig:particle_tabs}. These particle tabs are leptons, bosons, and quarks, and contain each respectively\footnote{A technical note: in gauge field theories such as the SM, one has a certain freedom when setting up the theory, known as `choosing the gauge', where this choice does not affect physical quantities such as scattering cross sections. Depending on the choice of gauge, one may also require additional `ghost particles' when computing loop corrections in the SM, which run around the loops. We do not include these ghosts or the corresponding diagrams in \FeynCraft{}; one may regard our graphs as corresponding to a suitable `physical gauge', in which such ghost particles do not appear.}. We have also included in the lepton tab the possibility to draw a general lepton ($l$) or general neutrino ($\nu_l$), and in the quark tab the possibility to draw a general up-type quark ($q^u$) or down-type quark ($q^d$). These are supposed to represent any lepton, neutrino or up/down-type quark (e.g. $l = \left\{e,\mu,\tau\right\}$). One reason for their inclusion is that in loop-level Feynman diagrams, there are often diagrams that are only distinguished by the type of lepton, neutrino or up/down-type quark flowing around some internal loop, and we can represent all of these diagrams in one using the general particle labels. Another example use of the general particle labels is to illustrate that a process can produce any species of lepton by using a general lepton line attaching to the final state.

We note that there are no anti-particles listed in \figref{fig:particle_tabs}, these are instead also drawn using the base particle, which can be selected by clicking the corresponding particle button. We can then draw the base particle with a click and drag from left-to-right, and its anti-particle from right-to-left. The ends of drawn particles create interactions, particles that share an interaction are connected, and the interaction degree (see Section \ref{sec:feyncraft_diagrams}) is shown by the number of dots inside the interaction box. Interactions that are not valid in the standard model are coloured red by \FeynCraft{}; the user can find out why the interaction is invalid by clicking on it, as is  discussed in more detail in Section~\ref{sec:validation}.

\begin{tcolorbox}[breakable]
For our example $\text{n} \rightarrow  \text{p} + e^- +\overline{\nu}_e$ process, let us start by drawing the electron, anti-electron-neutrino, and the $W^-$. For the electron, we first click on the lepton particle tab to open it and then click on the electron particle button to select the electron to draw. We then click and hold from where we want the electron to start - left of the final state-line - to where we want it to end - the final state-line, releasing to place. For the anti-electron-neutrino we must draw an electron-neutrino from right-to-left and we would like it to connect to our electron. First, we select the electron-neutrino, then starting from the final state-line we draw to the start of the electron and release. We see that where we connected these particles, the interaction has turned red showing that it is invalid. We need to draw the $W$ boson to fix this. To draw the $W^-$ boson, we open the bosons tab and select the $W$. As in \FeynCraft{} we consider the $W^-$ to be the base particle and the $W^+$ the anti-particle, we draw from left-to-right. We draw from a point in the middle of the diagram to the invalid interaction. We see that not only does this make the interaction valid, but that the interaction now has a single dot, showing that it has gone from degree-0 to degree-1.\\

\begin{center}
\includegraphics[width=0.32\linewidth]{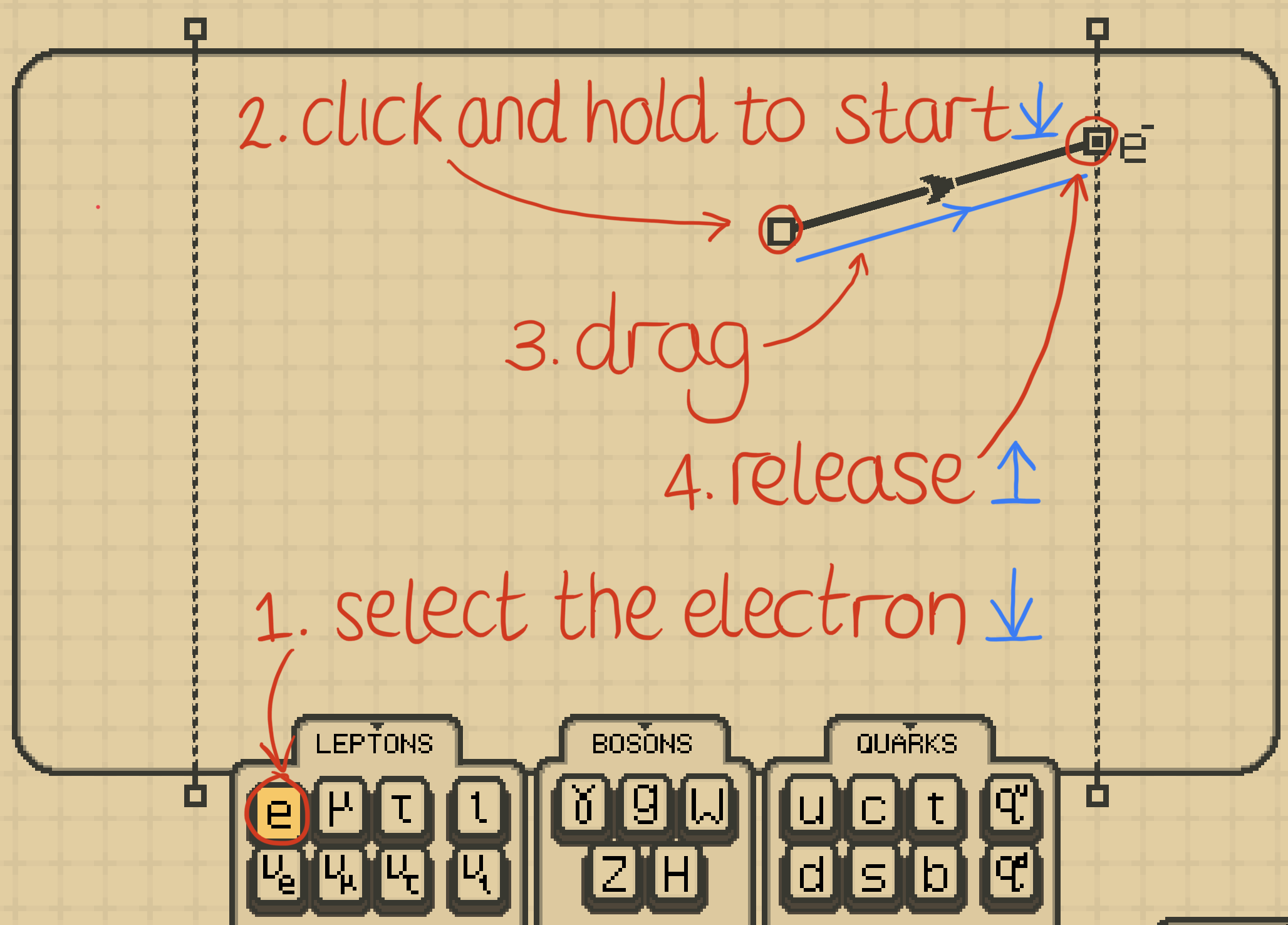}
\includegraphics[width=0.32\linewidth]{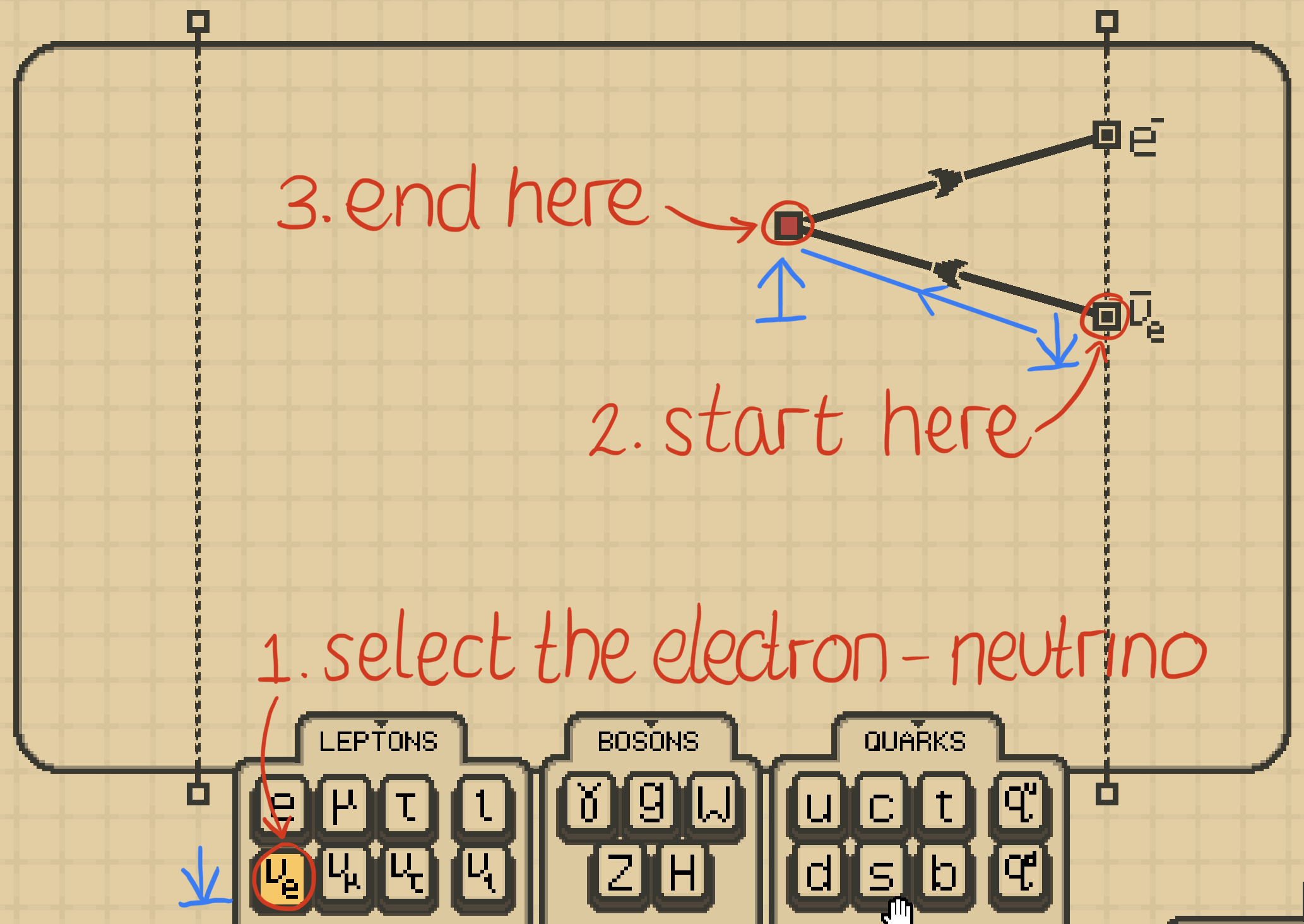}
\includegraphics[width=0.32\linewidth]{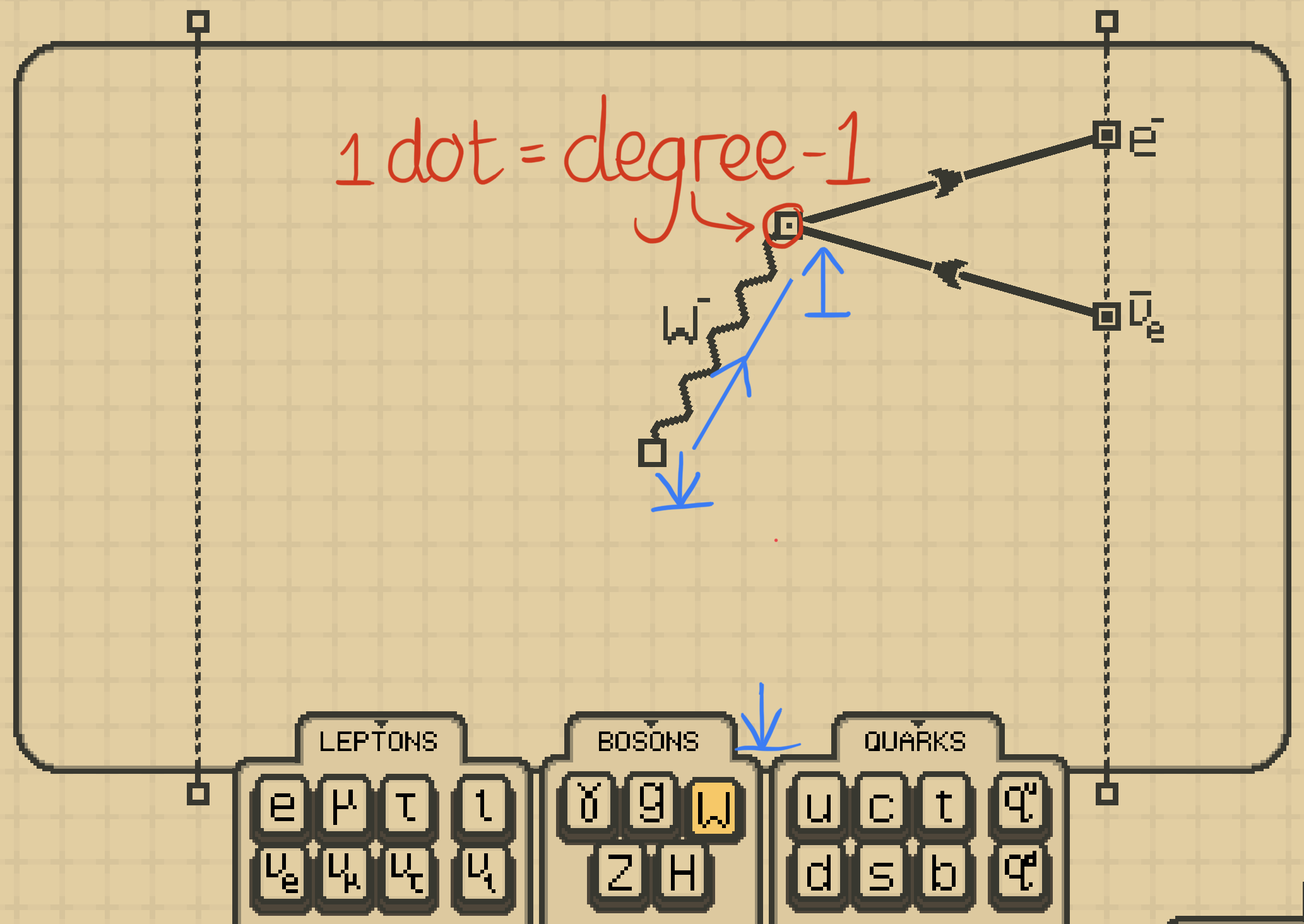}
\end{center}
\end{tcolorbox}

Hadrons can be created by drawing their quark content consecutively in any order on the same state-line. A gap must be left on both the top and bottom side between these consecutive quarks and any other particle, otherwise the hadron will not form.\\

\begin{figure}
    \centering
    \includegraphics[width=0.5\linewidth]{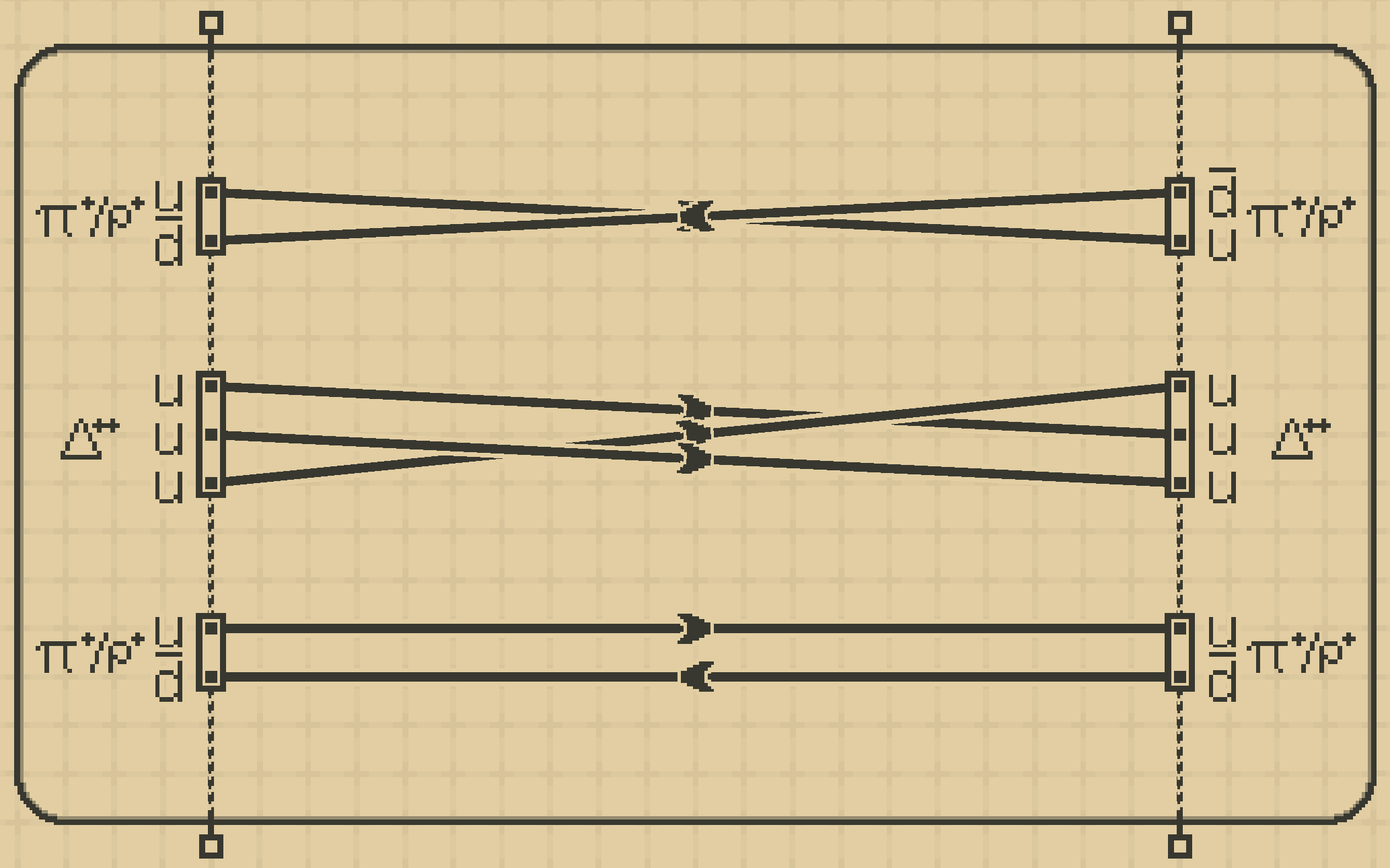}
    \caption{Examples of validly drawn hadrons. All quarks are consecutive and separated from other quarks.}
    \label{fig:hadron_configurations}
\end{figure}

\begin{tcolorbox}
Continuing the example, we shall now create the hadrons by drawing their quarks. After drawing a down-quark from the initial state-line to the lone $W^-$ interaction, we draw an up-quark from the same interaction to the final state-line, making sure to leave a vertical gap of at least two grid-spaces between the end point and the start of the anti-electron-neutrino. We then draw the remaining quarks directly from the initial to final state-lines, making sure that all three quarks take up three consecutive grid spaces, and are at least two grid spaces separated from any other particles. Examples of valid quark configurations are shown in \figref{fig:hadron_configurations}.\\

\centering
\includegraphics[width=0.4\linewidth]{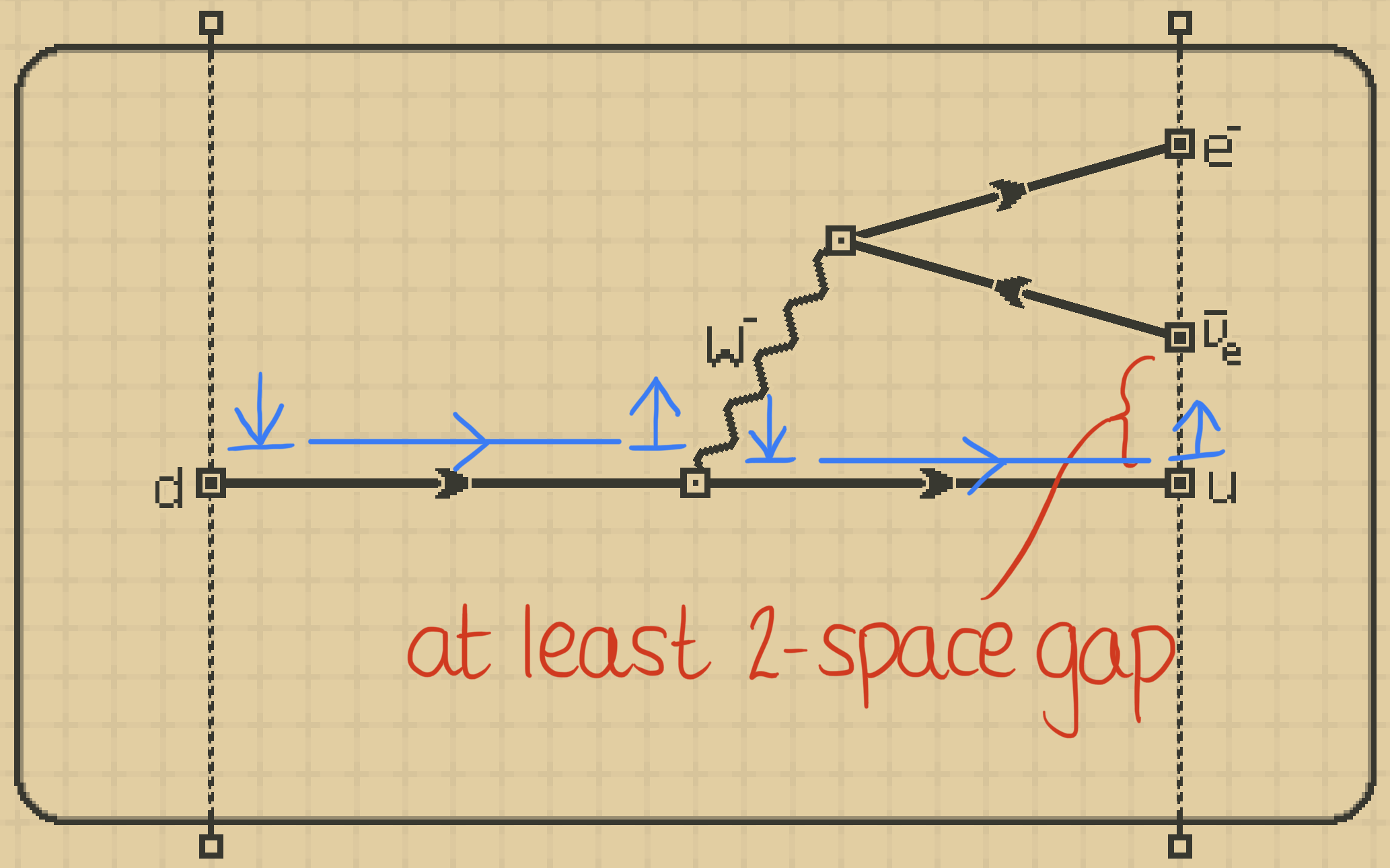}
\includegraphics[width=0.4\linewidth]{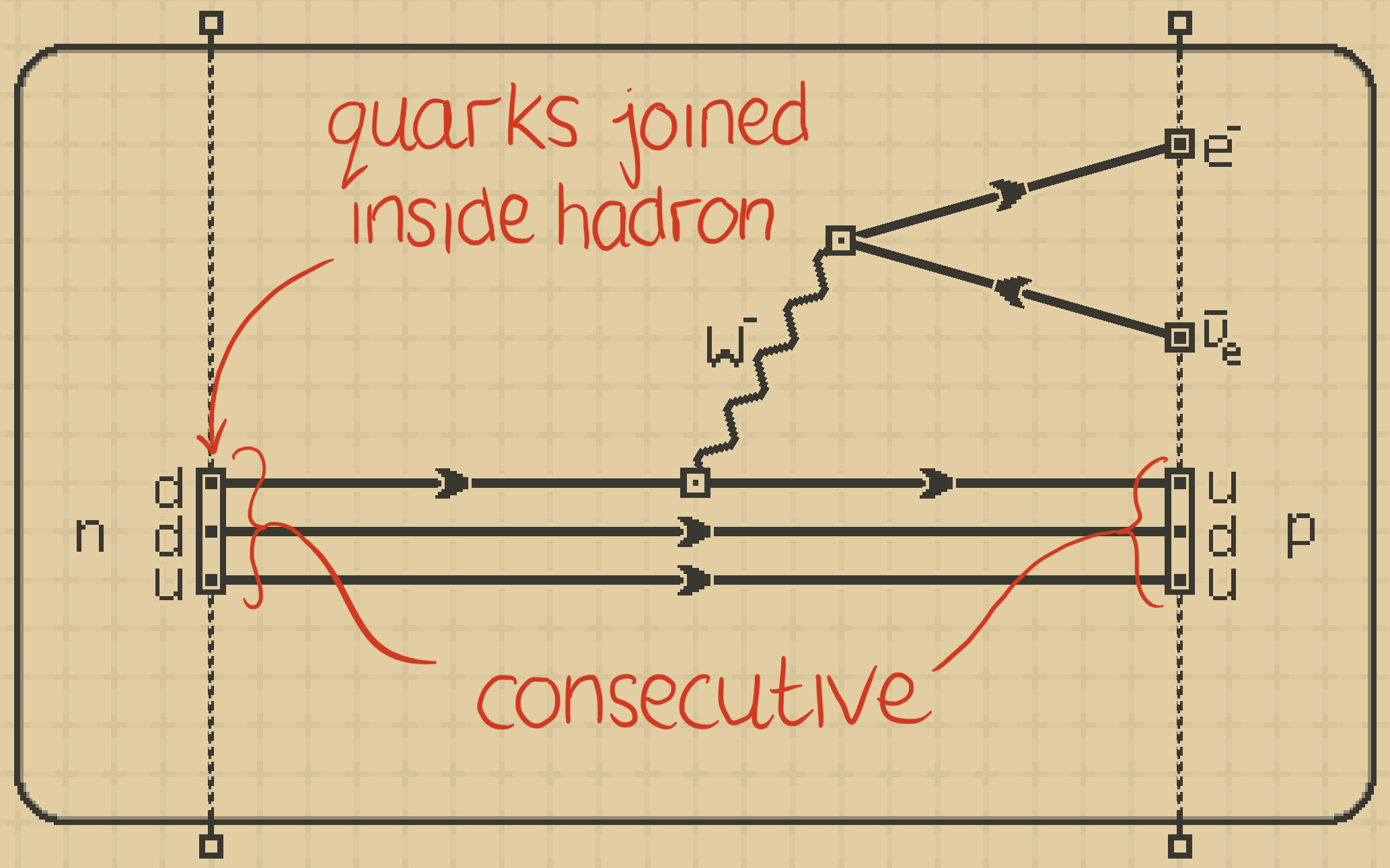}

\end{tcolorbox}

Say we placed a particle incorrectly or wanted to modify our drawn diagram. By holding \lstinline{X} on the keyboard and clicking on either a particle or interaction we can delete it; a deleted interaction will also delete any particles connected to it. By holding \lstinline{W}, we can click and drag to move an interaction, which will also move any connected particles. By instead holding \lstinline{SHIFT}, we can click and drag on an interaction to `split' it apart. This will choose one particle out of those connected to the interaction to move, leaving the other connected particles behind. We can also press \lstinline!C! to clear the diagram, and \lstinline!CTRL-Z! or \lstinline!CTRL-Y! to undo and redo changes, respectively. Touchscreen users are given access to the delete, drag and split functionality via dedicated buttons in the bottom left of the screen, which appear only when \FeynCraft{} detects that the device is a touchscreen device such as a tablet or phone. Such users can also access the undo, redo, and clear functionality via the controls tab \includegraphics[scale=0.8]{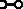}, where this tab also functions as a reminder for the keyboard commands for these functions outside of touchscreen mode.

Particle lines can be split into two segments. This happens if, when drawing a line, one ends up with an interaction intersecting a particle line. This could occur by either drawing a particle line across an existing interaction, or by drawing either the start or end of a particle on an existing particle line. Conversely, deleting a particle line that is connected to two particles, where those two particles are identical and the particle lines are in the same direction but on opposite sides of the interaction, will delete the interaction and join the two line segments together into a single particle. These functions are handy when drawing diagrams -- for example if we want to draw a photon emission off an electron line, we can simply draw a long electron line that intersects at least one of the grid points somewhere in the middle, and then draw a photon line starting from (or ending on) that grid point on the electron line, rather than laboriously attaching together three particles to create the $[e,e,\gamma]$ vertex.

As an interaction with more than one connected particle cannot exist on a state-line, there are a few drawing limitations to prevent this. In particular, one cannot draw or move a particle to an existing state-line interaction, and cannot move an interaction with more than one connected particle to a state-line.\\

\begin{figure}
    \centering
    \includegraphics[width=0.5\linewidth]{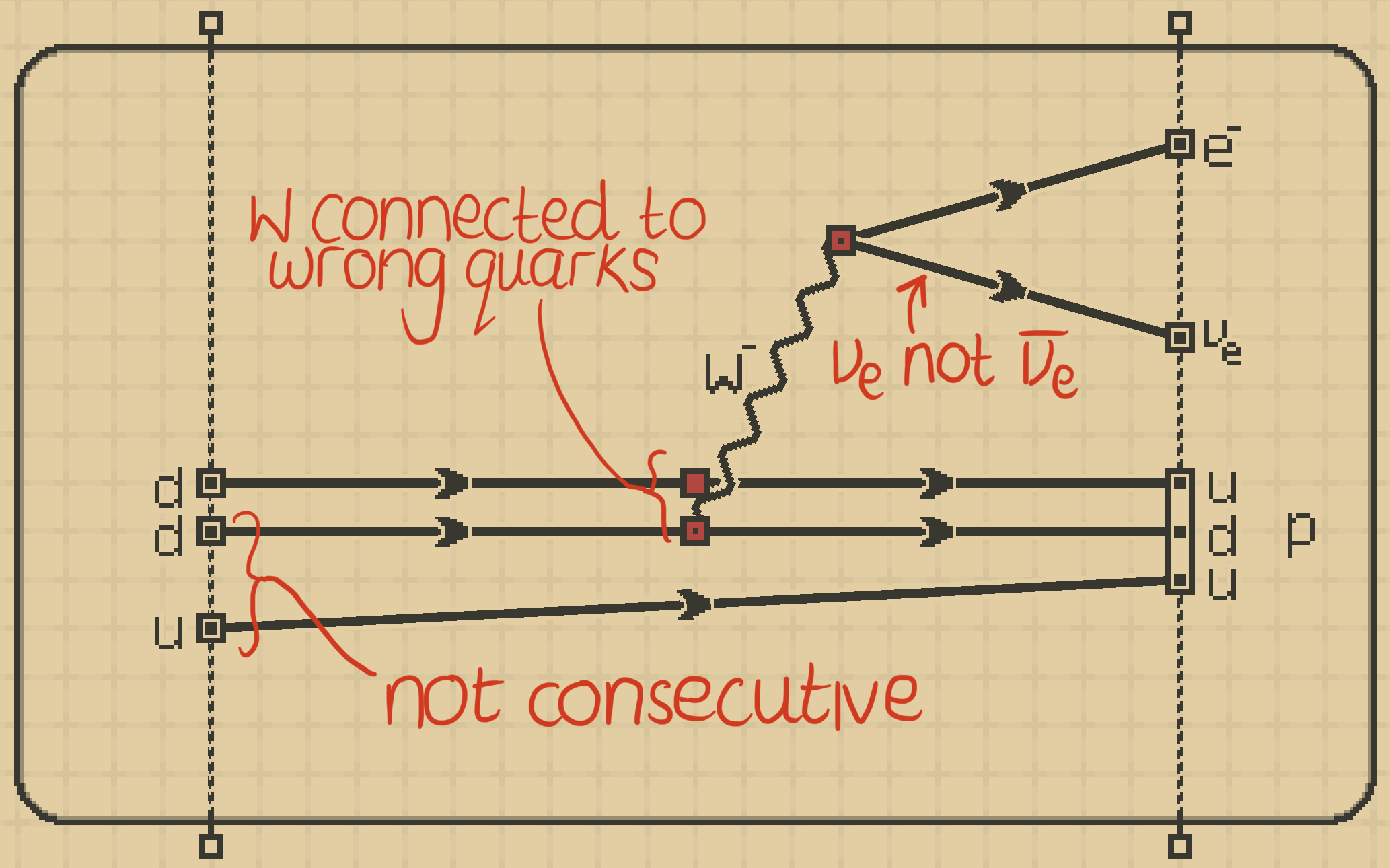}
    \caption{An invalid diagram to demonstrate the controls to edit drawn diagrams.}
    \label{fig:misdrawn_diagram}
\end{figure}

\begin{tcolorbox}
Let us say that we misplaced a few particles when attempting to draw the above example, leaving us with the diagram shown in \figref{fig:misdrawn_diagram}. We have drawn an electron-neutrino instead of an anti-electron-neutrino, an up-quark on the initial state-line such that it is not consecutive with the other quarks, and connected a $W^-$ to the wrong quarks. Let us start correcting this by holding \lstinline{X} and clicking on the electron-neutrino line to delete it. We can then draw it the correct way round, by selecting the $\nu_e$ in the lepton tab and dragging right-to-left. To fix the misplaced up-quark, we hold \lstinline{W} to click and drag its interaction on the initial state-line upwards, such that it is one grid space away from the other quarks. Finally, we can hold \lstinline{SHIFT} to click and drag the $W^-$ away from where it is connected to the up-quarks. When we do this, the two up-quarks will join into a single up-quark line (dropping the $W^-$ in the same place would split the up-quarks into two line segments again). We then move this interaction to where we see the down and up-quarks connected.\\

\centering
\includegraphics[width=0.5\linewidth]{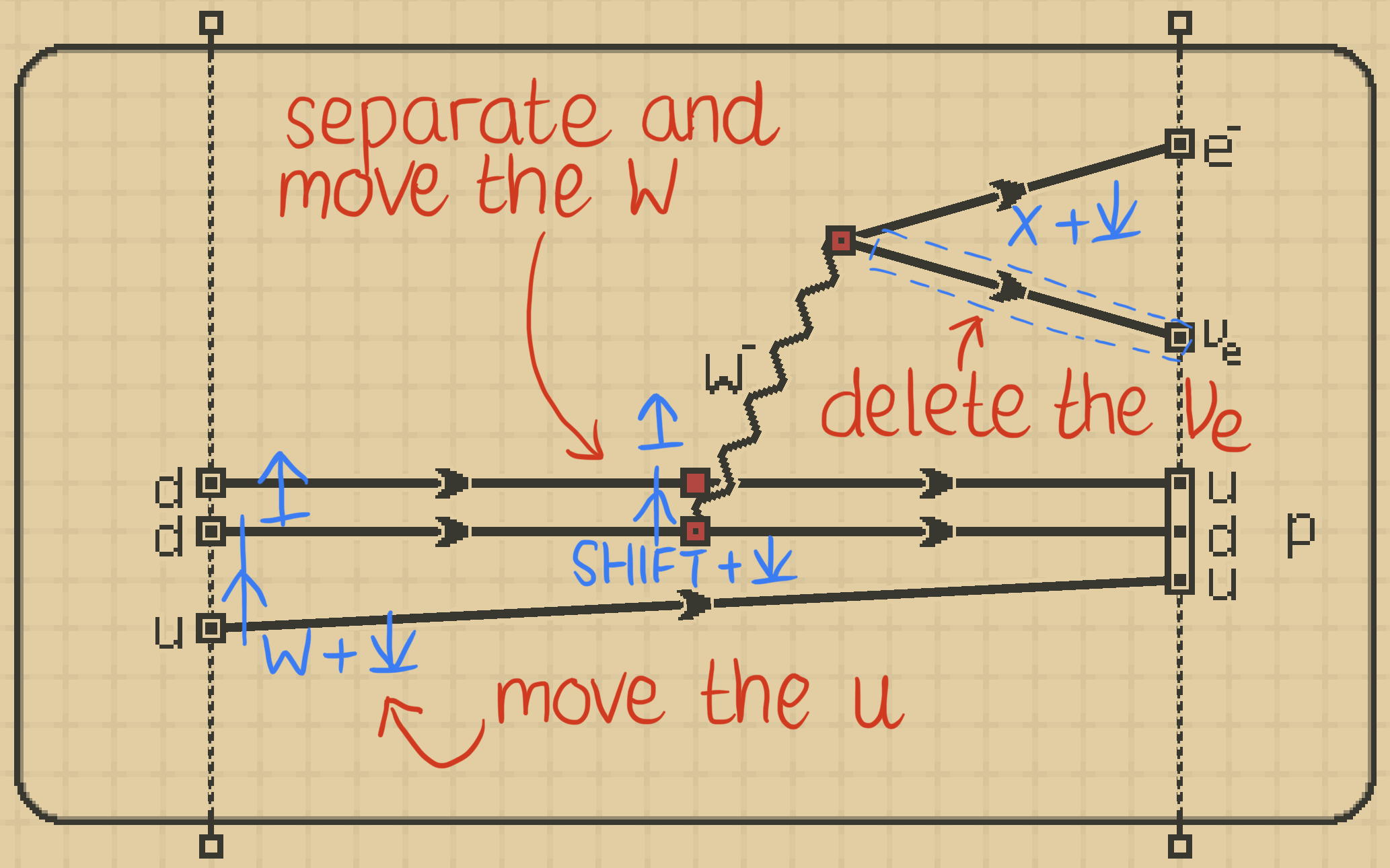}

\end{tcolorbox}

\section{Validation}
\label{sec:validation}
A diagram can be invalid on the level of an interaction, or on the level of the diagram as a whole; we'll discuss both possibilities in this section. 

\subsection{Interaction Validation}

Undergraduates typically learn the allowed vertices of the Standard Model by simply rote learning the list of vertices, which can be challenging (especially when trying to learn the plethora of vertices allowed in the electroweak sector). By contrast, \FeynCraft{} does not compare against an interaction against a list of known SM vertices. Instead, an interaction is passed through a series of checks to confirm its validity. The user can see the result of these checks by clicking on a vertex, in order that they may see specifically why a particular interaction they have drawn has been marked as invalid, and how to rectify it. As mentioned in the introduction, students may find this list of checks useful even outside \FeynCraft{}, for example to systematically and quickly check a vertex they have drawn in an exam is valid.

Most of the checks relate to the conservation of some quantum number. For those checks, we generally split the connected particles into those `before' the vertex on the left, and those `after' the vertex on the right, comparing the quantum numbers of the particles before to those after. As mentioned before, the concept of `before' and `after' is not in fact particularly meaningful for vertices inside a Feynman diagram, and all that fundamentally matters is the net flow of some quantum number into or out of a vertex (to illustrate this: moving a valid vertex in a Feynman diagram around to create an equivalent diagram can actually change the individual `before' and `after' numbers, but not the equality of the before and after numbers). However, we still present the check in this way as we believe that it is easiest for students to see the balance of quantum numbers at the vertex when presented like this. Given this presentation, one needs to make a prescription for particles entering the vertex in a vertical direction. We treat these as if the line has been tilted in the direction that yields a base particle traveling forwards in time. For example, let us say we have a vertex with an electron line entering vertically, with the fermion line pointing into the vertex. Since the electron is the base particle, we  treat this as if there is an electron line entering the vertex from the left hand side, and add the electron quantum numbers to the `before' side. Conversely, if the fermion line is pointing away from the vertex, we treat it as an electron leaving the vertex from the right hand side, adding the electron quantum numbers to the `after' side.

The most basic checks that are done are \textbf{conservation of electric charge, total lepton number and total quark number}. These quantities are always preserved, even in the presence of the $W$ boson which is the only boson able to change the identities of particles (it can change quarks into quarks of other types, and convert charged leptons into neutrinos). Further, since $W$ bosons can only convert charged leptons into neutrinos of the same family, and can never convert between lepton families, we can also check for \textbf{conservation of lepton family number}. In \FeynCraft{} the shorthand used for this family number is `electron/muon/tau num.', such that for example the `electron num.' value counts the number of electrons plus electron neutrinos (with antiparticles of these counting as $-1$, of course). \textbf{When the vertex does not involve the $W$, we also check for conservation of the numbers of individual quark types} i.e. conservation of the number of up quarks, down quarks, etc.

We now introduce a further rule limiting the number of particles that can attach together at the vertex. We can associate mass dimensions to all of the particle fields (note we use natural units here, as is typical in particle physics) -- in particular we associate a dimension of $3/2$ for fermions and $1$ to the bosons. Then, the rule is that \textbf{the sum of the mass dimensions of all particles attaching to the vertex must be $\le 4$}. We give a rough overview of why this is the case in the SM below; however, the explanation is best understood by students who have started doing some quantum field theory, so students below that level may wish to skip the explanation in the paragraphs below and simply take this as a rule that applies. Much of the discussion below is adapted from Section 3 of \cite{TongQFT}.

\vspace{2mm}

The fundamental `starting point' of a quantum field theory, from which all of its properties, particles and interactions can be derived, is the Lagrangian of that quantum field theory, or rather the Lagrangian density $\mathcal{L}(x)$, since we have fields that are a function of spatial position and time. The action of the quantum field theory is obtained by integrating this Lagrangian over spacetime, $S = \int d^4 x \mathcal{L}$, and since $S$ is dimensionless (in natural units), this implies that $\mathcal{L}$ has mass dimensions of $4$. The Lagrangian contains kinetic terms, from which one may deduce that the mass dimensions of the fermions are $3/2$ and the bosons are $1$, and then interaction terms. Each interaction term yields a vertex in the Feynman diagram, and is composed from a product of particle fields (where this product of fields determines what particles connect to the vertex in the Feynman diagram), some coupling factor associated with the vertex, and potentially also derivatives. Each derivative has mass dimension $+1$, and these derivatives lead to factors of momentum in the Feynman diagram vertex expression. Let us consider an interaction term that has a product of fields and derivatives with dimension $+4$ -- then the coupling constant associated with that vertex is just a dimensionless number with no associated energy scale, and the probability amplitude for the interaction associated with that vertex to occur (or dimensionless `strength' of that interaction) is the same whether we are at low or high energy scales. 

On the other hand, let us consider an interaction where the product of mass dimensions for the fields is $> 4$. Then the associated coupling constant $g$ must necessarily have some (fixed) energy scale associated with it, $\Lambda$, where this energy scale must be raised to a negative power to ensure that the overall dimension of the interaction term is $4$, $g \sim 1/\Lambda^\Delta$, $\Delta > 0$. In this case, the dimensionless strength of the interaction, if the particles taking part in the interaction are at an energy scale $E$, has to be $(E/\Lambda)^\Delta$. In this case, the interaction strength grows with energy, and one starts to experience issues with the theory as $E \gtrsim \Lambda$ -- for example, at high energies the coupling can become so strong that we break the unitarity limit (conservation of probability). Such theories cannot be fundamental, but must be considered to be some low energy effective description of some (more) fundamental theory, that has some additional dynamics to it at the mass scale $\Lambda$ (e.g. additional particles with masses $\sim \Lambda$). An example of such a theory is Fermi theory, where the four fermion $[n,p,e,\nu_e]$ interaction has dimension $6$, and this is only an effective low energy description of the full weak decay, where here the additional particle involved is the $W$ boson, $\Lambda \sim M_W$.

An inclusion of terms with dimension $>4$ in the SM Lagrangian would thus necessarily imply that this theory is incomplete, and that some additional particles or dynamics lurks at the scale $\Lambda$. Since we have not yet seen evidence for these new particles (up to the usual caveats), we do not include these terms in the SM Lagrangian. It is worth noting, however, that because the SM Lagrangian + additional terms with dimension $>4$ (the so-called Standard Model Effective Field Theory, or SMEFT) is a suitable effective field theory for heavy new physics at the mass scale $\Lambda$, this is a framework that can be used, and is being used, to search for heavy new physics at the LHC (see e.g.~\cite{Brivio:2019ius, Ellis:2020unq, ATLAS:2022xyx, Celada:2024mcf, CMS:2025ugn}).

An alternative way round to think of this is: let us imagine that there is a theory appropriate to a high energy regime $E \lesssim \Lambda$, with all of these dimension $>4$ interactions included. As we reduce the energy scale of interactions, the strength of the dimension $>4$ interactions reduces, until they become totally irrelevant and at sufficiently low energies we are left with the SM Lagrangian (of course the definition of sufficiently low depends on $\Lambda$, and can be rather large if $\Lambda$ is large enough!).

\vspace{2mm}

We now discuss some checks that are particular to the different types of interaction, or force-carrying boson. Since the photon couples to electric charge, \textbf{a vertex with a photon attached must also have electrically charged particles attached to it}. This check is referred to as `photon from neutral?' in \FeynCraft{}, and if the check returns `yes' then the vertex is forbidden. For the strong force, \textbf{an interaction is forbidden if it connects one or more gluons to one or more colourless particles, or if it has a single gluon connected, and that gluon is found to be colourless}. The former is forbidden since the colourless particles are (by definition) not able to couple to the gluon(s), and the latter is connected to the key property of QCD that the force-carrying gluons themselves must carry a colour charge. We mark a gluon as colourless if it couples into a `colour isolated' sub-graph, for which the colour injected by the gluon cannot escape (either out of the diagram via a state interaction, or back through the other end of the attached gluon). In this case the gluon is forced to be colourless by colour conservation. Some explicit examples of such `colour isolated' systems are given in section \ref{sec:algorithms-vision}. All of these checks on the gluon vertex are referred to as `gluon from colourless?' in \FeynCraft{}, and if the check returns `yes' then the vertex is forbidden.

\textbf{For interactions involving the $Z$, but not involving the Higgs, the relevant rule is that at least some of the other particles involved in the interaction must be able to have a weak `charge'}. We shall refer to this weak charge later as `shade', and in technical terms this weak charge is the weak isospin $T_3$. Particles that can hold this weak charge are the ones involved in the charged-current weak interactions: all the fermions, and the $W$ bosons themselves. The $Z$ boson in fact couples to a combination of $T_3$ and the electric charge $Q$, and so can interact with particles that have a zero $T_3$ but nonzero $Q$ (which is the case for the right-handed versions of the fermions). However, there is no particle that has a nonzero $Q$ and always has a zero $T_3$, so this simple check works to determine if an interaction involving $Z$ bosons (but not a Higgs) is allowed. This check is referred to as `$Z$ from shadeless?', with `yes' corresponding to the vertex being forbidden.

Finally, we have two checks that are done if the vertex involves a Higgs boson. \textbf{For a vertex involving a Higgs, all particles involved must be massive} -- the strength of the Higgs interaction with a particle is proportional to that particle's mass, so massless particles do not couple to the Higgs. In \FeynCraft{} we neglect the tiny masses of the neutrinos and thus have no coupling between the Higgs and the neutrino (although experimentally we know that at least some of the neutrinos have masses (see e.g.~\cite{BILENKY2003395}), the mechanism by which neutrinos acquire mass is unknown, so we refrain from including these Higgs-neutrino couplings). This check is referred to as `Higgs from massless', with the result `yes' corresponding to a forbidden vertex. \textbf{The other important property that a valid vertex with Higgs particles must satisfy, is that if you were to remove all the attached Higgs boson(s) from the vertex, one should be left with either nothing, or a two-particle degree-0 vertex corresponding to a particle simply propagating along.} The Higgs particle is the only SM particle with a non-zero vacuum expectation value (vev), and for every interaction vertex with Higgs particles, there is an equivalent where those particles are replaced by their constant vev. The interactions with the vev should be responsible for giving the particles their masses (recall the famous analogy of the Higgs vev to a syrup, filling the universe and slowing all the particles down) -- thus the nature of these interactions should be the vev interacting with a particle simply propagating along (and `slowing it down'). Thus, when we have a vertex involving the Higgs and remove the Higgs particles, we should just get a particle propagating along. The exception to this in when we are looking at the pure Higgs self-interactions, where if we remove all the Higgs bosons from the vertex there is nothing left. This check is referred to as `H-less is particle/0?' (i.e. do we get a particle line or nothing if we remove the Higgs), and if the check returns `yes' then the vertex is allowed.

To summarize, the set of rules for a valid vertex are:
\begin{enumerate}
    \item Electric charge, lepton number, and total quark number must be conserved.
    \item Lepton family number must be conserved.
    \item For vertices not involving the $W$: The numbers of individual quark types must be conserved.
    \item The sum of the dimensions of all fields attached to the vertex must be $\le 4$.
    \item For vertices involving the photon: there must be other electrically charged particles attached to the vertex.
    \item For vertices involving the gluon: gluons cannot be emitted from a vertex that has no other coloured particles attached, and a lone gluon cannot be emitted/absorbed by a `colourless' subgraph in a Feynman graph, which has no coloured attachments to the rest of the graph other than that lone gluon.
    \item For vertices involving the $Z$ but not the $H$: there must be other particles attached to the vertex that are able to carry the weak charge (`shade', or $T_3$).
    \item For vertices involving the $H$: all of the particles attaching to the vertex must be massive.
    \item For vertices involving the $H$: removing the $H$ bosons from the vertex should either yield nothing, or a two-particle degree-0 vertex corresponding to the particle simply propagating along.
    
\end{enumerate}

\begin{figure}
    \centering
    \includegraphics[width=0.4\linewidth]{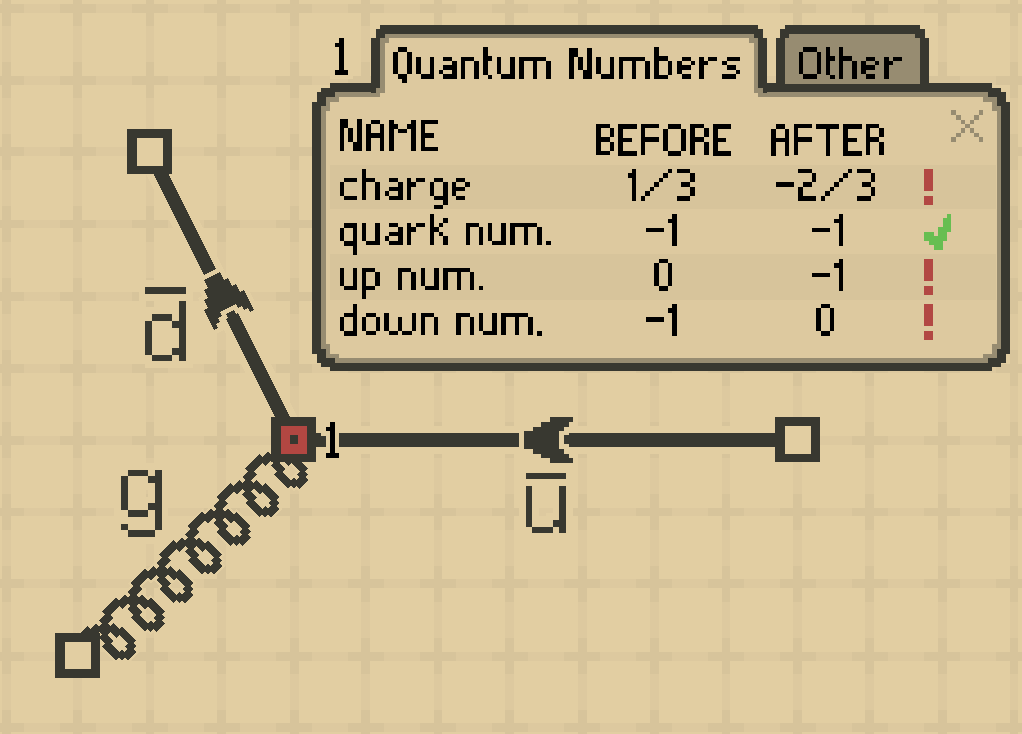}
    \includegraphics[width=0.4\linewidth]{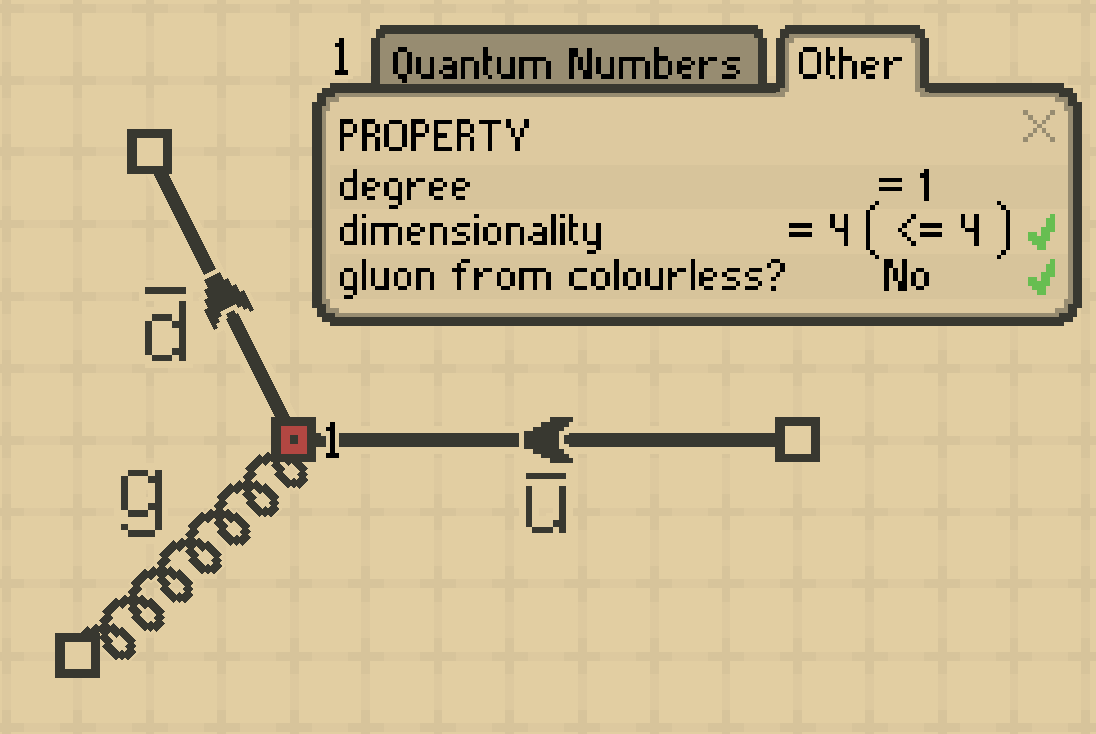}
    \caption{An example of the validation information for an interaction. Quantum number information on the left, in the first tab, and other information on the right, in the second tab. Invalid checks are shown by an exclamation mark.}
    \label{fig:interaction_information}
\end{figure}

As mentioned before, invalid interactions are highlighted red. When an interaction is clicked on, an information box will appear that displays the checks and lets the user know which have failed, an example of this is shown in \figref{fig:interaction_information}. The left (\lstinline!BEFORE!) and right (\lstinline!AFTER!) quantum number totals that are relevant to the interaction are shown in the first tab of the box: charge is always shown, lepton and quark numbers are shown if there are leptons or quarks, respectively, and lepton family numbers and individual quark numbers are shown when those lepton families or quarks are connected and when those numbers should be conserved. The second tab will show the other validation checks; again only the relevant ones are shown. For example, the colourless gluon check will only show when there is a gluon connected to the interaction, and the Higgs check only displays if there is a Higgs connected. Using this information, a user will be able to diagnose why the interaction they have drawn is invalid.

\subsection{Diagram Validation}
There are various reasons why a diagram can also be invalid on the level of the whole diagram, even if the individual interactions are all valid. We require that valid diagrams be \textit{fully connected}, such that all particles on the state lines are connected to all other particles via internal lines in the Feynman diagram, and there are no internal loops inside the Feynman diagram that are disconnected from the rest of the diagram (`vacuum bubbles'). This is the appropriate prescription when we are computing the decay rate of one particle to several, or the scattering rate of two (stable) incoming particles to one or more particles (see e.g. \cite{Peskin:1995ev}). Vacuum bubbles are not linked to the decay/scattering, but rather are associated with the definition of the vacuum. Likewise, diagrams in which some subset of the state particles are not connected to the rest of the state particles also do not correspond to the scattering/decay process (they might for example correspond to the initial state particles propagating to the final state unscattered, or they might be kinematically forbidden if they have a subset of final state particles only connected to each other). Of course, one also cannot have `dangling' one-particle interactions inside a Feynman diagram, which corresponds to a particle simply appearing or disappearing from nothing -- we also count this scenario as the Feynman diagram being disconnected.

A diagram may also be forbidden if energy cannot be conserved between the initial and the final state (we assume that the state particles are all physical with positive energy, as will of course be the case in a real scattering process). Of course if there are only particles in the initial state, or only particles in the final state, energy cannot be conserved. Another scenario where energy cannot be conserved is when we have one particle that decays into multiple particles, and the sum of the masses of the final-state particles exceeds the mass of the initial-state particle. This is straightforward to verify by boosting into the rest frame of the decay particle. A similar statement holds for the time reversed process where multiple particles `fuse' into one.

\begin{figure}
    \centering
    \includegraphics[width=0.4\linewidth]{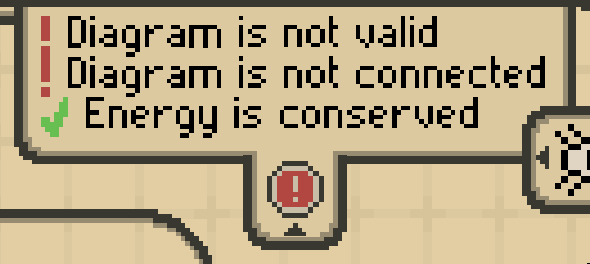}
    \caption{The health tab, showing diagram validation information. The exclamation mark is replaced by a green tick if all checks are passed.}
    \label{fig:health_tab}
\end{figure}

In \FeynCraft{} we check if the diagram is disconnected and if the diagram corresponds to one of the scenarios above where energy cannot be conserved. If one of these issues is detected, then it is flagged up in the `health tab' illustrated in \figref{fig:health_tab}. The health tab also keeps track of whether all vertices in the diagram are allowed (`Diagram is valid/not valid'). If all interaction and diagram checks are passed, the health tab icon is a check mark in a green circle, but if any checks are failed, it becomes an exclamation mark in a red circle (as shown in \figref{fig:health_tab}). This allows a user to see at a glance, even without opening the tab, if an issue has been detected with the diagram drawn.

\section{The Sandbox}
\label{sec:sandbox}
Inside the sandbox, the user has access to all features of \FeynCraft{}, and we shall go through them in the order that a user may encounter them. We start in section \ref{sec:sandbox-problems} with how to create and solve an random problem where the user has to draw Feynman diagrams for a given process. Then, in section \ref{sec:sandbox-generating_solutions}, we explain how to generate and view the Feynman diagrams for a specified initial and final state. In section \ref{sec:vision_tab} we introduce the `vision overlays' that provide additional information about a drawn Feynman diagram. Finally in section \ref{sec:exporting_diagrams} we describe how to export drawn diagrams into \LaTeX{} code.

\subsection{Creating and solving a problem}
\label{sec:sandbox-problems}

\begin{figure}
    \centering
    \includegraphics[width=0.4\linewidth]{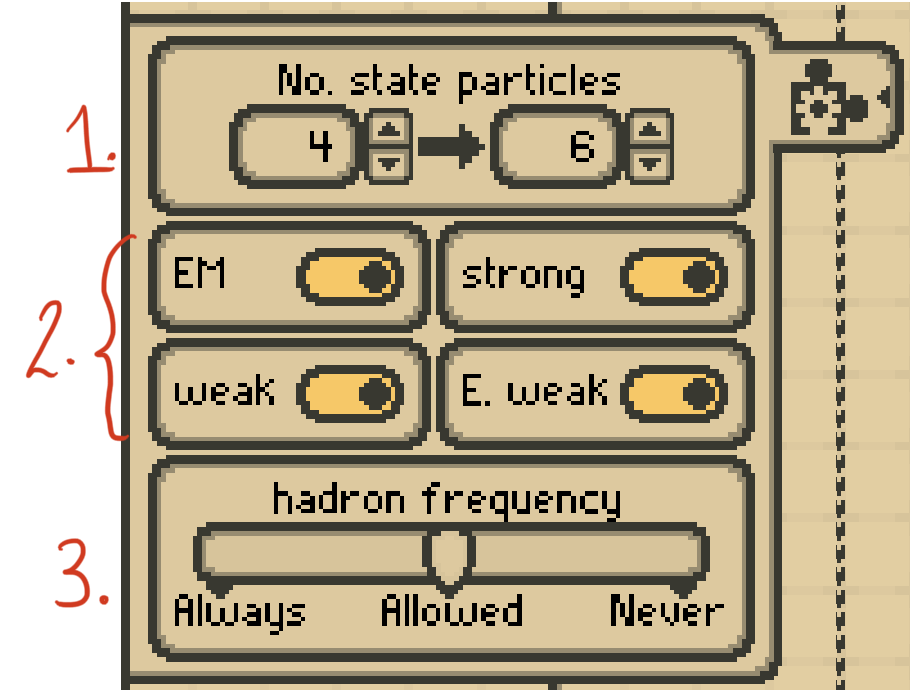}
    \caption{The problem options tab, which can be used to  configure problem generation.}
    \label{fig:problem_options_tab}
\end{figure}

In the sandbox, problems can be randomly generated. We can control this generation using the problem options, found in the problem options tab \includegraphics[scale = 0.6]{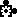}, shown in \figref{fig:problem_options_tab}. From top-to-bottom these options are:

\begin{enumerate}
    \item \textit{Number of state particles}: Sets the minimum and maximum total number of state particles the problem can have, spread randomly on either the initial or final state. For example, let us say that we restrict this to $4$; then we may have problems with one initial-state particle and three final state particles, two initial and two final, or three initial and one final.
    \item \textit{Fundamental force toggles}: Alters which particles and forces may appear in the problem and solution. Enabling \lstinline!electromagnetic! enables the photon, charged leptons, and quarks; \lstinline!strong! enables the gluon and quarks; \lstinline!weak! enables the $W$ boson and fermions; and finally \lstinline!electroweak! enables the $Z$, Higgs boson, and fermions, which cannot be enabled unless both \lstinline!electromagnetic! and \lstinline!weak! are enabled. Particles that are not allowed in the solution are `greyed out' and cannot be clicked when the problem is eventually generated. (If one wishes to re-enable all the particles  afterwards for the other sandbox functionalities, one should simply turn all the force toggles here to `on', and generate a further problem using the `next problem' button in the problem tab, as discussed below.)
    \item \textit{Hadron frequency}: Controls how often problems will have at least one hadron: \lstinline{never} and \lstinline!always! will either disallow or force the appearance of hadrons, respectively, whereas \lstinline!allowed! causes hadrons to only in some problems, see \ref{sec:algorithm-problem_generation}.
\end{enumerate}

\begin{tcolorbox}
Let us follow the process of settings these options with an example. Say are trying to generate problems similar to that seen in section \ref{sec:drawing}, neutron decay. In that case, there are four state particles in total but we don't mind this changing slightly, so let us set the minimum to three and the maximum to five by either typing or pressing the arrows in the state particle count boxes. Now for the force toggles: let us for now enable only the \lstinline!weak! and \lstinline!strong! toggles to focus on these two interactions. Finally, as we only want problems containing hadrons, we set the hadron frequency to \lstinline!always!. Let us leave these options here as we discuss the problem tab.
\end{tcolorbox}

\begin{figure}
    \centering
    \includegraphics[width=0.75\linewidth]{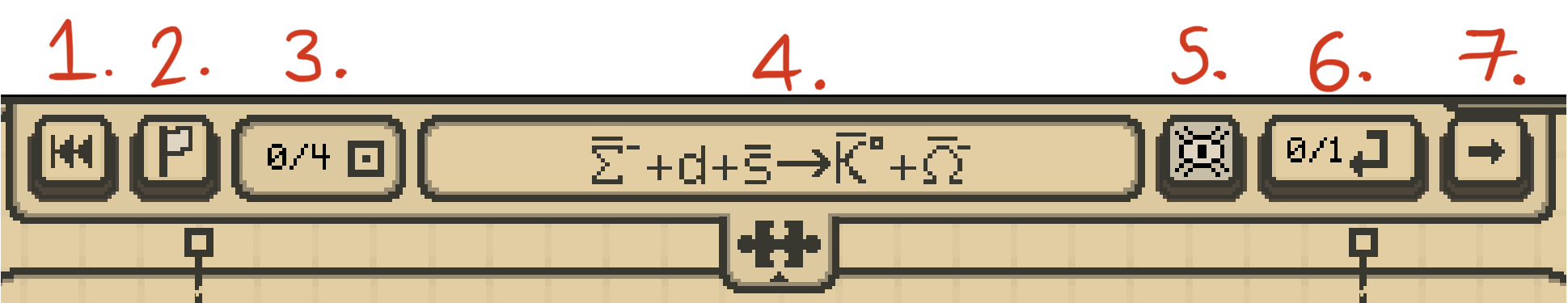}
    \caption{The problem tab, labelled, showing an example problem.}
    \label{fig:problem_tab}
\end{figure}

Whilst the settings for problem generation are accessed via the problem options tab \includegraphics[scale = 0.6]{icons/problem_options.png}, the generation of problems themselves, as well as the submission of solutions, is done via the problem tab \includegraphics[scale = 0.6]{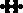}, illustrated in \figref{fig:problem_tab}. The elements of this tab, from left-to-right (and numbered in \figref{fig:problem_tab}) are:
\begin{enumerate}
    \item \textit{Previous problem button}: Returns to the previous problem, only enabled when at least one previous problem has been generated and solved/skipped.
    \item \textit{Show solution button}: Generates and draws an example solution to the current problem. 
    \item \textit{Degree counter}: Shows the total degree of the currently drawn diagram, and the total degree required in solutions.
    \item \textit{Problem equation}: Where the current initial to final state problem is shown. Hovering over hadrons in the equation will show their quark content. This is empty to begin with in sandbox until we generate a first problem.
    \item \textit{Show submissions button}: Shows all previously submitted solutions inside of a `mini-diagram viewer', from which users can view, load, delete, change and re-save past submissions. More details on this mini viewer are given in section \ref{sec:sandbox-viewing_submissions}.
    \item \textit{Submit button}: Shows the current number of submitted solutions, and the total required to complete the problem. Clicking the submit button will attempt to submit the currently drawn diagram and, if accepted, add it to the submitted solutions. If it is not accepted, submission feedback will be displayed to show why.
    \item \textit{Next problem}: Generates a new problem according to the rules specified in the problem options tab; this is both used to generate the first problem and also to move on to subsequent problems. In sandbox mode one may use this button to `skip' problems without entering all requested solutions, but in other modes the button may be disabled until the user fully completes the problem by submitting the required number of solutions.
\end{enumerate}

To generate the first problem to solve, one thus uses the `next problem' button (button 7 in \figref{fig:problem_tab}). Here, \FeynCraft{} always asks for the lowest order solutions (i.e. with smallest possible total degree) to a given problem, with the degree required being shown in the degree counter box (box 3 in \figref{fig:problem_tab}). If there are $\le 4$ diagrams for the process at the lowest order, then \FeynCraft{} asks the user to find them all, otherwise it simply asks them to find $4$ diagrams (this is designed to avoid creating problems that are too laborious). The exceptions to this are problems involving hadrons, or problems where solutions at degree 5 or greater are required. In these cases, \FeynCraft{} only asks for one solution. In undergraduate courses, Feynman diagrams drawn with hadrons are used mostly for illustrative purposes to establish the forces/interactions that can be at play in various processes, and are not used for calculations (in any case the strong interactions drawn in such diagrams are typically at a low energy scale where the strong coupling constant is large, and perturbative calculations via Feynman diagrams can no longer be used for quantitative predictions). Generally Feynman diagram questions involving hadrons only ask students to draw one illustrative diagram, and we also only ask for this in \FeynCraft{}. For the case of problems which require degree 5 or higher, the diagrams involved are so complex that just finding and drawing one diagram will already be a non-trivial task.

 A submission is valid if it has the same initial and final states as the problem, it is not a duplicate of a previous submission, and passes the validation checks discussed in section \ref{sec:validation}. The method for checking if two diagrams are duplicate is somewhat non-trivial and is discussed in section \ref{sec:algorithms-duplicate_diagrams}.

\begin{tcolorbox}
In-order to generate our first hadron problem, we open the problem tab by clicking on it and then click on the next problem button. This would generate a problem at random, but for this example it gives us the problem

\begin{equation}
    K^+ + K^- \rightarrow J/\psi
\end{equation}

We can hover over each meson in the equation to get a reminder of their quark-contents, revealing that $K^+ = \{u, \overline{s}\}$, $K^- = \{\overline{u}, s\}$, and $J/\psi = \{c, \overline{c}\}$, which are the state particles we have to draw. Looking at the degree counter, we see that we will need to draw a diagram of degree-4, meaning a total of four interaction dots. From the counter on the submit button, we also see that we will only need to submit one valid solution. Using this information, we draw an example solution shown below, and after checking it is valid we submit it. Our submission is accepted and the submission button now shows 1/1, we have completed the problem.\\

\centering
\includegraphics[width=0.48\linewidth]{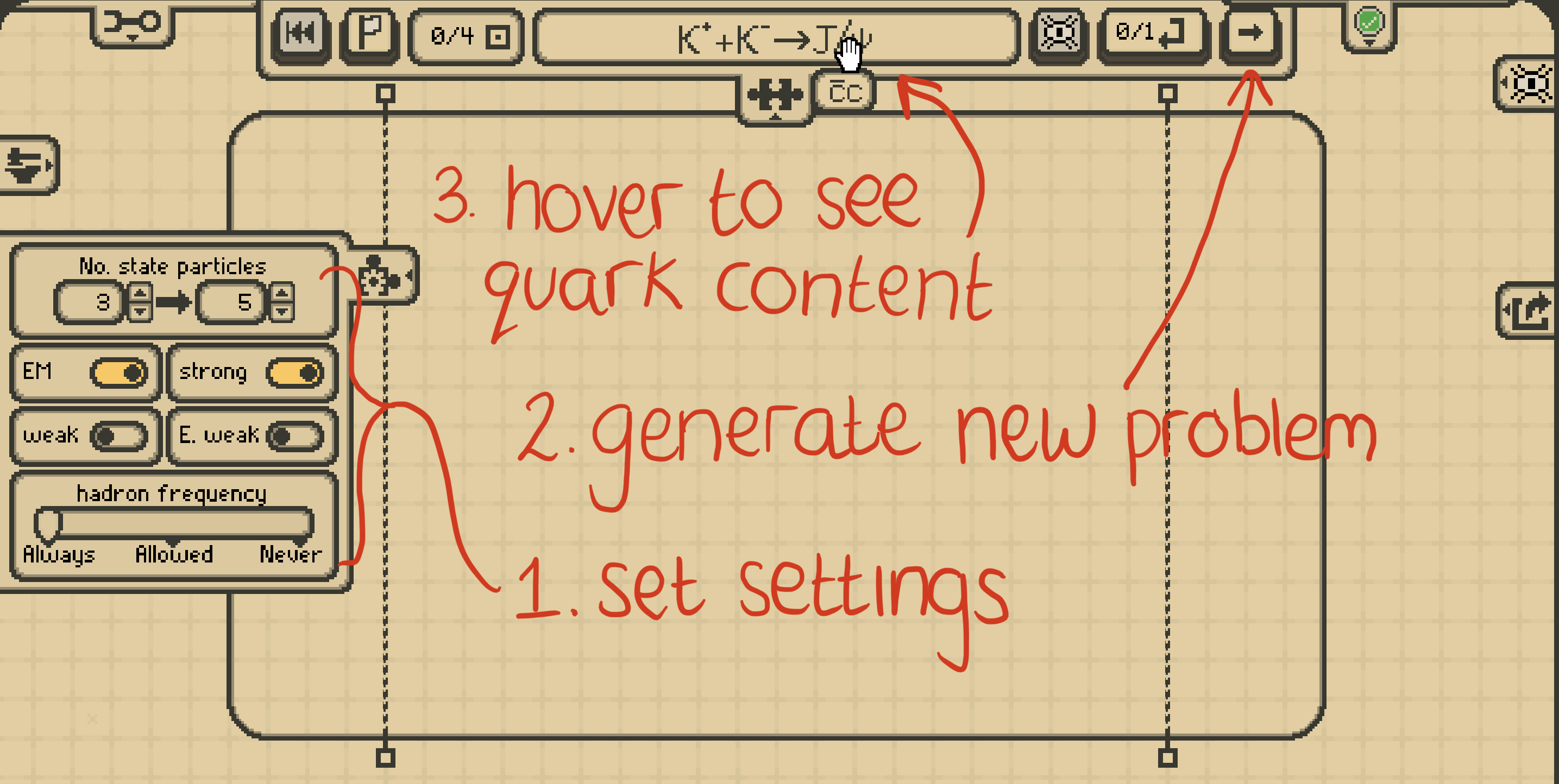}
\includegraphics[width=0.48\linewidth]{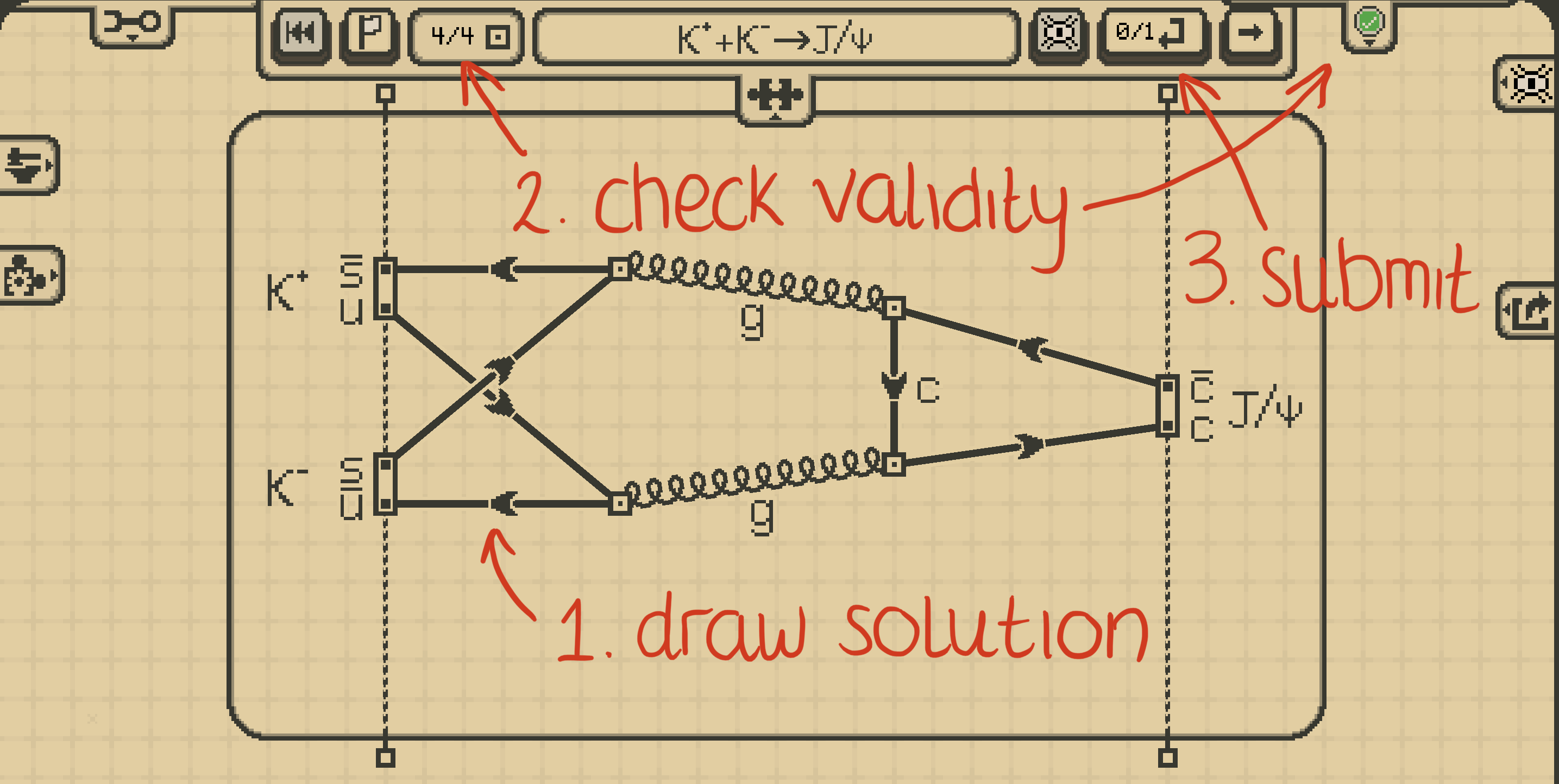}
\end{tcolorbox}

\subsubsection{Viewing submissions}
\label{sec:sandbox-viewing_submissions}

\begin{figure}
    \centering
    \includegraphics[width=0.53\linewidth]{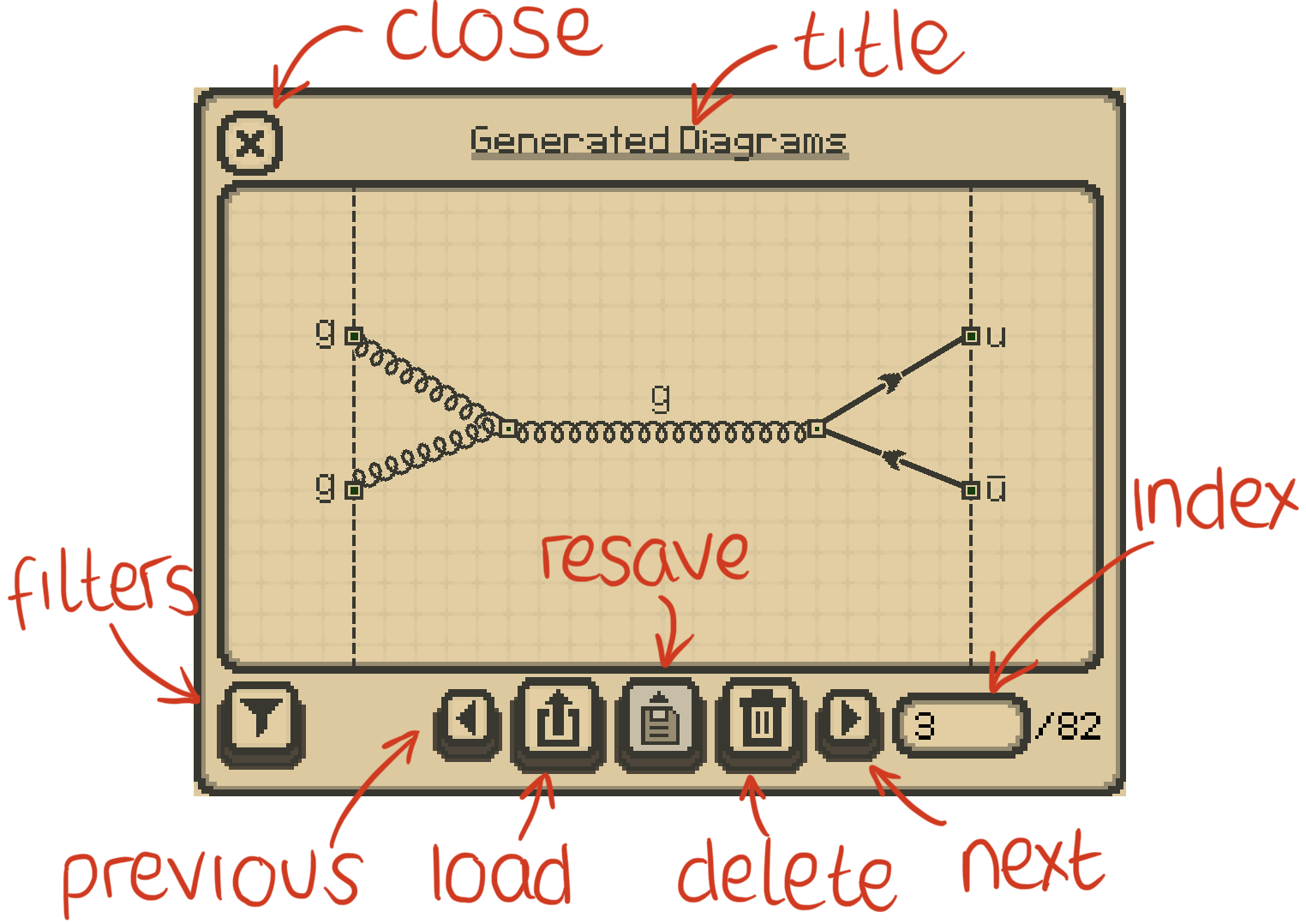}
    \includegraphics[width=0.3\linewidth]{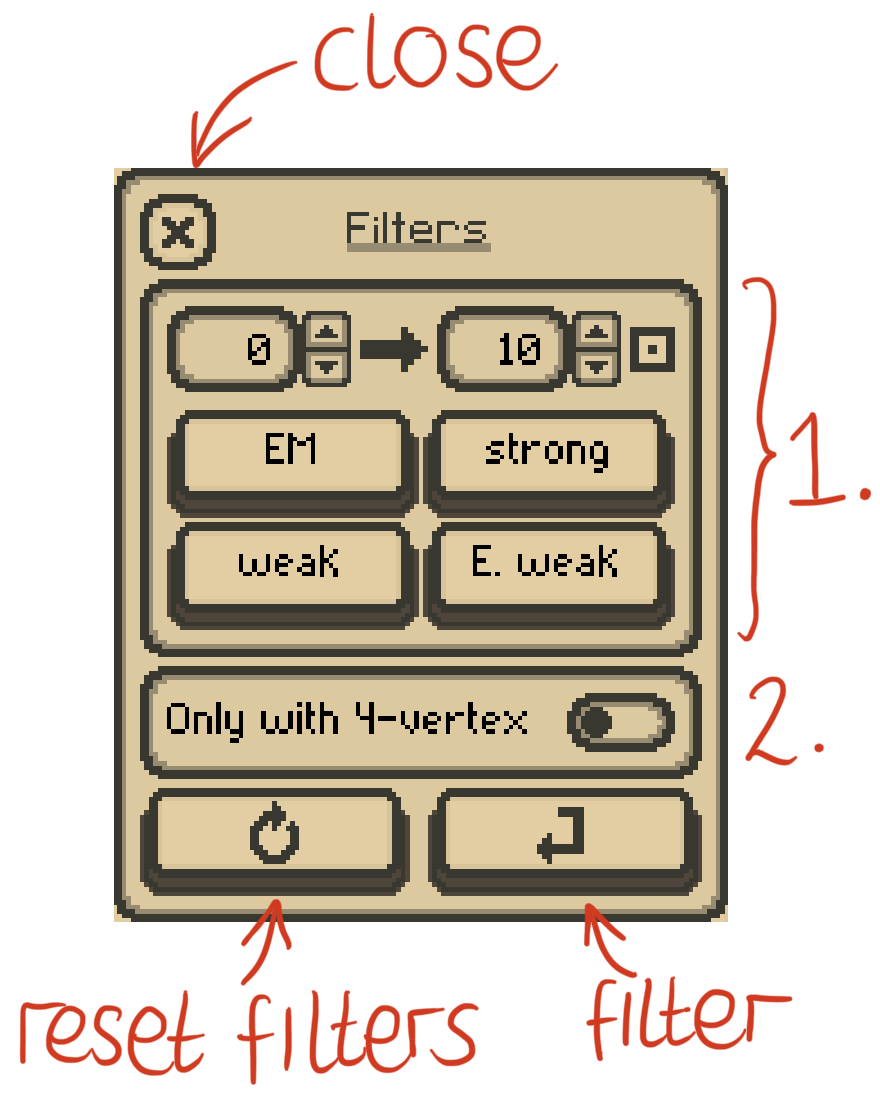}
    \caption{\textit{left}: the mini-diagram viewer, used to view submitted diagrams and those generated during solution generation. \textit{right}: filters that can be used to filter diagrams stored inside the mini-diagram viewer.}
    \label{fig:mini_diagram_viewer}
\end{figure}

When the view submissions button is pressed, button 5 in figure \figref{fig:problem_tab} (or when viewing the Feynman diagrams  \FeynCraft{} generates for a user-specified process, as discussed in section \ref{sec:sandbox-generating_solutions})
 the user is greeted with the `mini-diagram viewer', labelled in \figref{fig:mini_diagram_viewer}. Stored diagrams can be loaded into the main diagram window, deleted, or re-saved (replaced by the currently drawn main diagram). Re-saving is only allowed when the diagram drawn in the main window is a duplicate of (or technically, `topologically equivalent to a') current stored mini-diagram -- by this we mean that the `wiring' of connections in the diagrams is the same, just some of the interactions have been moved to different positions in one diagram compared to the other (see section \ref{sec:algorithms-duplicate_diagrams}). This re-saving functionality is useful when one wants to move around the vertices within a stored diagram to make it easier to read. It is especially useful when examining the diagrams generated by \FeynCraft{} itself (although \FeynCraft{} does attempt to generate diagrams that look broadly reasonable -- see section \ref{sec:algorithms-diagram_layout} -- the algorithms are not perfect, and a user may wish to `smarten up' \FeynCraft{}'s output using this re-saving function). One may also filter the diagrams stored in the mini-viewer via the `filters' button. The filtering options available are:
\begin{enumerate}
    \item \textit{Force filters}: Pressing one of the force buttons, and then specifying a range using the two numbers above the force buttons, limits the mini viewer to only show diagrams in which the interaction degree of that force is in the given range. For example, if one presses `strong' and then specifies the range to be $2 \to 2$, then one can only have diagrams in which we have two three-particle QCD interactions (each of these have degree 1), or one four-gluon vertex (with degree 2), with the other interactions not being restricted. Adjusting the degree range without a force selected will set the total degree range of the diagrams displayed in the mini viewer. The `EM', `weak' and `strong' buttons control the number of interactions involving the photon, $W$ and gluon respectively, whilst the `EW' button controls the number of interactions involving the $Z$ and $H$. Adjusting the degree range without a force selected will set the total degree range of the diagrams displayed. One can set multiple force filters at the same time -- one first selects a range for the first force, and then presses the button for the next force and selects the range for that (and so on); the choices for each force are remembered.
    \item \textit{Only with 4-vertex}: Only displays diagrams with at least one 4 particle vertex.
\end{enumerate}

Once filters are set they can be applied by pressing the `submit filters' button, and reset with the `reset' button. Filters are only temporary, and do not delete any stored diagrams.

\subsection{Generating the Feynman diagrams for a given process}
\label{sec:sandbox-generating_solutions}

\begin{figure}
    \centering
    \includegraphics[width=0.35\linewidth]{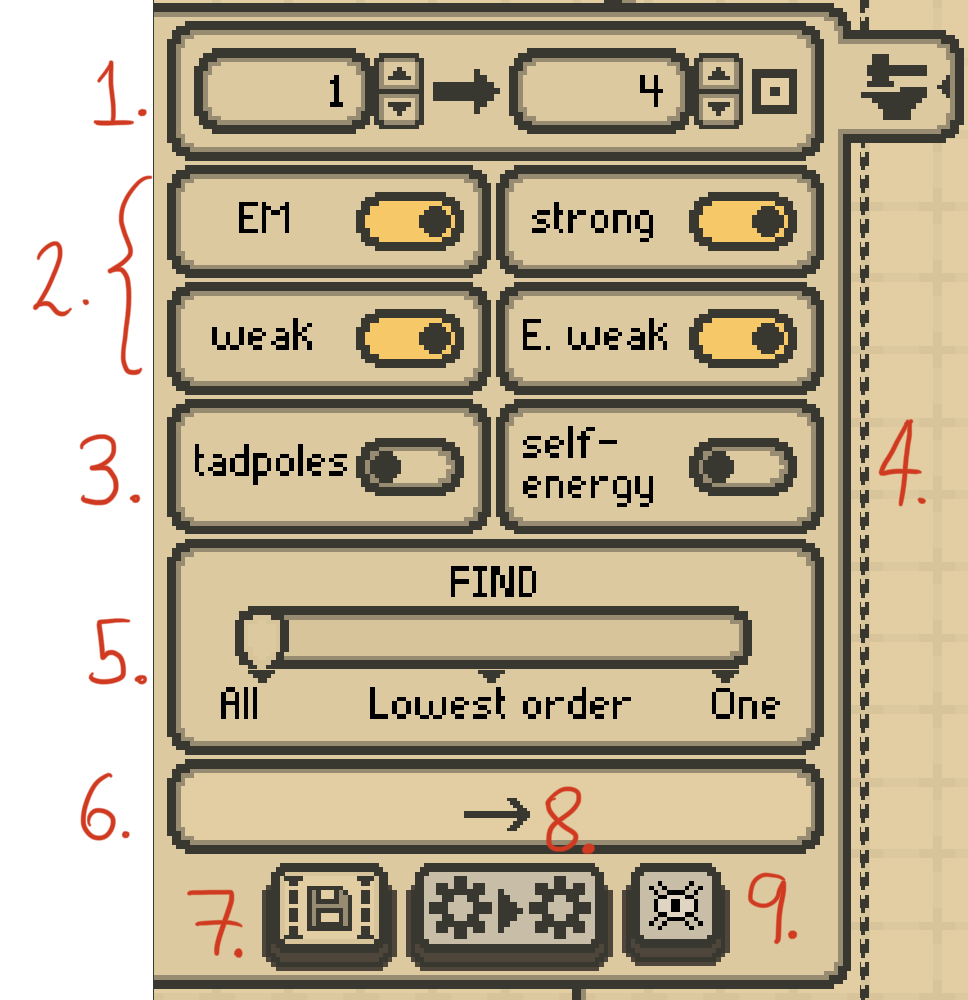}
    \caption{The solution generation tab, used to generate solutions between an initial and final state.}
    \label{fig:labelled_solution_generation_tab}
\end{figure}

If we would like to generate the Feynman diagrams for a process with a specified initial and final state, we can do so using the solution generation tab \includegraphics[scale = 0.6]{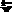}, labelled in \figref{fig:labelled_solution_generation_tab}. The options in this tab let us control which diagrams are generated. From top-to-bottom, and left-to-right, these are:

\begin{enumerate}
    \item \textit{Degree range}: This will set the minimum and maximum degree to look for solutions. The hard minimum is degree 1 and the maximum is degree 10.
    \item \textit{Force toggles}: These will toggle which particles are able to be used to generate a solution; these operate in a similar way to the force toggles found in the problem generation feature (see section \ref{sec:sandbox-problems}, where the particles associated with the toggles are also listed).
    \item \textit{Allow tadpoles toggle}: This determines if diagrams with tadpoles are to be drawn or not. A diagram is said to have a tadpole if there is a loop subgraph within the diagram that is only connected to the rest of the graph via a single particle line - the tadpole is that loop subgraph (the term `tadpole' is used since one has a loop subgraph `blob', with a single particle `tail' emerging from it, resembling a tadpole). The tadpole graphs are linked to the vacuum expectation value (vev) of the single `connecting' particle; in fact, the vev is computed from exactly these tadpole diagrams. The vev of any SM particle with spin has to be zero from Lorentz invariance considerations (e.g. for a spin 1 field, $\langle 0  | A^\mu | 0 \rangle = 0$ as the right hand side would have to be a Lorentz vector, but there is no `direction' in the vacuum available to construct this vector from). This just leaves the Higgs, for which the tadpole diagrams are not in general zero -- although with a suitable choice of scheme, one can arrange that their contribution vanishes (see for example \cite{Martin:2016xsp, Martin:2019lqd}). By default, diagrams with tadpoles are not drawn.
    \item \textit{Self-energy toggle}: This determines if diagrams with self-energy processes on the external legs will be drawn. A self-energy process is a loop diagram with two external particle lines, so diagrams with self-energy processes are ones in which one (or more) of the external particles has some quantum fluctuation and then recombines into a single particle before (or after) the main scattering process. In computing the scattering amplitude $\mathcal{M}$ for a given process, one does not include diagrams with the full scattering process plus self energy processes on the external legs. Rather, one computes all the \textit{amputated} diagrams without self-energies. Then, one multiplies the result by a factor `$\sqrt{Z}$' for each external particle that is computed via the self-energy diagrams for that particle -- for details see e.g. Chapter 7 of \cite{Peskin:1995ev}. One can make some rough intuitive sense of this procedure: if one thinks about a scattering diagram with a self energy process, then that self energy process is not really related to the scattering, rather it should be associated to the nature of the external particle itself (recall that in QFT, a single particle is never alone, but is always surrounded by a `cloud' of virtual particles). Thus, if we are interested in the scattering process, we should compute only the amputated diagrams. We then perform a separate procedure to `dress' each of the external states in this calculation into the full QFT external particle including the quantum fluctuations (this is the multiplication by $\sqrt{Z}$ factors). By default, diagrams with self-energies on the external legs are not drawn. 
    \item \textit{`Find' setting}: This will change how many diagrams to generate. When set to \lstinline!All!, all diagrams in the degree range are generated; when set to \lstinline!Lowest Order!, the code will draw all the lowest order (smallest degree) diagrams possible given the degree range specified, and when set to \lstinline{One}, it will generate one diagram.
    \item \textit{State equation}: This will display the process that solutions will be generated for.
    \item \textit{Save states button}: This saves the initial and final states that are currently drawn in the main \FeynCraft{} window into the state equation. This represents the mechanism by which the user specifies the process for which Feynman diagrams should be drawn. Here only the drawn state particles matter, it does not matter if the drawn states are connected or would even have a valid solution.
    \item \textit{Generate button}: Generates the diagrams for the process specified in the `state equation', using the settings set above. Once finished, generated diagrams are stored and can be viewed using the view solutions button.
    \item \textit{View solutions button}: Pressing this button will show the generated solutions inside a mini-diagram viewer (as discussed in section \ref{sec:sandbox-viewing_submissions}). If no diagrams are found, this button is disabled.
\end{enumerate}

The time taken to generate solutions will depend on the degree range and the find setting. If the find setting is set to \lstinline{One}, then it will be fast no matter the degree, but otherwise it will slow down trying to generate diagrams above degree-6. This could either be if \lstinline!Lowest Order! is set and the minimum degree is degree-6 or higher, or if \lstinline!All! is set and the maximum degree is degree-6 or higher. A load-time warning will show above this point in either case. If there are state hadrons that can directly connect to each-other, this will further slow down generation, as discussed in section \ref{sec:algorithms-solution_generation}. 

The method to specify a process and draw the diagrams for that process is thus as follows: first, draw the initial and final states in the main diagram window. Only the attachments to the initial and final state lines matter here so do not worry about drawing a valid diagram or even if any state particle is connected to another. After this, open the generation tab and save your drawn states by pressing the `save' button. Then, specify your desired settings for the drawn diagrams using the buttons and toggles discussed above. After setting these, press the generate button, wait for generation to finish, and view the solutions by pressing the view solutions button. This will open the solutions inside of a mini-diagram viewer which is described in section \ref{sec:sandbox-viewing_submissions}.

\begin{tcolorbox}
Let us illustrate the process of generating the leading-order diagrams for the example process

\begin{equation}
g + g \rightarrow u + \overline{u}.
\end{equation}

First, we draw these states, not worrying if we draw a valid or connected diagram (as illustrated in the figure below). Then, we open the solution generation tab and click the save button to use these states for the generation, seeing them appear inside the equation. Now, we'll adjust the generation settings. The default degree range is from 1-4, but we only want to generate leading-order diagrams, so let us change this to 2-2. For the the same reason, let us let us disable all but the strong force toggle by clicking on each one. We are happy with the default setting that diagrams with tadpoles and self-energies are not generated, so let us leave those toggles disabled as well. We could change the find setting to lowest order, but as we have already limited the degree we can leave this on ``find all''. Finally, to start generation we click the generate button, and wait for the pop-up to tell us that generation has finished (very quickly in this case). To see the generated solutions, click on the `view solutions' button, which will open the stored solutions inside the mini-diagram viewer (see section \ref{sec:sandbox-viewing_submissions}). For our example, this will show the three leading-order diagrams that are also shown in \figref{fig:completed_diagrams}.\\

\centering
\includegraphics[width=0.5\linewidth]{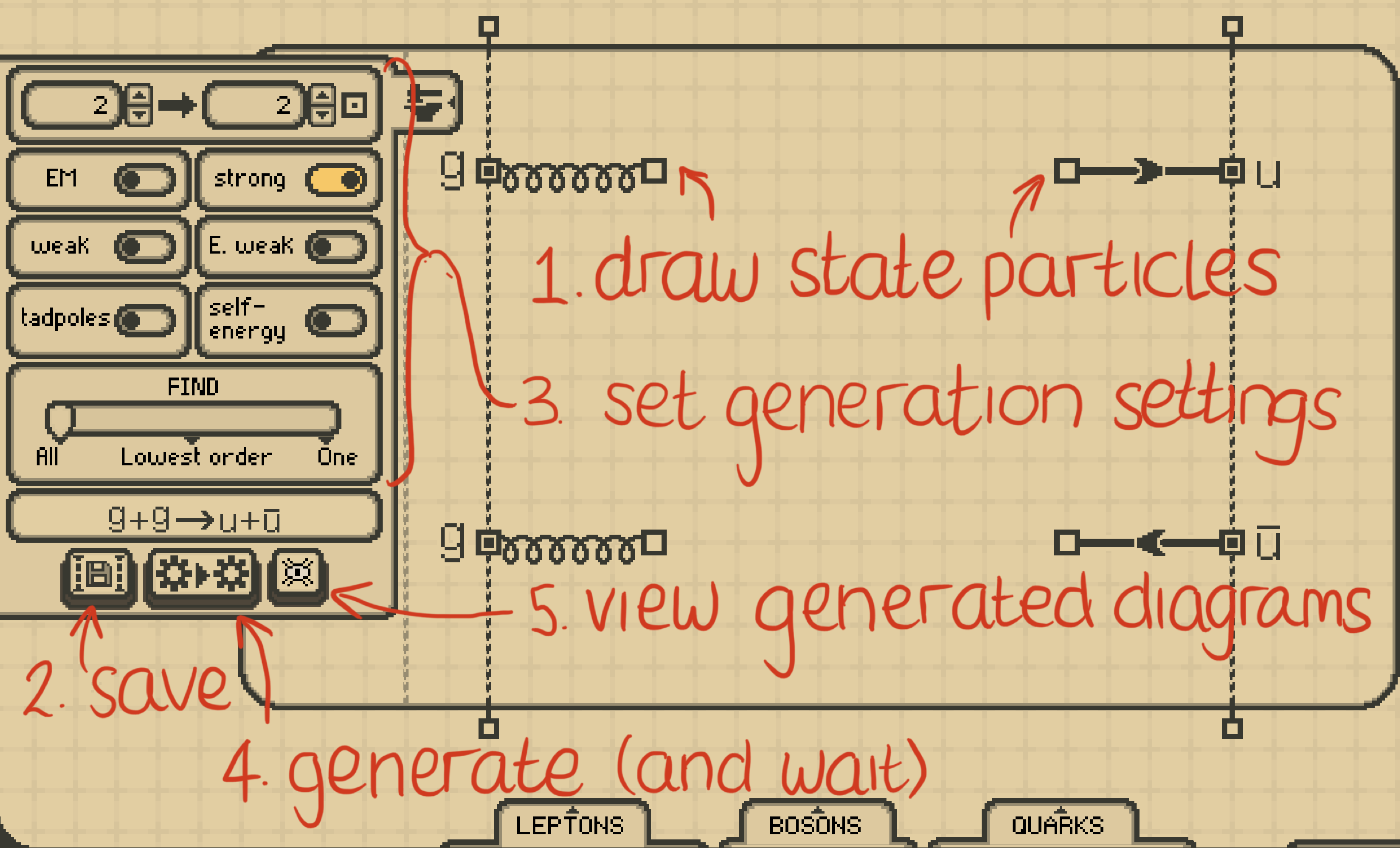}
\end{tcolorbox}

\subsection{Vision overlays}
\label{sec:vision_tab}

\begin{figure}
    \centering
    \includegraphics[width=0.4\linewidth]{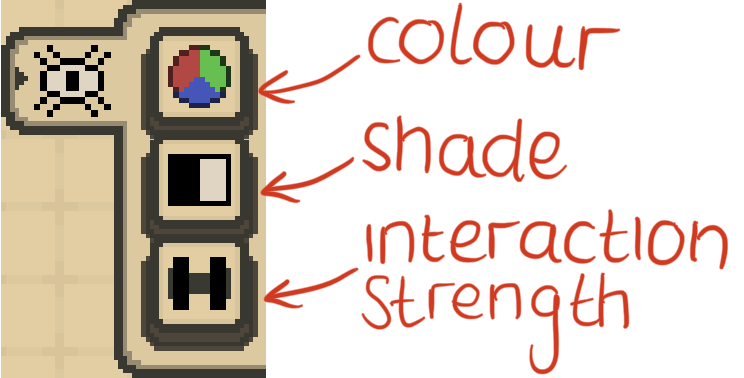}
    \caption{The vision tab, used to display visual overlays. These are colour flow, shade flow, and interaction strength.}
    \label{fig:vision_tab}
\end{figure}

The user can also gain more information on a drawn diagram through visual overlays, accessible from the vision tab shown in \figref{fig:vision_tab}. There are currently three available vision overlays:

\begin{figure}
    \centering
    \includegraphics[width=0.32\linewidth]{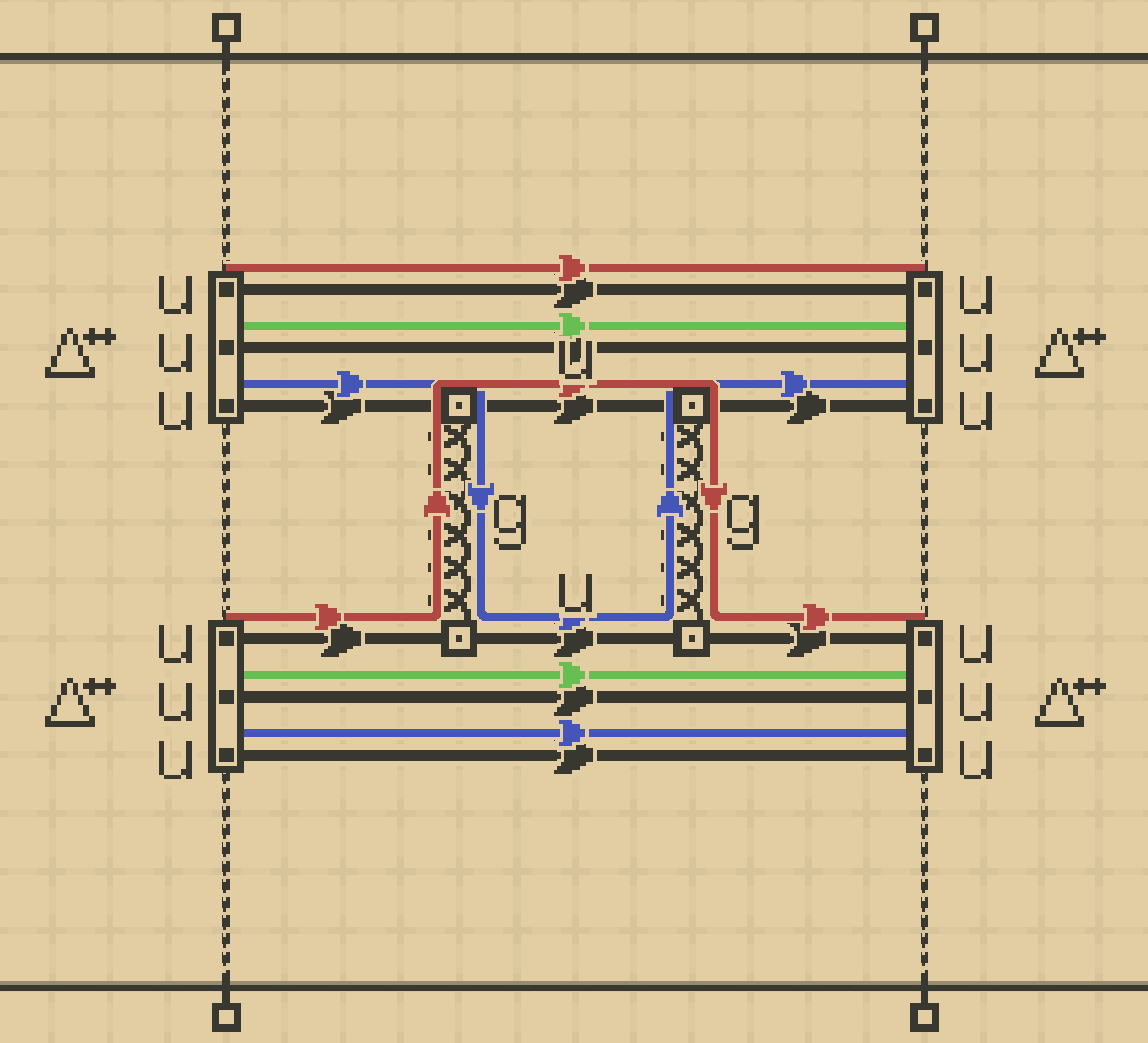}
    \includegraphics[width=0.32\linewidth]{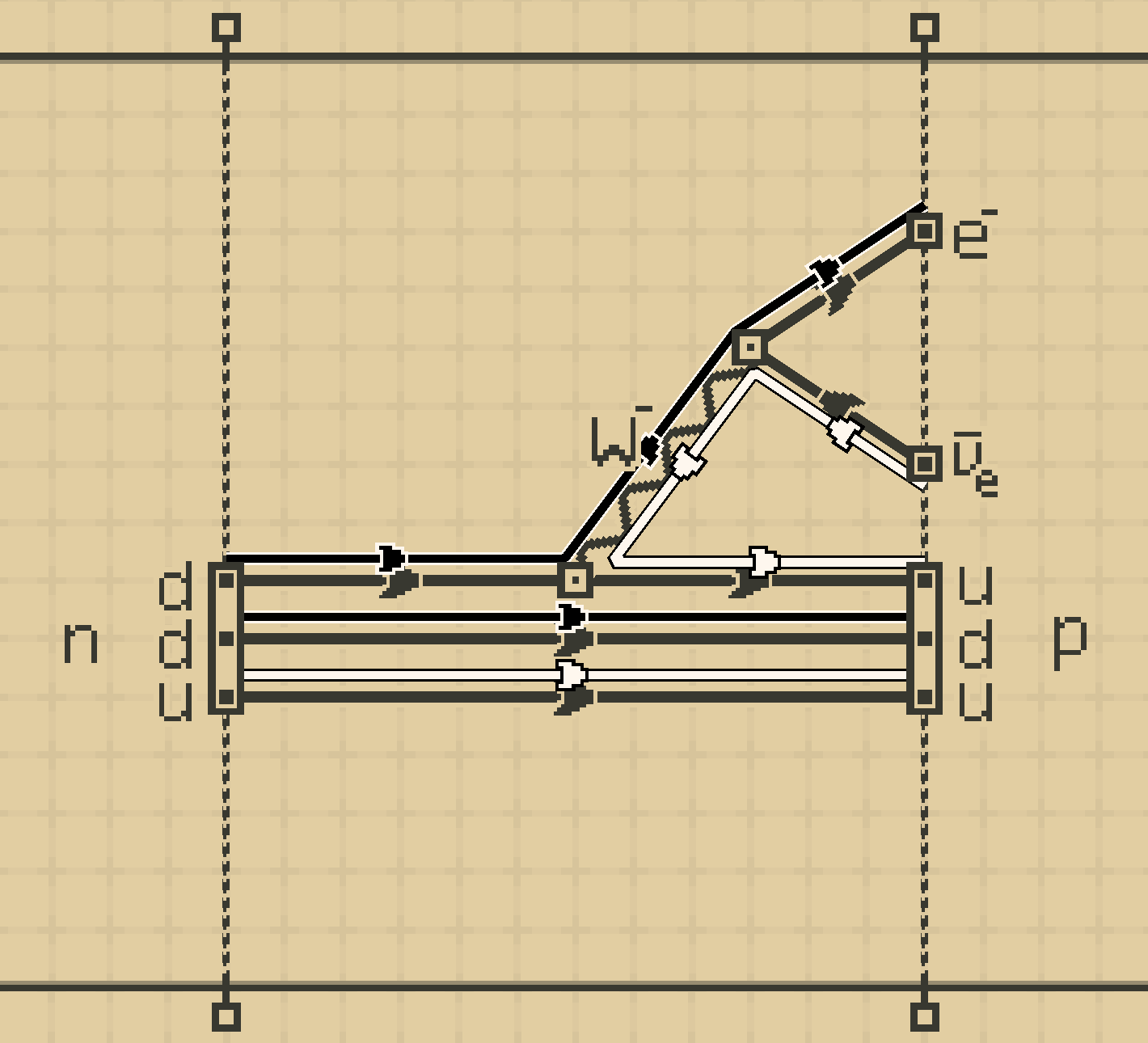}
    \includegraphics[width=0.32\linewidth]{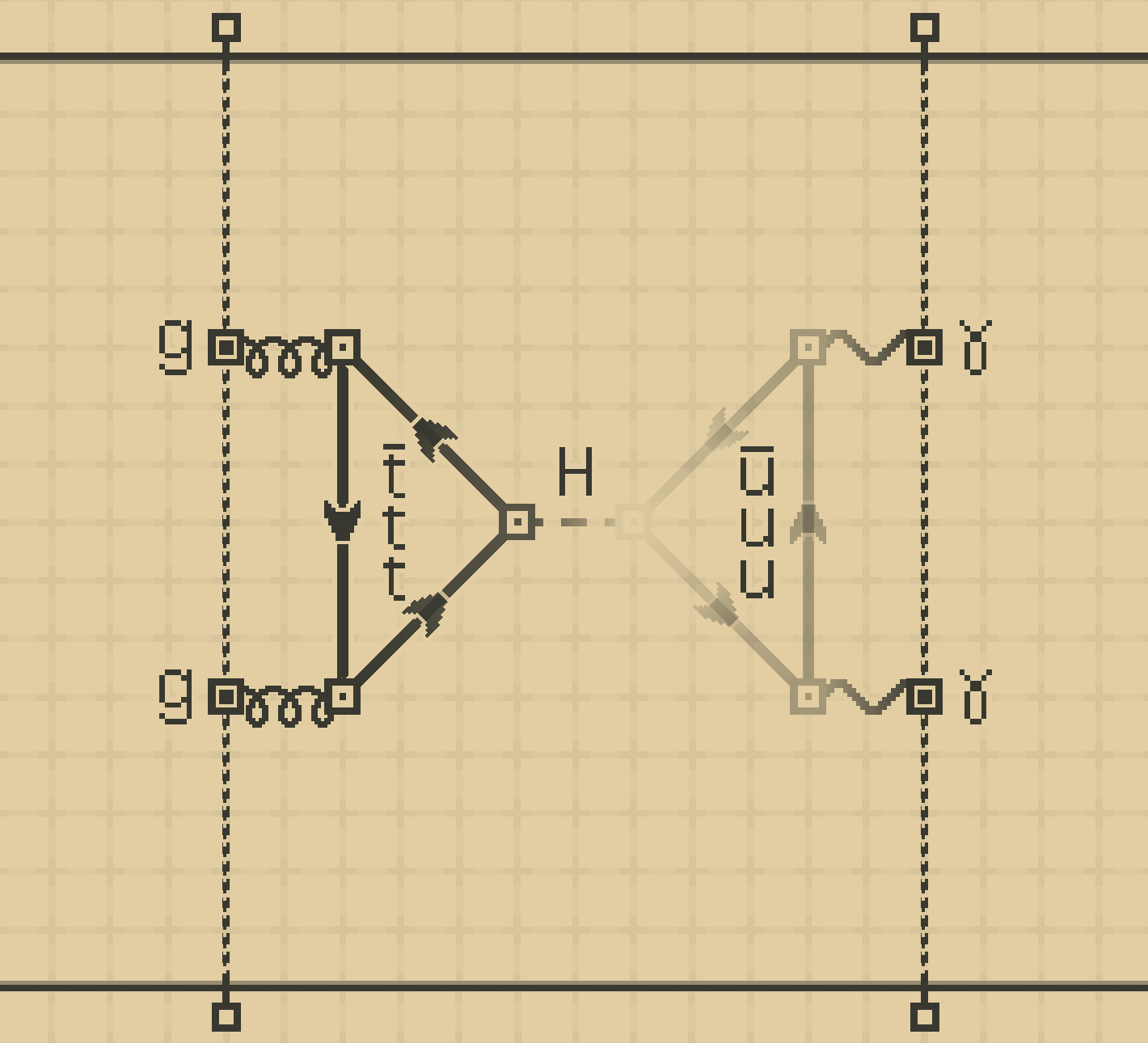}
    \caption{From left-to-right, examples of the vision colour flow, shade flow, and interaction strength visual overlays, respectively.}
    \label{fig:vision_examples}
\end{figure}

\begin{itemize}
    \item \textit{Colour flow}: This vision overlay will generate a illustrative QCD colour flow configuration on top of the current diagram, with an example shown on the left in \figref{fig:vision_examples}. The flows drawn take into account the fact that hadrons must overall be colourless (so, baryons are always drawn with one red, one green and one blue particle, whilst mesons are drawn with a colour and anti-colour). Whether a particular path is coloured red, green, or blue is decided arbitrarily, which might cause the colours to switch when a change in the diagram is made. Quark lines have a single colour flowing through them, whilst gluon lines have a colour and an anti-colour. Conservation of colour implies that a flows must enter and exit via the external particles, or flow round in loops inside the diagram. A gluon being drawn in which the colour and anti-colour match (e.g.~ blue and anti-blue, $b\bar{b}$) does not necessarily mean that the diagram is forbidden, since this does not necessarily correspond with the gluon being colourless (this is essentially because only the fully symmetric combination $\propto r\bar{r} + b\bar{b} + g\bar{g}$ is actually colourless, 
    see e.g. Chapter 10 of \cite{Thomson:2013zua}). Each colour flow diagram that is drawn effectively corresponds to a diagram of the `colour flow representation' discussed in \cite{Kilian:2012pz}, omitting the flows that correspond to the colour singlet `phantom gluon' discussed there.
    \item \textit{Shade flow}: This vision overlay will display an electroweak analogue of colour flow that we refer to as `shade'. An example of the shade overlay is shown in the middle in \figref{fig:vision_examples}. This was originally introduced into \FeynCraft{} as a technical device to assist with the generation of diagrams involving the $W$ boson, but we have added this flow as one of the visualization options as we believe it might be interesting and useful to students.

    Let us briefly overview how shade works. The flows are similar to the colour flows discussed above, but now we only two colours, or rather `shades', black and white. The $\{u, c, t, \nu_e, \nu_\mu, \nu_\tau\}$ fermions have a white shade, whilst the $\{d, s, b, e, \mu, \tau\}$ fermions a black shade. The force-carrying $W^-$ has a black and anti-white shade, which then immediately implies that the $W^+$ has a white and anti-black shade (so the force-carrying particles carry a shade and anti-shade, analogous to the gluon carrying colour and anti-colour in QCD). All other particles are `shadeless'. Similar to QCD, shade flows are conserved, either entering or exiting via external line, or flowing round in loops inside the diagram.

    In technical terms, shade corresponds to weak isospin, or rather the third component of weak isospin $T_3$. The left-handed $\{u, c, t, \nu_e, \nu_\mu, \nu_\tau\}$ fermions have $T_3 = +1/2$, and the left-handed $\{d, s, b, e, \mu, \tau\}$ have $T_3 = -1/2$, thus white corresponds to $T_3 = +1/2$ and black to $T_3 = -1/2$. The $W^-$ has $T_3 = -1$, consistent with the allocation black and anti-white, and the $W^+$ has $T_3 = +1$, again consistent with the shade allocation. Conservation of the shade flow would correspond to the conservation of $T_3$. Interactions with the Higgs field do lead to non-conservation of $T_3$ -- importantly, interactions with the Higgs field (either the vacuum expectation value or the Higgs particle itself) flip fermions between the left-handed state with $T_3 \neq 0$ and the right-handed state with $T_3=0$, where the latter does not interact with the $W$ boson. However, this interaction can't flip the $T_3$ value of the fermions over to the opposite sign (such that the interaction with the $W$ bosons would be different), it can only flip it into an `inert' state in terms of the weak interaction, and back again. Thus, in terms of potential allowed weak interactions in the diagram we can ignore this subtlety, and treat shade as if it is conserved. The shade assignments then correspond to potential nonzero values of $T_3$ for particles within the diagram. See e.g.~Chapter 15 of \cite{Thomson:2013zua} for more information on weak isospin.

    \item \textit{Interaction strength}: With this overlay activated, the opacity of drawn vertices scales with the size of the coupling constant, as demonstrated on the right in \figref{fig:vision_examples}. This is helpful (for example) to qualitatively compare the relative contribution of different mechanisms to a given process or decay. For example, one can easily see that the most significant contribution to the decay $H \to \gamma \gamma$ is from the process with an intermediate top quark loop (rather than the processes with loops of lighter quarks). This scale is truncated on the weaker end so that stronger interactions are more easily distinguishable.
\end{itemize}

\subsection{Exporting Diagrams}
\label{sec:exporting_diagrams}

\begin{figure}
    \centering
    \includegraphics[width=0.3\linewidth]{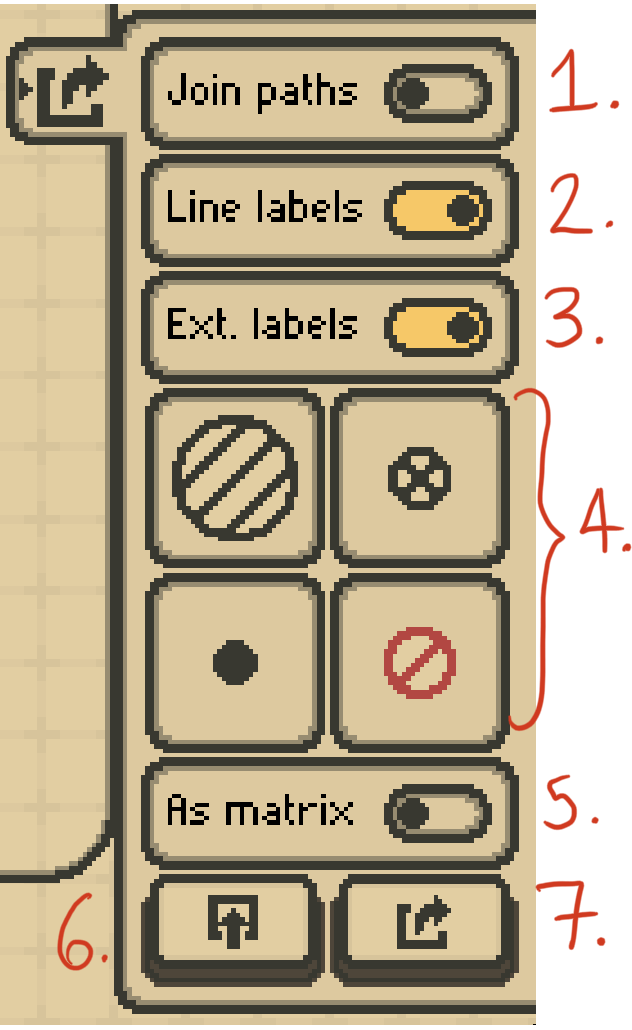}
    \caption{The export tab, used to export drawn diagrams either to be shared in \FeynCraft{}, or as \LaTeX{} code.}
    \label{fig:export_tab}
\end{figure}

\begin{figure}
    \centering
    \includegraphics[width=0.48\linewidth]{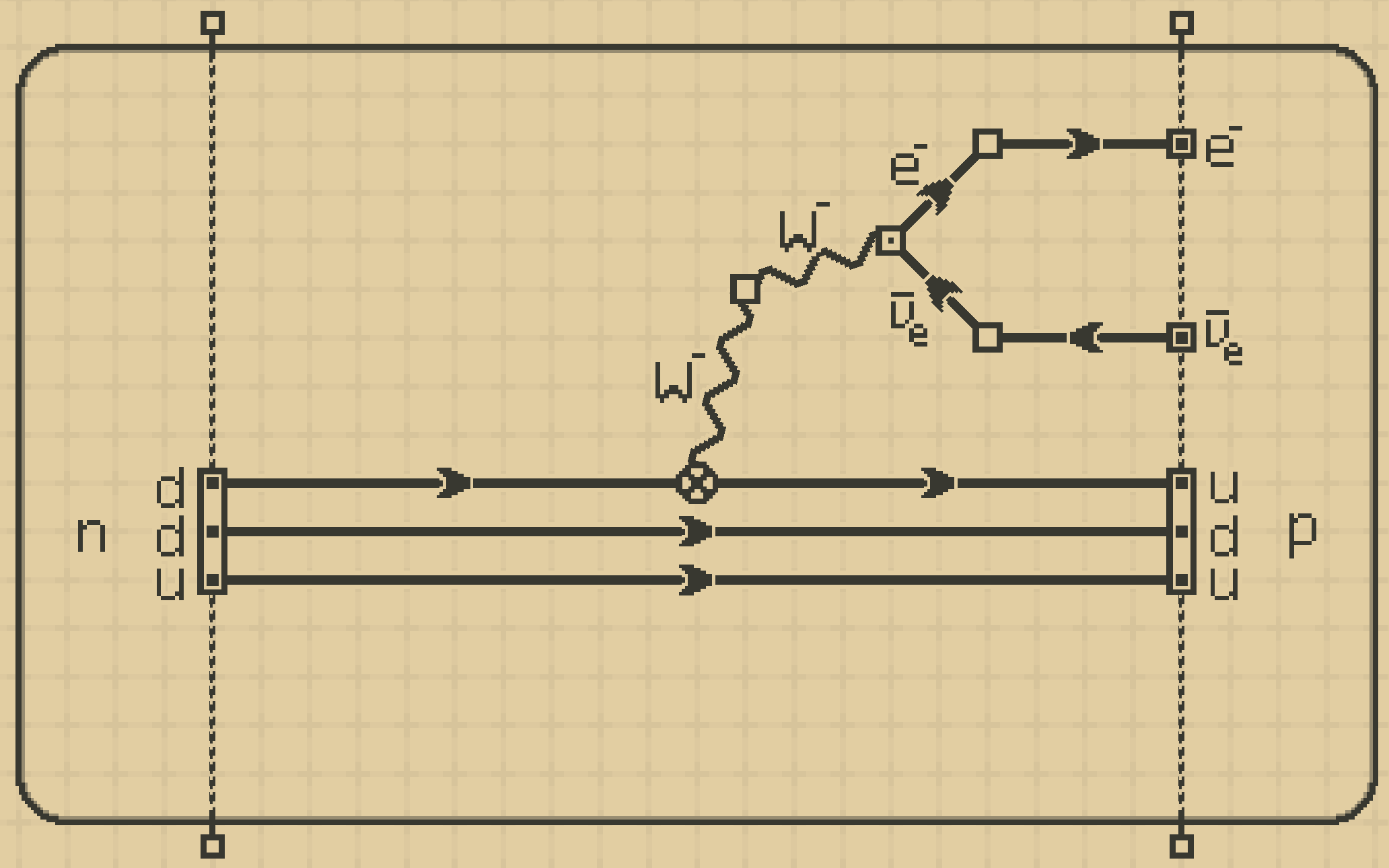}
    \includegraphics[width=0.48\linewidth]{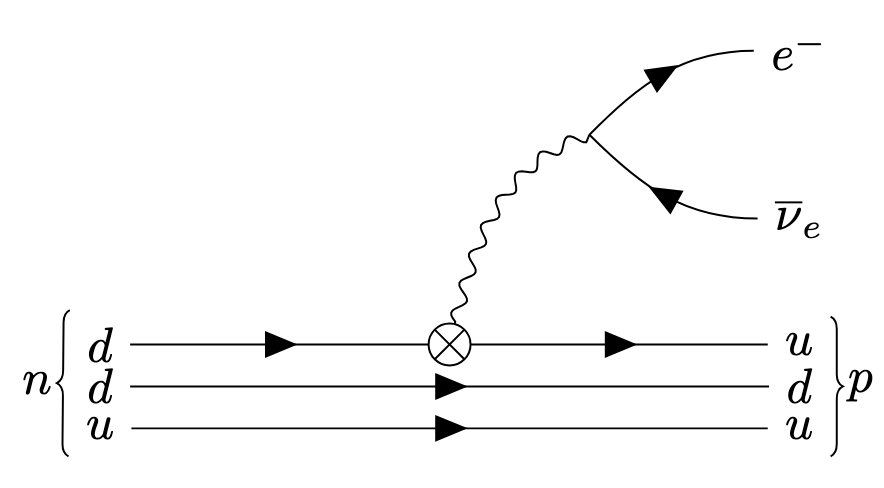}
    \caption{A example of a drawn diagram and how it looks after exporting into \LaTeX{}. The $W$ label is not visible as line labels have been toggled off.}
    \label{fig:export_example}
\end{figure}

Drawn diagrams can be exported into \LaTeX{} code used by the \lstinline{tikz-feynman} package \cite{Ellis:2016jkw} using the export tab \includegraphics[scale = 0.7]{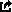}, labelled in \figref{fig:export_tab}. Note the drawn diagram does not need to be valid or connected. Inside the export tab the user will also find several export options and decorations, and from top-to-bottom these are:

\begin{enumerate}
    \item \textit{Join paths toggle}: If enabled, will ensure there is only one arrow per fermion path and one label per internal fermion path. Intial- and final-state fermions will still keep their labels.
    \item \textit{Line labels toggle}: Toggles whether internal lines will have particle labels.
    \item \textit{External labels toggle}: Will toggle whether the initial- and final-state particles will be labelled. For hadrons, if this is turned on then a hadron label and brace symbol (showing that the quarks are collected together inside that hadron) will be drawn, as depicted in \figref{fig:export_example}. Turning it off will remove the quark and hadron labels, as well as the brace.
    \item \textit{Interaction decorations}: There are currently three decorations that can be added to the diagram: the blob, dot, and crossed-dot. These can be moved from their position in the export tab onto the diagram in the same manner as moving interactions (see section \ref{sec:drawing_particles}) -- i.e. one has to hold down \lstinline!W! and drag the decoration onto the diagram. These may be placed as a new interaction or on an existing interaction to change its appearance. Decorations are treated slightly differently to normal interactions inasmuch as if one has a decoration vertex that is connected to two of the same particle, the two particles will never be `rejoined' into one (and the vertex deleted) when moving the vertex around. Similarly, dragging a decoration onto a grid point on a line is allowed, and splits that line into two segments joined by the decoration. To remove a decoration, either drag the cancel symbol in the same way onto the decoration or delete the decoration similar to an interaction.
    \item \textit{Export as matrix}: This will export the diagram as the internal matrix representation used by \FeynCraft{} to the clipboard. This allows diagrams to be drawn, saved, and loaded again within the program.
    \item \textit{Load diagram}: This will attempt to load a diagram from the clipboard; the format needs to be \FeynCraft{}'s internal matrix representation of diagrams discussed above.
    \item \textit{Export button}: Pressing this button will convert the drawn diagram into \LaTeX{} code and copy it to the clipboard.
\end{enumerate}

Diagrams are exported `as is', not making use of the package's automatic layouts, and placing hadron braces and labels where hadrons are drawn. Users are free to modify the exported \LaTeX{} code further using \lstinline{tikz} or the \lstinline{tikz-feynman} package, for example by adding particle momentum labels.

\begin{figure}
    \centering
    \includegraphics[width=0.48\linewidth]{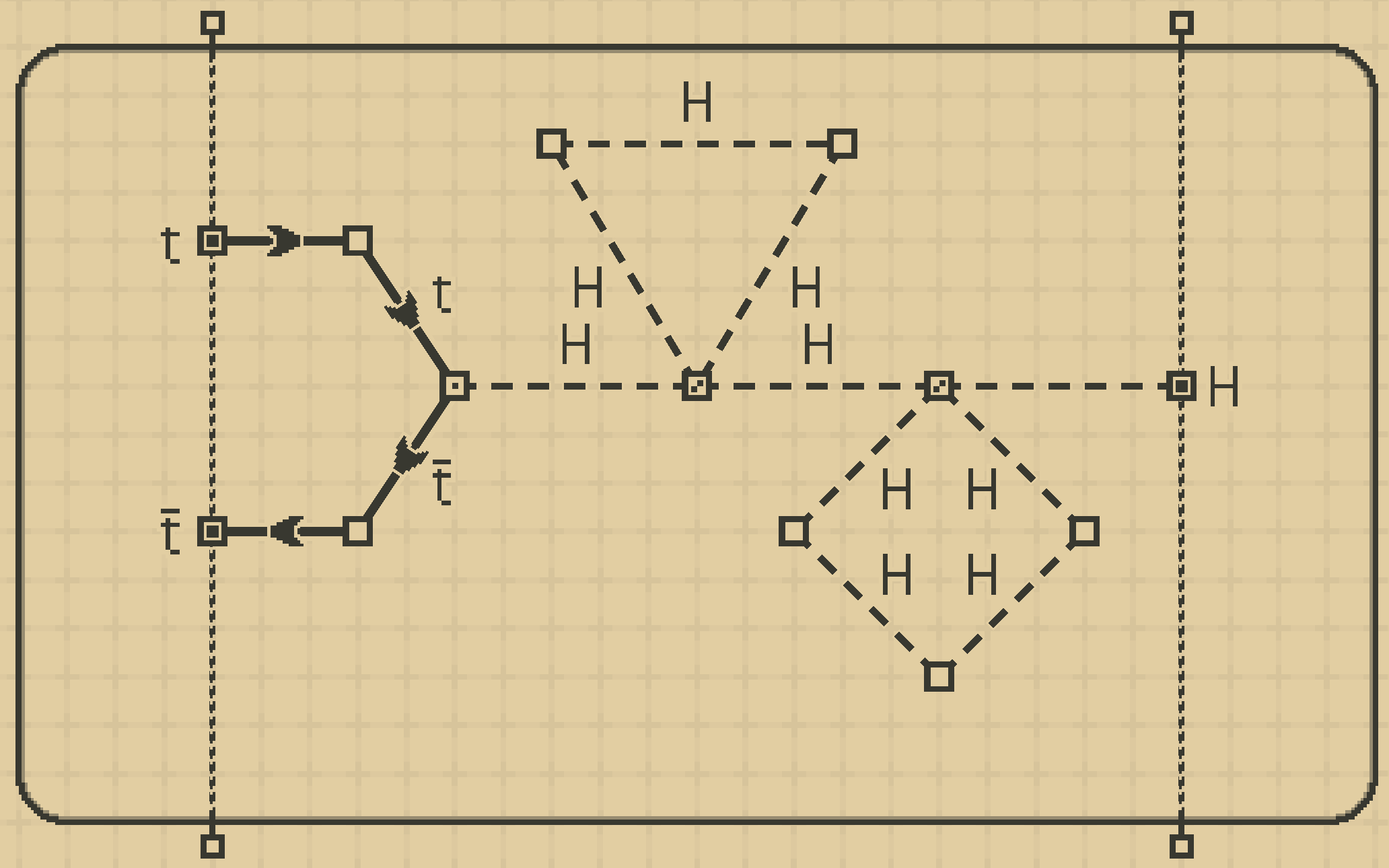}
    \includegraphics[width=0.48\linewidth]{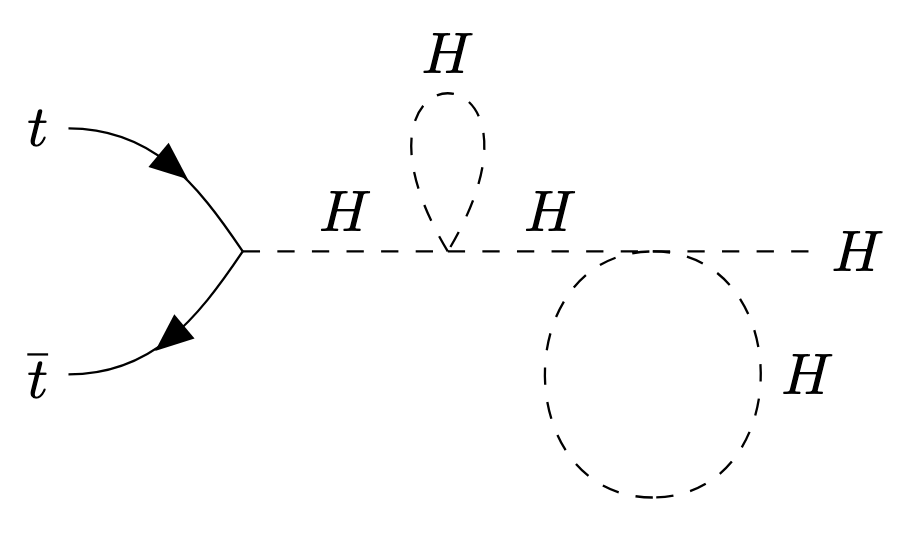}
    \caption{A side-by-side comparison between drawn `intermediary' interactions (left) and how they get exported as curves (right).}
    \label{fig:export-curves}
\end{figure}

Curves and loops can be drawn by drawing intermediary two-particle interactions, which are removed upon export and the path converted into a curve or loop. Placing a single intermediary interaction between two different points will create a curve when exported, the bend of this curve depending on the placement of the interaction. Placing two intermediary interactions on a path starting and ending from the same point creates a tadpole, and three intermediary interactions placed similarly creates a more circular loop. A side-by-side between these drawn in \FeynCraft{} and their exports are shown in \figref{fig:export-curves}. 

\begin{tcolorbox}[breakable]
Say that we would like to export the diagram that we drew in section \ref{sec:drawing_particles}. We would also like the electron, anti-electron-neutrino, and $W^-$ paths to be curved; and for the sake of the example, we would like to add a crossed-dot to the $[d, u, W^-]$ interaction. We start by drawing the same diagram as before, but this time we place intermediary interactions along the paths of these particles we want to curve, the resultant drawn diagram is shown on the left hand side of \figref{fig:export_example}. We then open the export tab. To add a crossed-dot to the diagram, we hold \lstinline!W! and drag and release the crossed-dot over the desired interaction, changing its appearance. Looking at the export options, we'd like both the down to up-quark path, and the anti-electron-neutrino to electron paths to only have one fermion arrow, so we toggle on \lstinline!join paths!. As we can discern the nature of the internal particle from the external ones, we do not need the $W^-$ label, and so we disable \lstinline!line labels!. We do want to keep our hadron and state particle labels, so we keep \lstinline!ext.! \lstinline!labels! on. Finally, we click export which copies the diagram to our clipboard, and paste it into \LaTeX{} which gives

\begin{verbatim}
\begin{tikzpicture}
\begin{feynman}
\def\xscale{0.25} %change to stretch in x
\def\yscale{0.3} %change to stretch in y
\vertex (i0) at (0*\xscale, 9*\yscale) {\(u\)};
\vertex (i1) at (0*\xscale, 7*\yscale) {\(d\)};
\vertex (i2) at (0*\xscale, 8*\yscale) {\(d\)};
\vertex (i3) at (20*\xscale, 0*\yscale) {\(e^{-}\)};
\vertex (i4) at (20*\xscale, 4*\yscale) {\(\overline
\nu_{e}\)};
\vertex (i5) at (20*\xscale, 7*\yscale) {\(u\)};
\vertex (i6) at (20*\xscale, 9*\yscale) {\(u\)};
\vertex (i7) at (20*\xscale, 8*\yscale) {\(d\)};
\vertex (i9) at (14*\xscale, 2*\yscale);
\vertex (i12) [crossed dot] at (10*\xscale, 7*\yscale) {};
\diagram* {
(i0) -- [fermion] (i6),
(i1) -- [fermion] (i12) -- [fermion] (i5),
(i2) -- [fermion] (i7),
(i4) -- [fermion, out = 180, in = 45] (i9) -- [fermion,
out = -45, in = 180] (i3),
(i12) -- [boson, out = -76, in = 162] (i9)
};
\draw [decoration={brace}, decorate] (i1.south west) --
(i0.north west) node [pos=0.5, left] {\(n\)};
\draw [decoration={brace}, decorate] (i6.north east) --
(i5.south east) node [pos=0.5, right] {\(p\)};
\end{feynman}
\end{tikzpicture}
\end{verbatim}

and, when using the \lstinline{tikz-feynman} package, produces the diagram on the right hand side in \figref{fig:export_example}.

\end{tcolorbox}

\section{Problems}
\label{sec:problems}

\begin{figure}
    \centering
    \includegraphics[width=0.48\linewidth]{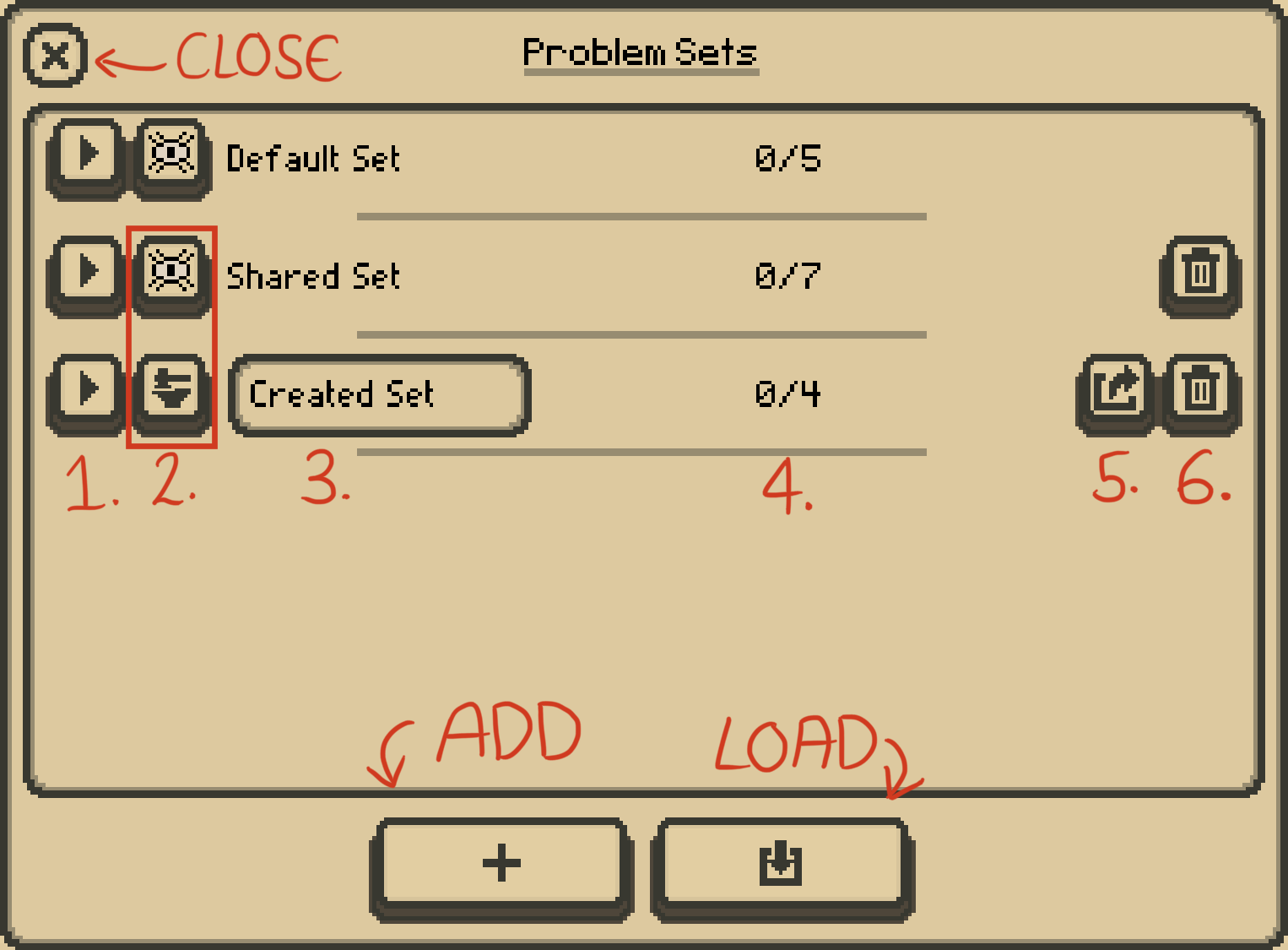}
    \includegraphics[width=0.48\linewidth]{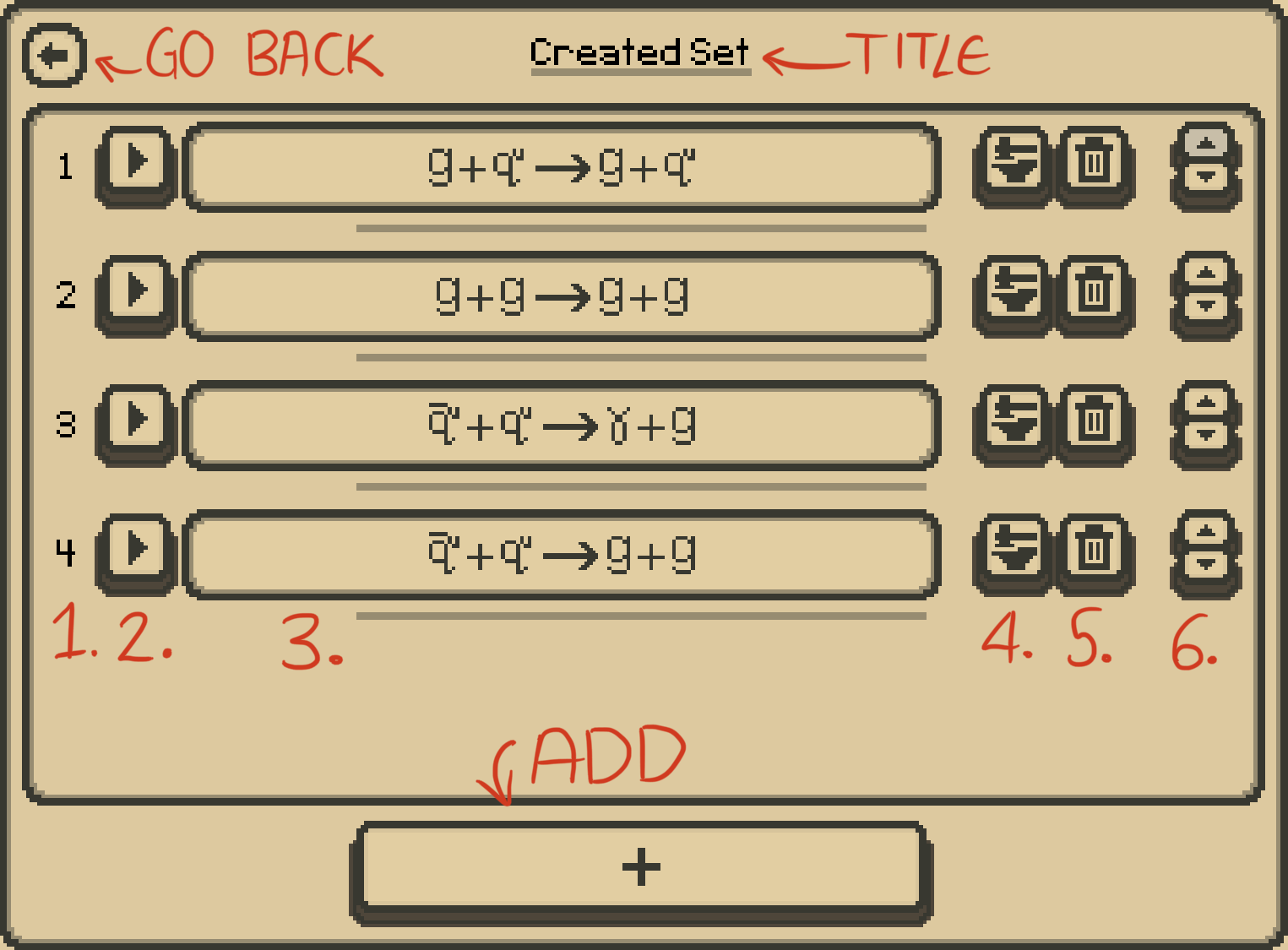}
    \caption{\textit{Left}: The list of problem sets, showing examples of a default, shared, and created set from top-to-bottom, respectively. \textit{Right}: The list of problems inside a created problem set named `Created Set'.}
    \label{fig:problem_list}
\end{figure}

Beyond solving generated problems, users are also free to create, solve, and share sets of problems, called \textbf{problem sets}. This is useful for lecturers who may want to create a problem set based on the latest lecture to be shared with the students, or undergraduates who may want to collect problems they have struggled with to practice before the exam. Problem sets fit into three categories: default (those supplied with \FeynCraft{}), shared (sets shared with the user that they have loaded into their instance of \FeynCraft{}), and created (those created by the user). Sets are accessed by pressing the `problems' button on the main menu which will load a list of problem sets; a labelled example is shown on the left hand side of \figref{fig:problem_list}. Here we can see examples of the three types of problem sets. Looking at a problem set, we can see from left-to-right:

\begin{enumerate}
    \item \textit{Play button}: `resumes' the problem set, loading the last problem the user has yet to complete, or the last problem in the set if all have been completed.
    \item \textit{View problems button/modify button}: pressing the view problems button on default and shared problem sets will list the problems inside the set, allowing the user to see all problems, and play those they've previously completed. On a created set, pressing the modify button will also list all problems inside the set, but this time the problems are able to be modified, switched, and deleted etc. This is discussed in section \ref{sec:problem_creation}.
    \item \textit{Title}: the title of the problem set, only modifiable on created sets.
    \item \textit{Completed/total problem count}: here we can see how many problems in the problem set we have completed, as well as the total number of problems. We will also see a green tick if we have completed the entire set.
    \item \textit{Share button}: pressing the share button, only available on created sets, will export the problem set as text and copy it to the user's clipboard. Sharing this text allows other users to load the set as a shared problem set into their own game.
    \item \textit{Delete button}: pressing this will delete the problem set. Only created and shared sets can be deleted.
\end{enumerate}

Below the list of problem sets we can also see the `add' button on the left, and the `load' button on the right. The `add' button creates a new, empty problem set for the user to modify. Pressing the load button will try to load a shared problem set from the text in our clipboard; therefore, the process to share a problem set would be to first press the share button, then paste the exported text where another user can access it. The user would then copy this text and press the load button to load it into \FeynCraft{}. 

As mentioned, pressing the view problems button on default or shared problems sets, and the modify button on created problem sets, will show a list of the contained problems in the set. A labelled example of this for a created set is shown on the right hand side of \figref{fig:problem_list}. The only difference between this list and that of a default or shared set is the visibility of certain buttons, which we will mention below. From left-to-right, a problem in this list will display:

\begin{enumerate}
    \item \textit{Problem index}: The index of the problem in the list, also the order problems must be completed in default or shared sets.
    \item \textit{Play button}: Starts the selected problem. In default and shared problem sets, which must be completed in order, this button will be disabled until the user has completed the previous problems.
    \item \textit{Problem equation}: displays the initial and final states of the problem.
    \item \textit{Modify button}: Only visible in created sets, allows the problem to be modified after it has been created; the modification of problems is discussed in section \ref{sec:problem_creation}.
    \item \textit{Delete button}: Only visible in created sets, deletes the problem from the set.
    \item \textit{Move up/down buttons}: Only visible in created sets, these buttons allow problems in the set to be reordered, which will change the order of completion if shared.
\end{enumerate}

In default and shared sets only, problems will also display a green tick when completed.

\subsection{Solving problem sets}
\label{sec:solving_problem_sets}
Solving problems in problem sets works similarly to those in the sandbox, described in section \ref{sec:sandbox-problems}. There are some differences though: the next problem button may not be available, completing the final problem in the set will return to the main menu, and individual particles may be disabled or hidden. Moreover, a set problem may require specific `custom' solutions and require a custom degree or solution count, as described in section \ref{sec:problem_creation}. These changes do not change the process of drawing, submitting, and completing a problem, however.

As mentioned, default and shared problem sets must be completed in order. This does not mean that a user can get stuck, as users may always ask \FeynCraft{} to display a solution to the current problem, using the `show solution' button in the problem tab. Since the problems have a fixed order, users generating problem sets are encouraged to order them in order of increasing complexity, with simple problems at the start (that might introduce a particular process/type of vertex), and more complicated problems towards the end (which might combine some of the lessons learned in earlier problems).

\subsection{Problem creation}
\label{sec:problem_creation}
Before creating problems, first an empty problem set must be created by pressing the `add' button in the problem set list, shown on the left in  \figref{fig:problem_list}. The new set can then be named by clicking on its title. To create the first problem in the problem set, press the `modify' button on the problem set to open the currently empty list of problems, and press the `add' button at the bottom of the frame, shown on the right in \figref{fig:problem_list}. This will enter the problem creation process, which is split into three steps:

\begin{enumerate}
    \item \textit{Particle selection}: In this step we select which particles will be available to the solver; all particles are available by default. Disabling a particle will prevent it from being used in the initial and final state, generated solutions, and prevent it being drawn. It will appear `greyed out' to the solver and can't be pressed. Instead of disabling a particle, we can also choose to hide it altogether.
    \item \textit{Problem creation}: Here we create the initial and final states for the problem using the particles we left available in particle selection. We do this by drawing the particles we want onto the initial and final state lines, similar to what is done in Sandbox when generating the Feynman diagrams for a given process (see section \ref{sec:sandbox-generating_solutions}). The degree for the problem by default is of lowest order, but we can specify a custom degree. When we try to go to the next step, the program attempts to generate a solution with the available particles, states, and chosen degree. We cannot progress until we have a setup with at least one solution found.
    \item \textit{Solution creation}: In this step one may draw some preferred example solutions to the problem, referred to in the code as `custom solutions'. The process of specifying the custom solutions is the same as is used to solve a problem, discussed in section \ref{sec:sandbox-problems}. If the `Custom solutions' toggle is on but the `Allow other solutions' toggle is off, then the solver can only pass the problem by submitting the specified custom solutions (in this case the user generating the problems would have to give some hints to the solver outside of \FeynCraft{} as to what kind of solutions they were looking for).  If `Custom solutions' is on but `Allow other solutions' is on, then the specified custom solutions are the `preferred model solutions' (and \FeynCraft{} will only display these model solutions if the solver presses the `show solution' button), but the solver is allowed to submit any valid diagram.
    
    In this step one may also set a custom required number of solutions, which cannot exceed the number of solutions \FeynCraft{} finds to the problem, or the number of custom solutions if the `Custom solutions' toggle is set to on. There is also a hard limit that the required number of solutions must be $\le 4$, to avoid generating problems that are too laborious (similar to the sandbox problems mode).

    One may choose to skip specifying any of these options, in which case users can submit any valid solution. Finishing this step creates the problem and returns the user to the list of problems in the set.
\end{enumerate}

Note that problems may also have a title, which we can set during any of these steps by typing in the title box above the diagram area. Once the problem is created, we can make further changes by clicking the `modify' button. This will enter problem creation again, but with our previous problem and settings loaded. In-order for any changes to be saved, we must pass through all problem creation steps; this is because, if we make a new particle unavailable during particle selection, say, it may invalidate a custom solution, or prevent the problem from being solved entirely.\\

\begin{tcolorbox}[breakable]
We shall now follow this process with an example. We will create a hadron-decay themed problem set with the example problem of neutron decay, shown in section \ref{sec:drawing_particles}.
To start, we press the `add' button on the problem set list to create an empty problem set. We'd like this set to contain solely hadron problems, so let us click on the empty title and call it `Hadron problems'. We then press the `modify' button to enter the empty problem set, and click the `add' button to enter particle selection.\\

\begin{center}
\includegraphics[width=0.40\linewidth]{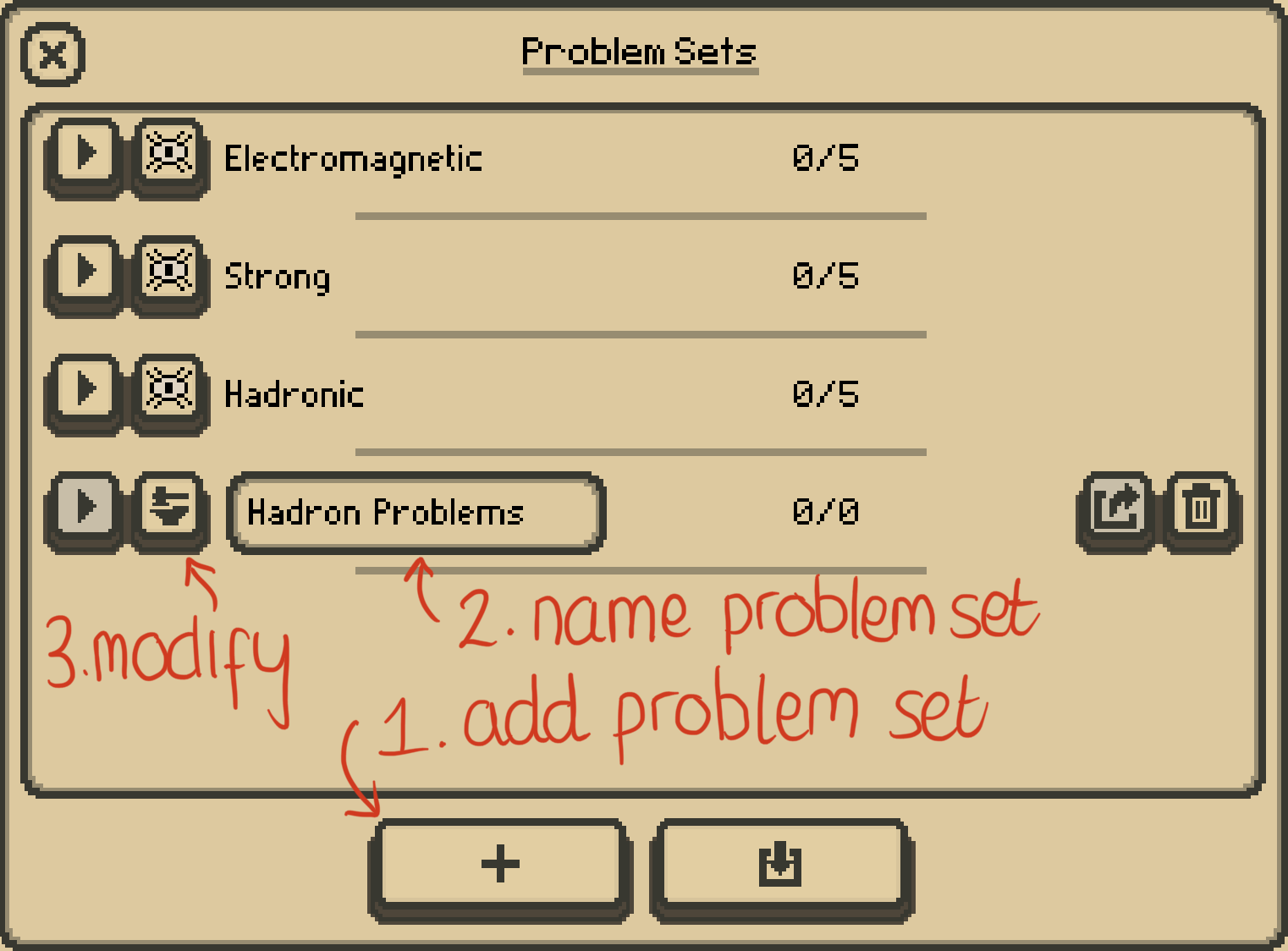}
\includegraphics[width=0.40\linewidth]{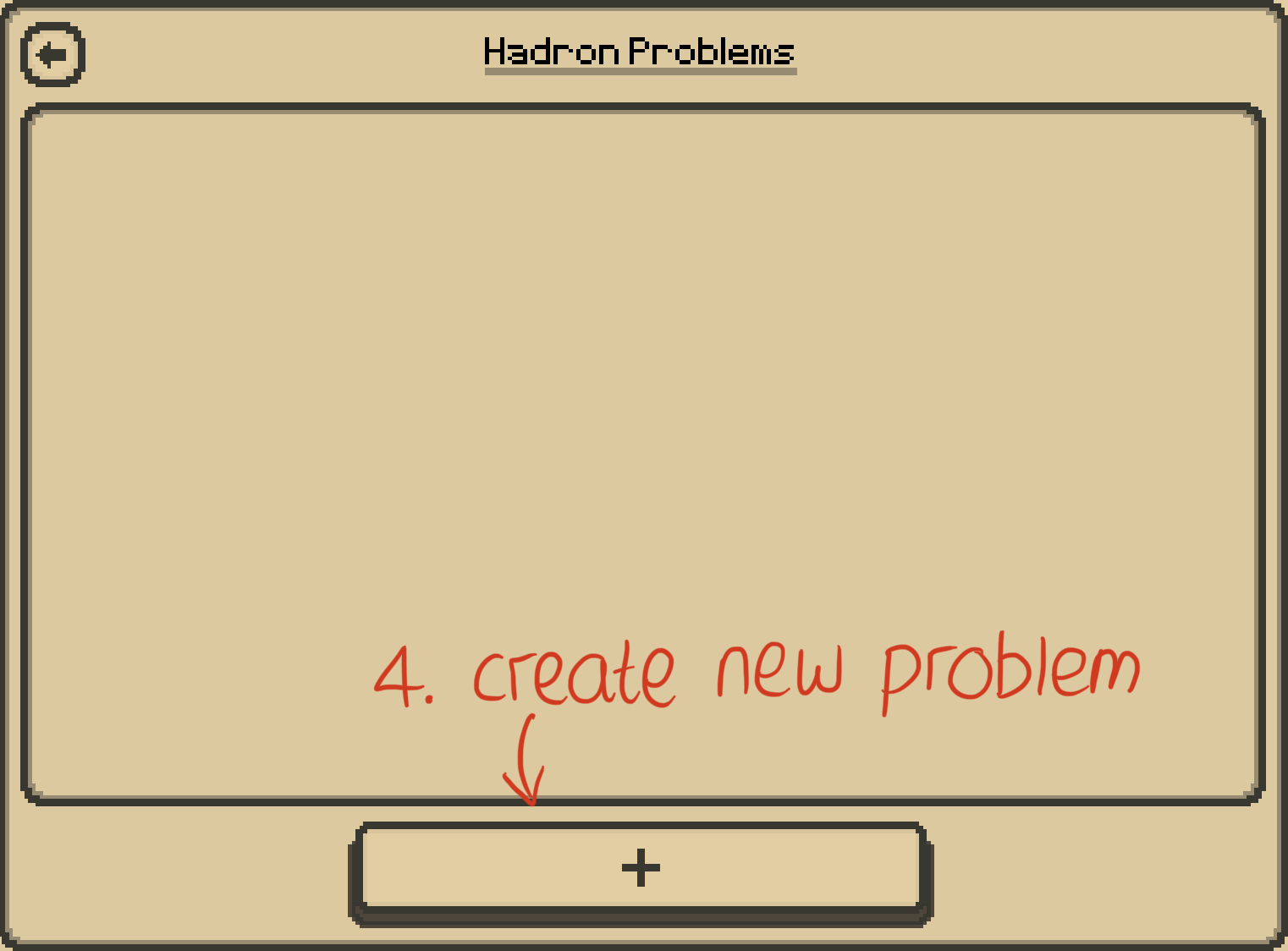}
\end{center}

Now we'll construct the neutron decay problem. In the first step we'll choose to disable some particles that don't feature in this process, the $Z$ and $H$ bosons. Since we'll specify the types of all the quarks, leptons and neutrinos, we'll also disable the general $l$, $l_\nu$, $q^u$, and $q^d$ fermion buttons. We would like these disabled particles to not be visible during solving, we enable the `hide unavailable particles' toggle. We can title this problem `Neutron Decay' by clicking on the title above the diagram area, and then click next to go to problem creation. 

During problem creation we only have to draw the initial and final state particles we want in the problem, and don't have to worry about connecting up the diagram (see the second pane of the figure below). We would like this problem to be lowest order (degree 2) so we do not set a custom degree, and press next to go to solution creation. 

As there is only one solution to this problem at this degree, we do not need to worry about setting a custom required number of solutions. We would like to ensure that this example solution is aesthetically pleasing, however, so let us toggle on `custom solutions', draw the solution, and submit it as we would to a problem (see section \ref{sec:sandbox-problems}). As there is only one solution, it makes no difference if we toggle on or off `allow other solutions'; we turn it on. We now press `submit' to finish creating our problem, which we can now see in the problem list.\\

\begin{center}
\includegraphics[width=0.48\linewidth]{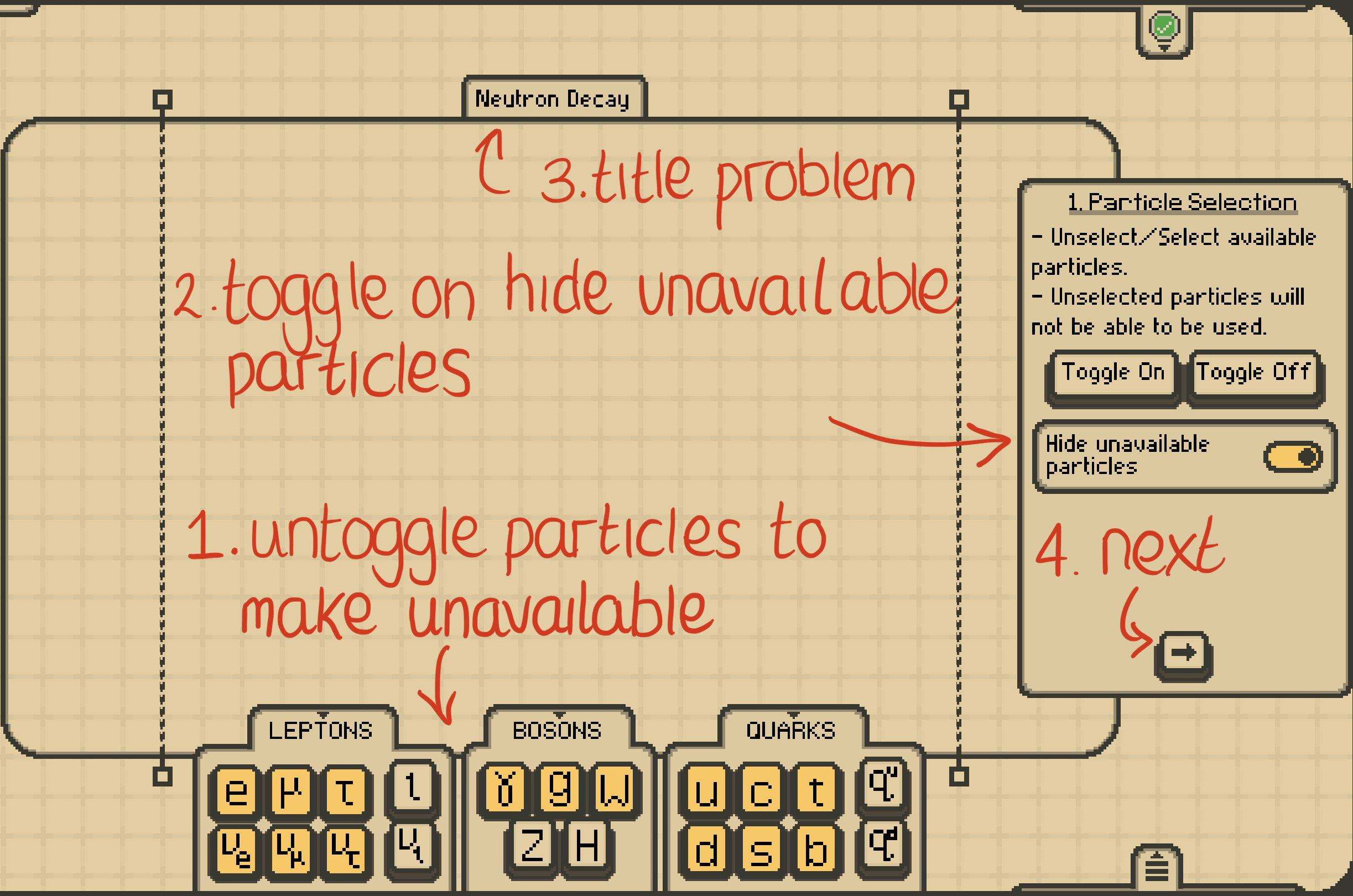}
\includegraphics[width=0.48\linewidth]{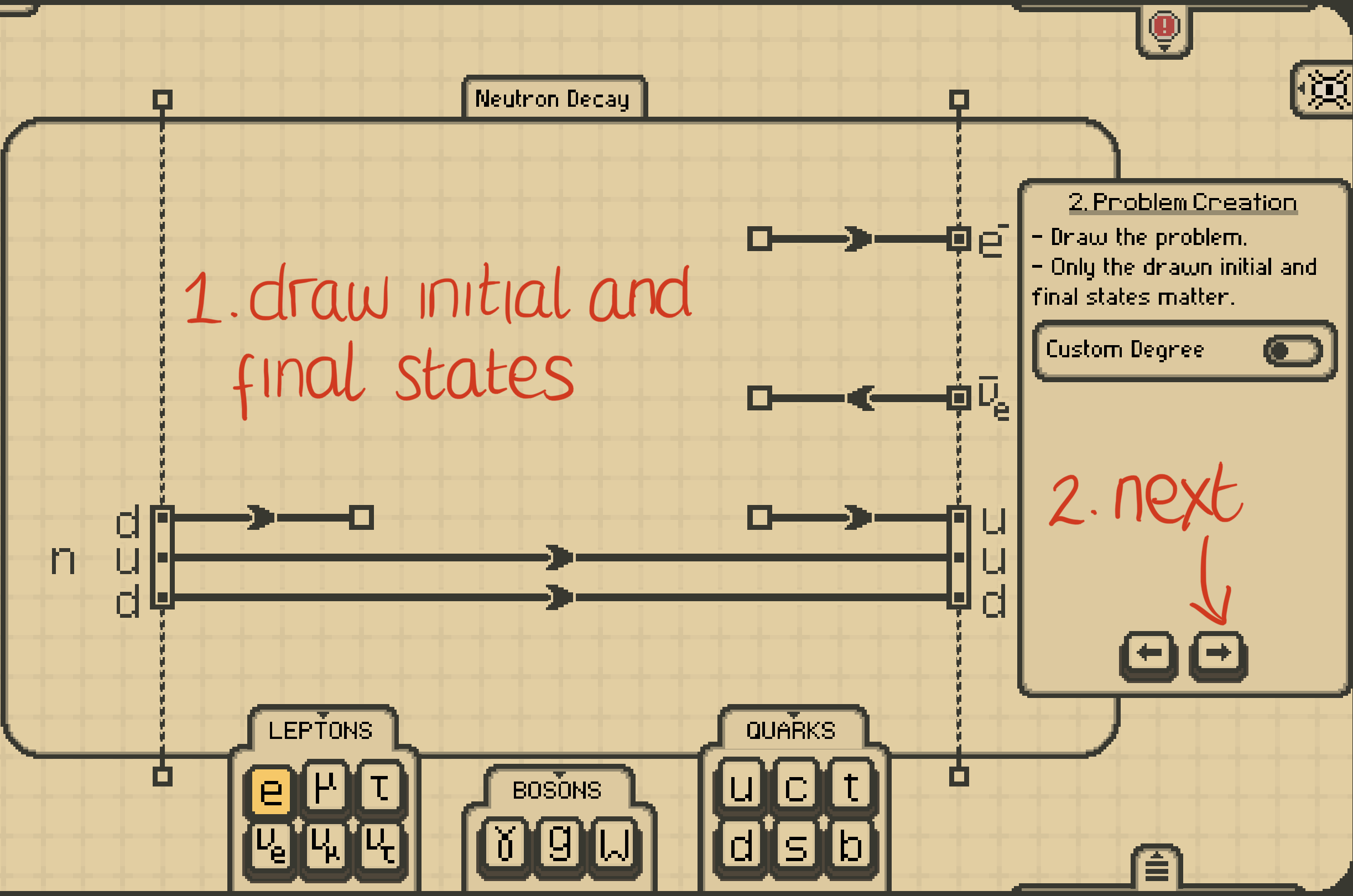}
\includegraphics[width=0.48\linewidth]{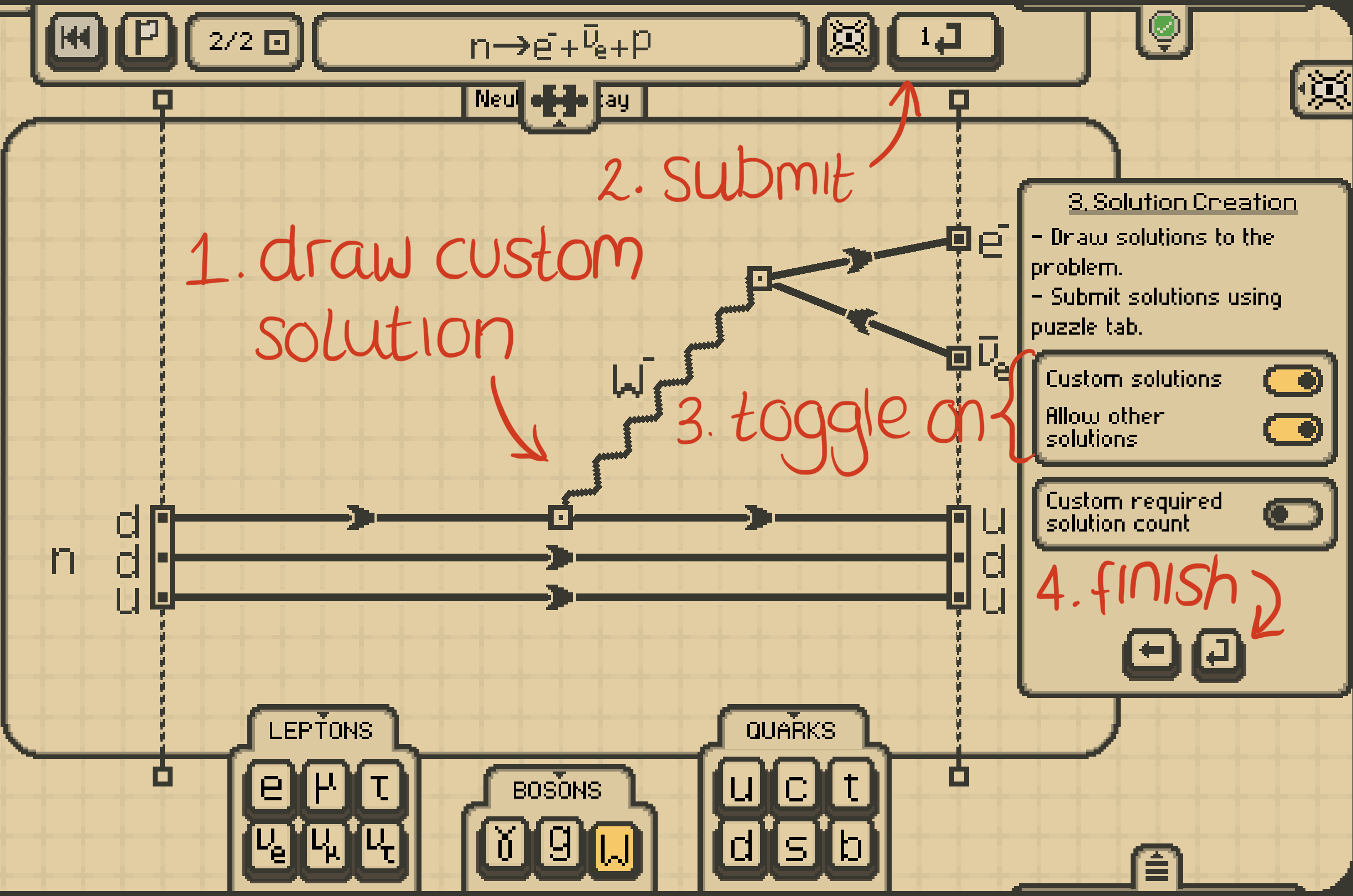}
\end{center}

If we want to modify this problem, say to re-enable the $Z$ boson, we would click on the `modify' button, re-enable the $Z$ boson by clicking on it, then skip through the steps without changing anything else. To share this problem set (after adding more problems), exit the problem list by pressing `back', and press the `share' button on the problem set. To demonstrate this working, we press the `load' button on the problem set list to load the same hadron set as a shared set.
\end{tcolorbox}

\section{Algorithms}
\label{sec:algorithms}
In this section, we will briefly look through the different algorithms used inside \FeynCraft{}. These are: problem generation, solution generation, vision and colourless gluon checking, checking duplicate diagrams, and diagram layout generation.

\subsection{Problem Generation}
\label{sec:algorithm-problem_generation}
Before starting problem generation, we supply the range of number of state particles (4 to 6 by default), the available particles (via the force toggles, see section \ref{sec:sandbox-problems}), and hadron frequency. Problems will always be generated as one-to-many or two-to-many, to align with both exam questions and calculations of scattering cross sections or decay rates. From the available particles we can filter out the hadrons that we aren't allowed to use. Next, we decide how many hadrons we are going to add, depending on hadron frequency. If \lstinline!Never!, we will not include any, if \lstinline!Always! we will add between one and the total number of state particles, and if \lstinline!Sometimes! then there is a 60\% chance we will add one or more hadrons. We then enter a tree recursive function. At each step we add a particle from those available to either the initial or final state, add its quantum numbers to the total (multiplied by a factor of -1 if on the final state), and check if the resultant total is valid. The total is considered invalid if, ignoring $W$ bosons for now, any individual quantum number cannot reach zero with the number of remaining particles. For example, if we have two particles left to add and the current lepton number total is $+2$, we cannot add a gluon (with lepton number zero) because then we could not reach a total of zero lepton number with the single remaining particle. Instead, we would need to add either an anti-lepton to the initial state (lepton number $-1$), or a lepton to the final state (also lepton number $-1$, as we multiply by -1 if on the final state line). We continue this process until we have reached the particle count we chose. We only add hadrons to begin with, until the chosen numbers of hadrons has been added. State $W$ bosons mean some quantum numbers (e.g.~number of up quarks) do not need to reach zero. We can say, however, that the total in these quantum numbers cannot exceed the total $W$ boson count (the difference between $W^-$s and $W^+$s). For example, if we added a single $W$ boson, we can turn one up-quark into a down-quark, and the final up quantum number total could be one higher. Quantum numbers conserved by $W$ bosons still have to reach zero, such as charge, lepton family numbers, or shade (described in section \ref{sec:vision_tab}). Once we find a combination of particles that has a solution, we quit the process.

\subsection{Diagram/Solution Generation}
\label{sec:algorithms-solution_generation}
To start solution generation, we first feed in the initial and final state, the particles we are allowed to use, and the degree range in which to look for solutions. We can then create a list of interactions we are able to use in the solution.

The generation of a solution will go through the following main steps:
\begin{enumerate}
    \item Choose the degree (minimum to maximum).
    \item Connect hadrons.
    \item Connect the state fermions that are not general fermions (i.e. are $e, \mu_e, u, d,...$ rather than $l, \nu_l, q_u, q_d$).
    \item Connect the rest of the diagram
    \item Remove solutions in which a single gluon couples into a `colour isolated' system (forbidden by colour conservation). See section \ref{sec:algorithms-vision} below.
\end{enumerate}

Each connection step will branch one diagram into many, creating a diagram tree. If we only want to find one solution, we will follow one branch at a time through all steps, moving down branches until we find a solution and then quit. If we are finding all solutions (or lowest order), we gather all diagrams from one step then move to the next. If finding lowest order, we quit if we find diagrams before moving to the next degree.

We deal with hadrons first because the quarks inside them can connect directly to the other state. When connecting hadrons we produce all unique combinations of these `directly connected' diagrams. This is why we mentioned in section \ref{sec:sandbox-generating_solutions} that solution generation can take a lot longer if there are hadrons that can directly connect; it only takes a few hadrons on each side to produce thousands of these combinations, without gaining in degree. Each of these diagrams will then branch through the rest of these steps.

Let us illustrate the solution generation procedure with an example:

\begin{equation}
g + g \rightarrow u + \overline{u}
\end{equation}

We will follow this diagram for degree-2. Particles are connected as paths -- for example a fermion path starts with a particle that enter the diagram, i.e. a particle with the fermion arrow pointing into the diagram, which ensures we connect particles the right way round. To make this easier in code we give all particles with arrows pointing into the diagram a +1 factor and particles with arrows pointing out a -1 factor; neutral bosons, which are neither, stay as +1. Therefore, as long as we ensure we are connecting from a +1 fermion to a -1 fermion, we will be connecting in the correct direction. This makes our states now:

\begin{equation}
\{g, g\} \rightarrow \{-u, +u\}
\end{equation}
as the anti-up-quark is entering the diagram, and the up quark is exiting. We notice we are also only now concerned with the base particle as, apart from those on state-lines, the location of interactions is irrelevant and only determined when a diagram is drawn and therefore whether a particle is an anti-particle is only determined when drawn. Now, starting with the entering (+) up-quark, we go through the list of interactions that it could create, which are are stored in a dictionary for each particle, e.g.:

\begin{lstlisting}
Particle.down : [
    ...
],
Particle.up : [
    [Particle.up, Particle.photon],
    [Particle.up, Particle.gluon],
    [Particle.up, Particle.Z],
    [Particle.up, Particle.H],
    [Particle.dark_quark, -Particle.W]
],
Particle.strange : [
    ...
]
\end{lstlisting}

\begin{figure}
    \centering
    \includegraphics[width=0.3\linewidth]{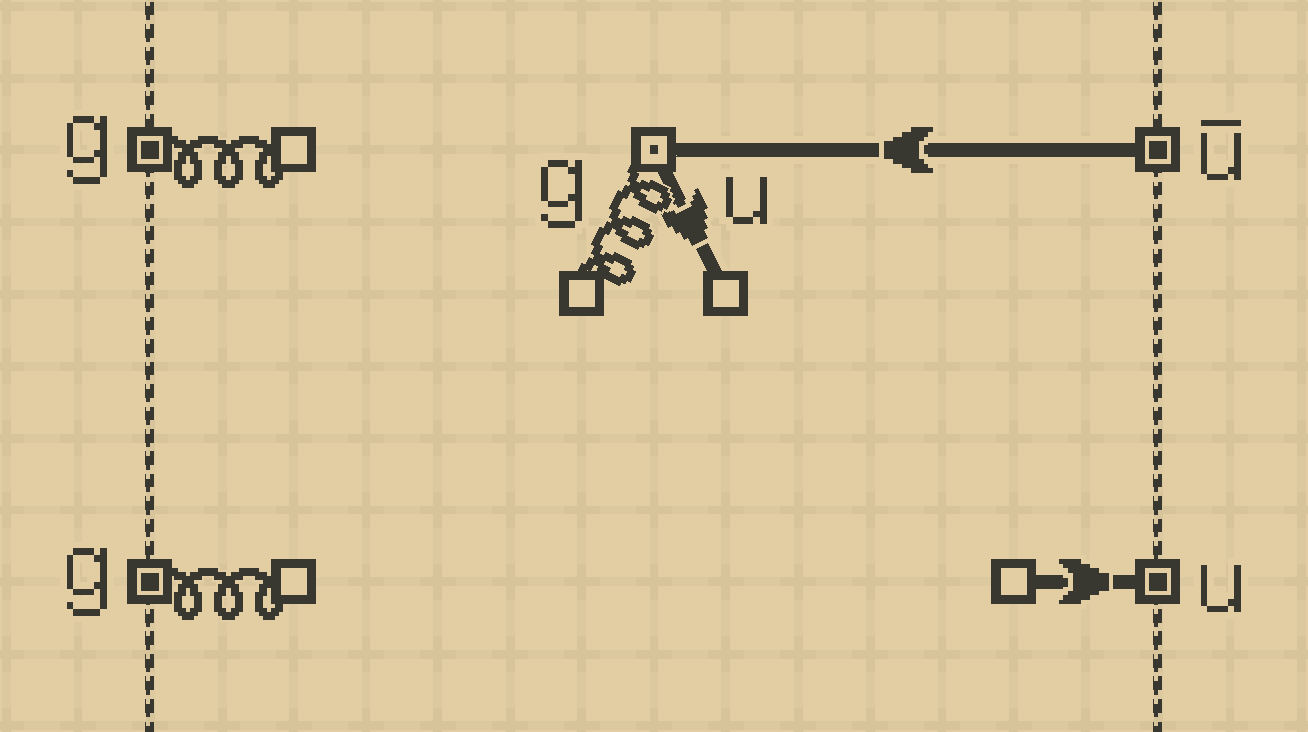}
    \includegraphics[width=0.3\linewidth]{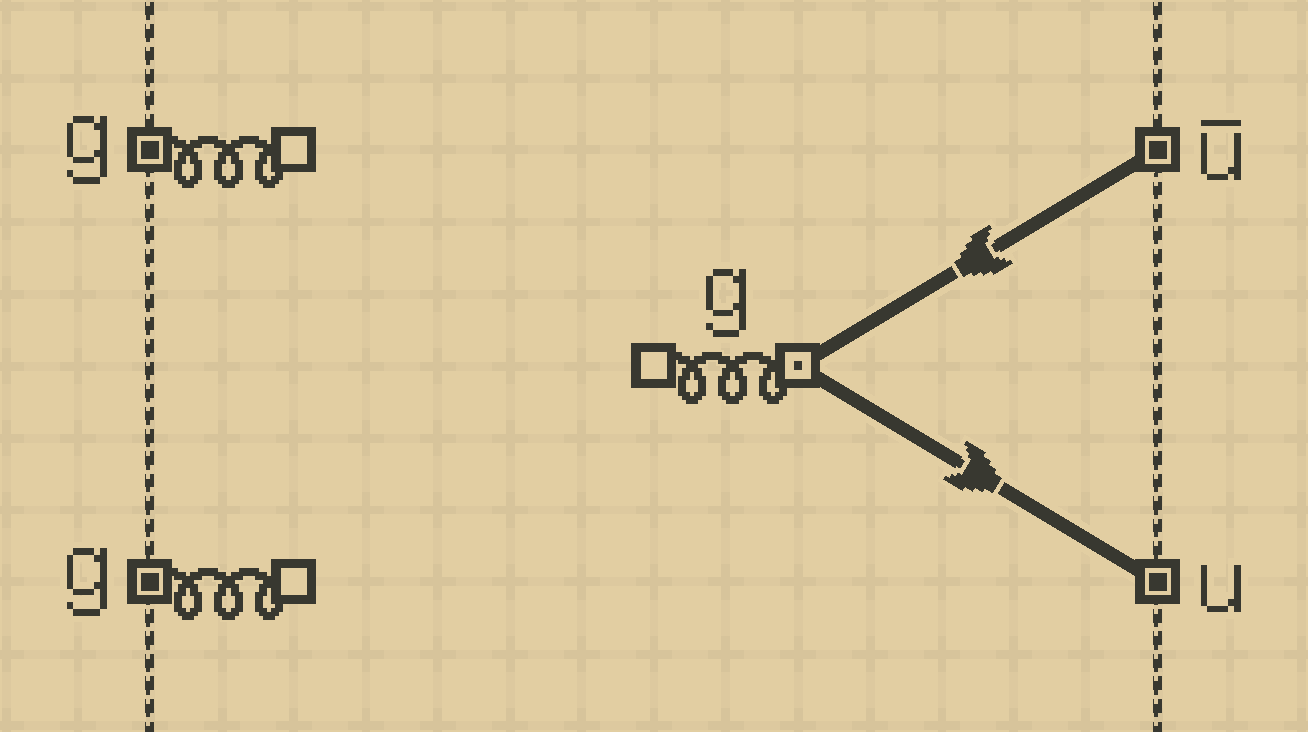}
    \includegraphics[width=0.3\linewidth]{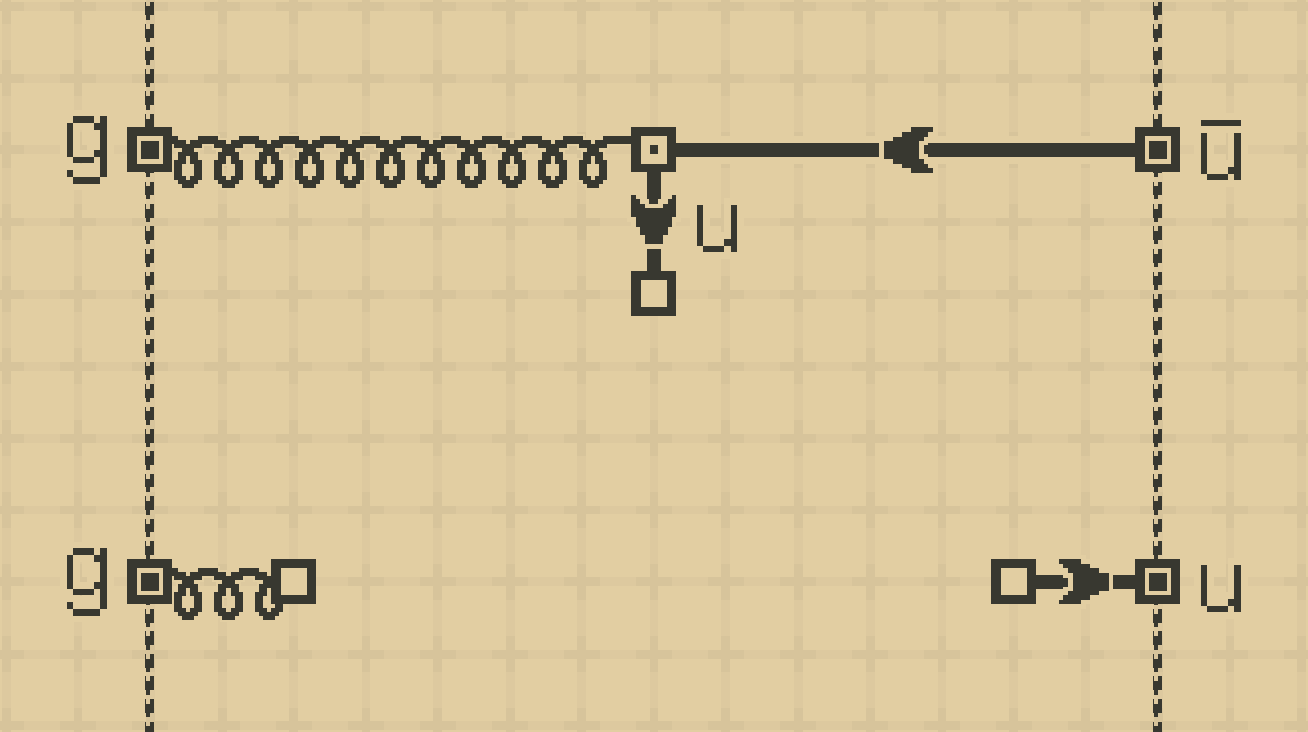}
    \includegraphics[width=0.3\linewidth]{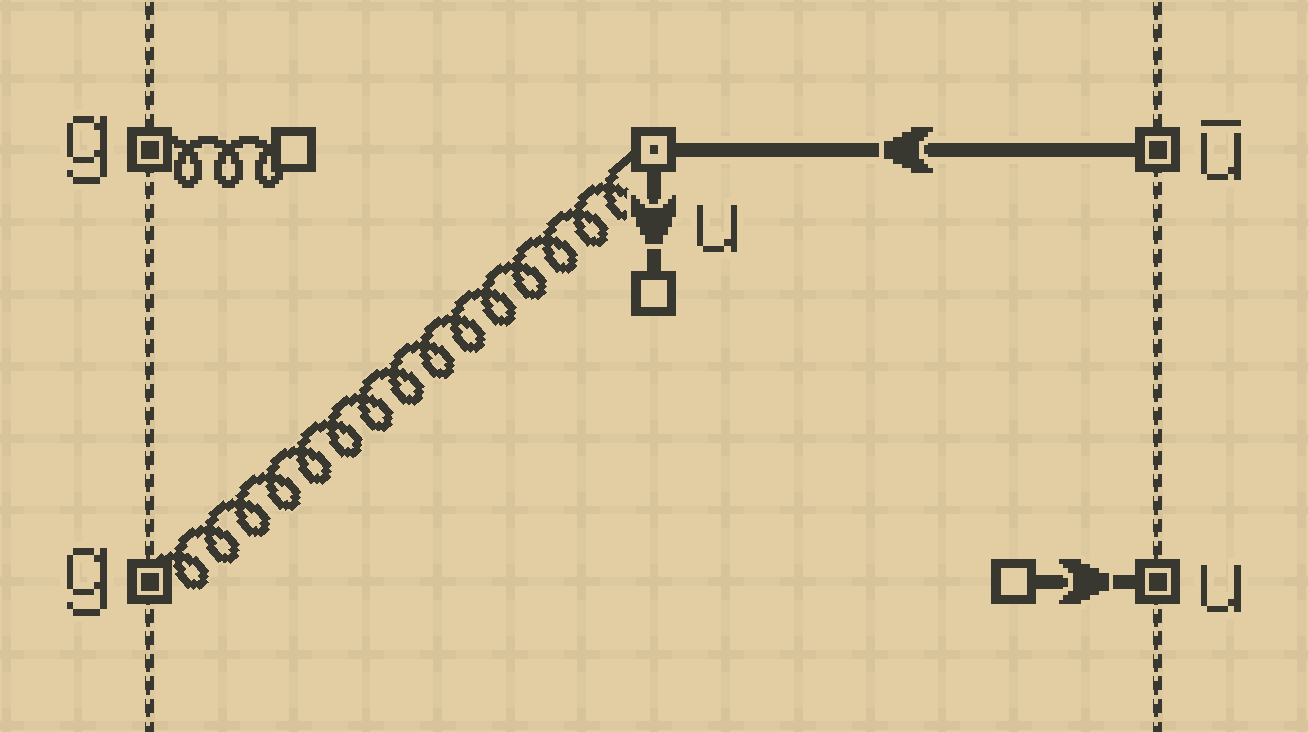}
    \includegraphics[width=0.3\linewidth]{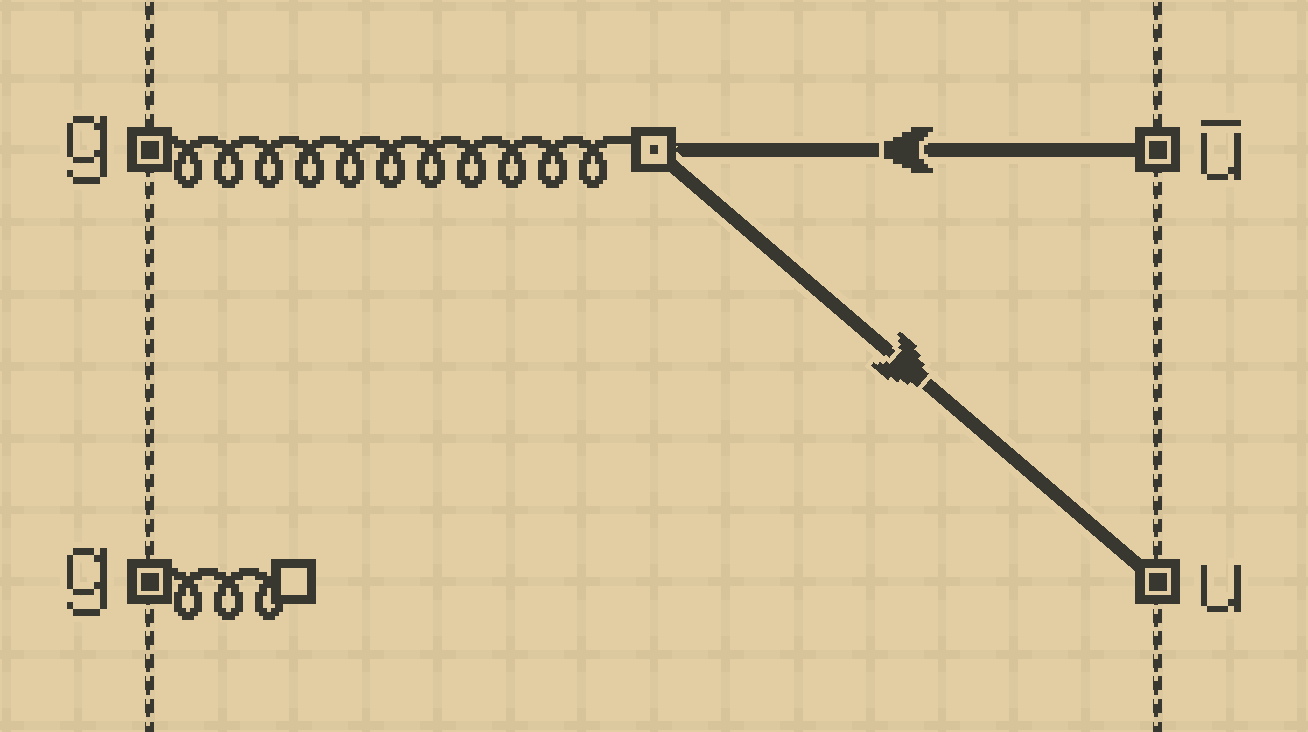}
    \includegraphics[width=0.3\linewidth]{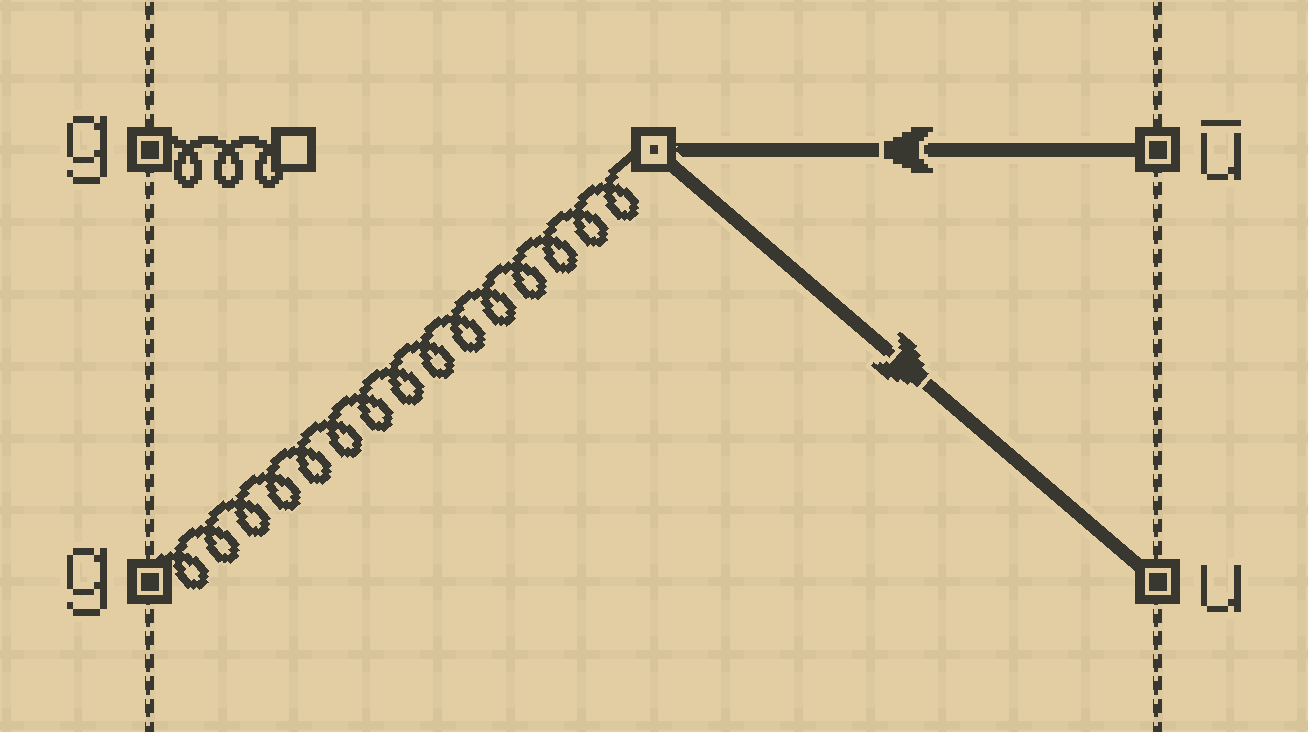}
    \caption{The different connection combinations that could occur for a $[u, u, g]$ interaction. Dangling particles are considered unconnected.}
    \label{fig:connection_combinations}
\end{figure}

In the code the up-\textit{type} quark and down-\textit{type} quark are instead referred to by their shade: bright and dark respectively, which we see in the last interaction. We also see that these interactions do not contain the entering particle. The +1 and -1 factors ensure we connect \textit{from} +1 particles and connect \textit{to} -1 particles, which we can see with the $W$ interaction, which will flow into this interaction by our convention discussed in section \ref{sec:feyncraft_diagrams} (that is, we must connect the base $W^-$ particle flowing \textit{into} this vertex, rather than out from the vertex). So we go through this list, and uniquely connect these particles to unconnected particles in our diagram. We must pay attention if the particle is charged (i.e. has a direction), and that we only connect a +1 particle to a -1 particle. In our example, we have three unconnected particles, as we count the particle we came from as connected, which are \lstinline![Particle.gluon, Particle.gluon, -Particle.up]!. If we create the interaction \lstinline![Particle.up, Particle.gluon]!, we can see that we could connect both particles, the up-quark and the gluon. The combination of connections that we choose here will lead to the different solutions, the different combinations in this step are shown in \figref{fig:connection_combinations}. Not all of these combinations are actually created because we can tell that they cannot lead to a valid solution, in this example these are the diagrams where we connect both particles (bottom centre and bottom right diagrams in \figref{fig:connection_combinations}) and the diagram where we do not connect either (top left diagram in \figref{fig:connection_combinations}). The former cases will lead to disconnected diagrams, since we have an unconnected state particle that can no longer connect to the rest of the diagram. In the latter case we now have too many unconnected particles (5) to connect with only one degree remaining (as we were looking at degree-2 and we have now used one).\\

\begin{figure}
    \centering
    \includegraphics[width=0.3\linewidth]{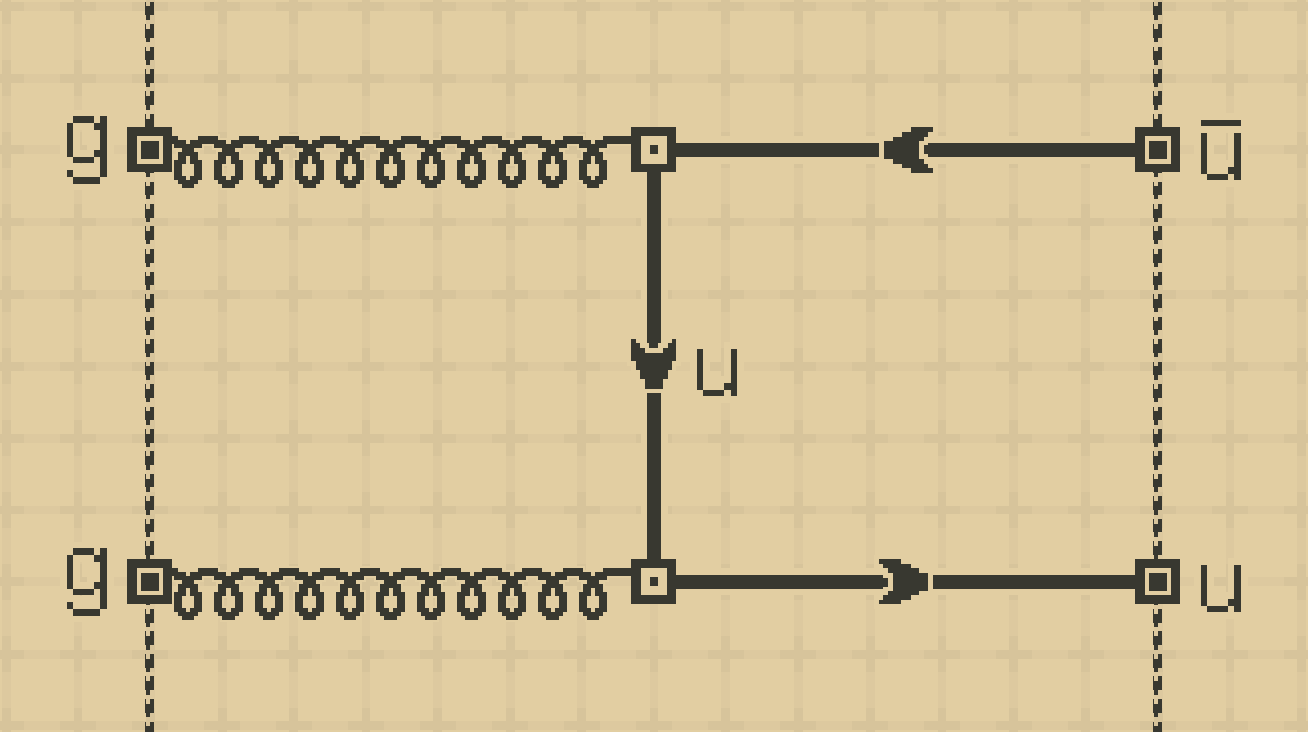}
    \includegraphics[width=0.3\linewidth]{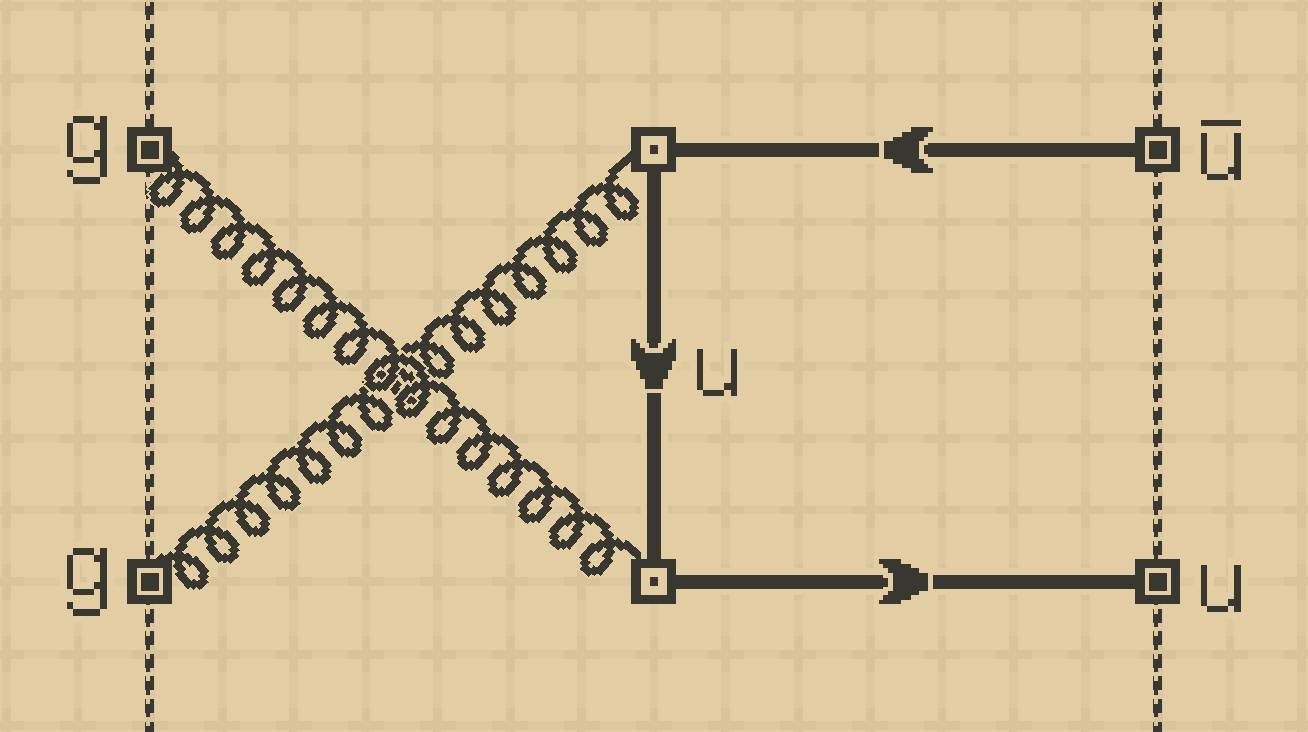}
    \includegraphics[width=0.3\linewidth]{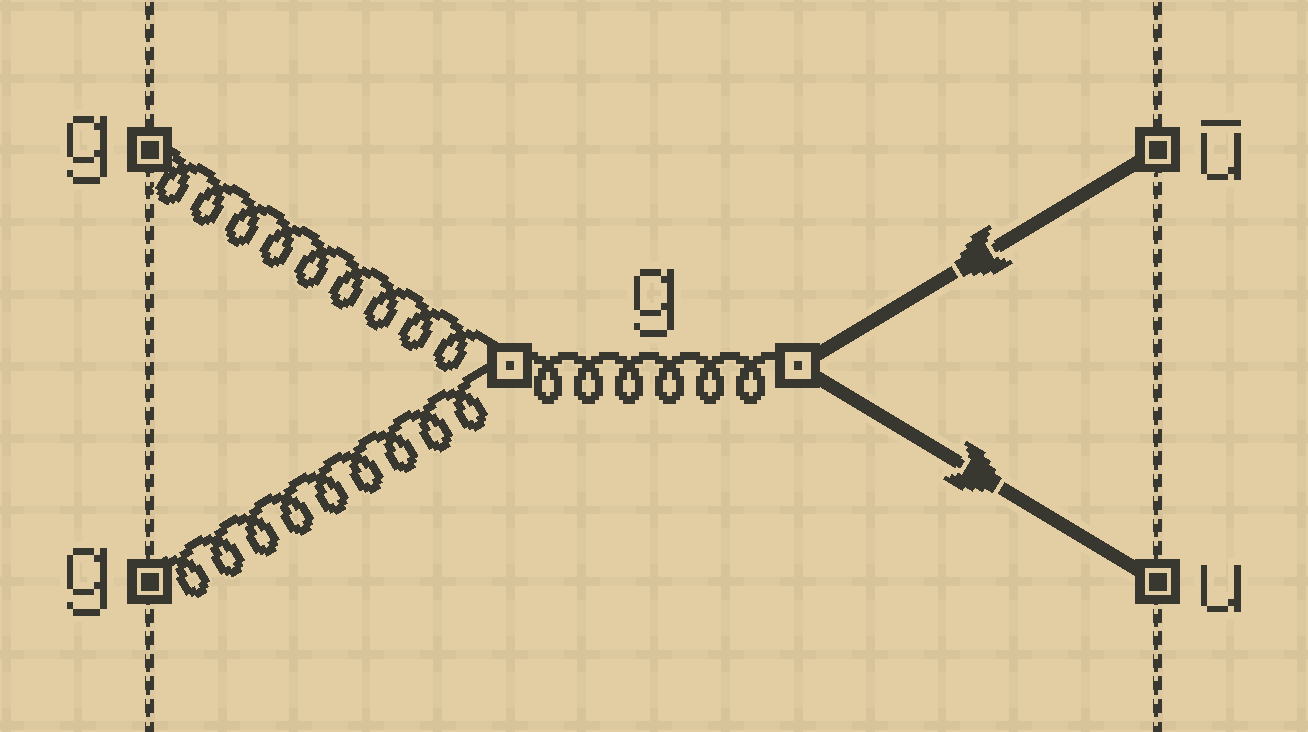}
    \caption{The completed diagrams produced by continuing to connect the diagrams in \figref{fig:connection_combinations}.}
    \label{fig:completed_diagrams}
\end{figure}

For the diagrams where we have not connected the up-quark, we now repeat the previous step from the newly created up-quark: going through the same list of interactions and checking connection combinations. As we will have no degree remaining, we must connect the remaining unconnected particles with this next interaction. In this example, it is only the top right and bottom left diagrams in \figref{fig:connection_combinations} that we do this with. We see we can only add the same interaction, \lstinline![Particle.up, Particle.gluon]!, again to leave no unconnected particles which leads to the first two completed diagrams shown in \figref{fig:completed_diagrams}. For those diagrams where the up-quark was connected, we now have no entering non-general state fermions, and by conservation no exiting non-general state fermions, remaining and move to the next step (connecting the rest of the diagram). In this step, we pick another entry particle, not necessarily starting from a state-line, and do the same process again. In this case, we would start with one of the remaining unconnected gluons whose interactions are:

\begin{lstlisting}
Particle.gluon : [
    [-Particle.bright_quark, Particle.bright_quark],
    [-Particle.dark_quark, Particle.dark_quark],
    [Particle.gluon, Particle.gluon],
    [Particle.gluon, Particle.gluon, Particle.gluon]
]
\end{lstlisting}

 We see that the only interaction that will lead to completed diagrams is \lstinline![Particle.gluon, Particle.gluon]!. This leads to the final completed completed diagram in \figref{fig:completed_diagrams}.

 Note that the gluon interactions (and indeed all boson and general fermion interactions) use the general fermions, for simplicity. If we did not connect the non-general state fermions first, this feature could potentially cause problems. We can illustrate this by considering the process $g \to u + \bar{c}$, which is forbidden in QCD. If we started connecting the gluon first, we would consider the interaction \lstinline{[-Particle.bright_quark, Particle.bright_quark]} shown above. We might then connect the bright quark to the outgoing up quark (specifying the bright quark now as an up), and connect the charm into the negative bright quark (specifying that as a charm line). This would result in \FeynCraft{} constructing an invalid $[u, c, g]$ QCD interaction. Such a potential pitfall can simply be avoided by connecting up the non-general state fermions first. The nature of the non-general fermion interactions guarantees that we don't generate forbidden interactions when dealing with these connections. For example, the gluon emission from an up quark is written above as \lstinline![Particle.up, Particle.gluon]! rather than \lstinline![Particle.bright_quark, Particle.gluon]!, ensuring that the correct $[u, u, g]$ vertex is generated.

 The $W$ boson is the only particle that is able to mix general and non-general quarks (since a $W$ emission from a specific up/down quark is able to produce down/up type quarks from any family). An example of this is seen in the up-quark interactions above: \lstinline![Particle.dark_quark, -Particle.W]!. Since this interaction is created when we are connecting state fermions, it may be that this dark quark actually has to be connected to a non-general state quark. It is then no longer general, but specified to be that particular quark. This is fine, and causes no issues. For example, if we were looking at the process $u \to W^+ + b$, we would insert the up quark $W$ emission interaction, and would convert and connect the outgoing dark quark to the bottom quark. This yields the permitted $[u, b, W]$ interaction. Note that the $W$ boson does not mix general and non-general \textit{leptons} -- for example, a $W$ emission from an electron is only able to produce an electron-neutrino, and not a neutrino from any other family.

\subsection{Vision and colourless gluon checking}
\label{sec:algorithms-vision}
For shade and colour, we have to find a configuration of paths through a given diagram. As each particle has a definite shade, the $4-W$ vertex is the only place that an entering shade flow has a choice of which particle to leave by. This choice also cannot lead to an incorrect configuration.
With colour, on the other hand, the path we chose could lead to an invalid colour configuration. For example, in a $[q, q, g]$ vertex, if we enter by a quark, we cannot exit by the other quark else we would leave the gluon disconnected and lead to the colour in the gluon looping back within itself, creating an invalid colour configuration. Once a complete configuration \textit{is} found, we can then use any present hadrons to colour it, which are restricted to being colourless. Paths left uncoloured are simply coloured in order of red, blue, green.
With this now coloured diagram, we can check to see if there are any colourless gluons.
This occurs when a single gluon attaches into some `colour isolated' system, where such a system has the property that the colour injected by the gluon cannot escape from the system (either to a state particle, or to the other end of the attached gluon). By colour conservation this forces the gluon to be colourless, and thus forbidden. Examples of such colour isolated systems are hadrons (with arbitrary numbers of gluon interactions internal to the hadron system, plus arbitrarily many emissions/absorptions of colour singlet particles from the hadron), and internal colour loops that have no other gluons connected (but arbitrarily many emissions/absorptions of colour singlet particles from the hadron). We can check if a gluon is colourless by temporarily removing it, and then seeing what interactions in the diagram we can reach from the two points that were at either end of the previously-connected gluon, only moving along particles with colour. If we are able to escape the diagram via a state interaction that is not a hadron, from both `ends' of the gluon, the gluon is not colourless (note that hadrons are colourless and thus not able to carry away the colour of the gluon). Similarly, if we are able to trace a coloured path from one end of the previously-connected gluon to the other via the remainder of the diagram, we do not mark the gluon as colourless either. Note that in this exercise we regard state interactions inside hadrons as being reachable from one another (the quarks inside hadrons may exchange colour via gluon exchanges internal to the hadron, which are usually left implicit). 

\subsection{Diagram layout}
\label{sec:algorithms-diagram_layout}
During solution generation, we do not care about an interaction's position other than if it appears in either the initial or final state. In order to load these diagrams visually we thus have to convert all our interactions and connections that we found during generation into an actual physical diagram. With the exception of state particles which are spread out on their respective state lines, this is done for the rest of the interactions through the spring-layout algorithm described in \cite{kobourov2013forcedirected}. Essentially, we place each interaction randomly, then apply attractive spring forces towards connected interactions, and repulsive forces from those that are not. After several iterations we would hope to reach a minimum energy, and visually pleasing, state. Post-adjustments are required to keep the graph within the bounds, prevent interactions from getting too close, and balance the forces. While this simple process does typically produce reasonable looking diagrams for processes of sufficiently low degree, further work needs to be done in order to ensure that \FeynCraft{} consistently produces `publication-ready' diagrams.

\subsection{Duplicate diagrams}
\label{sec:algorithms-duplicate_diagrams}

\begin{figure}
    \centering
    \includegraphics[width=0.45\linewidth]{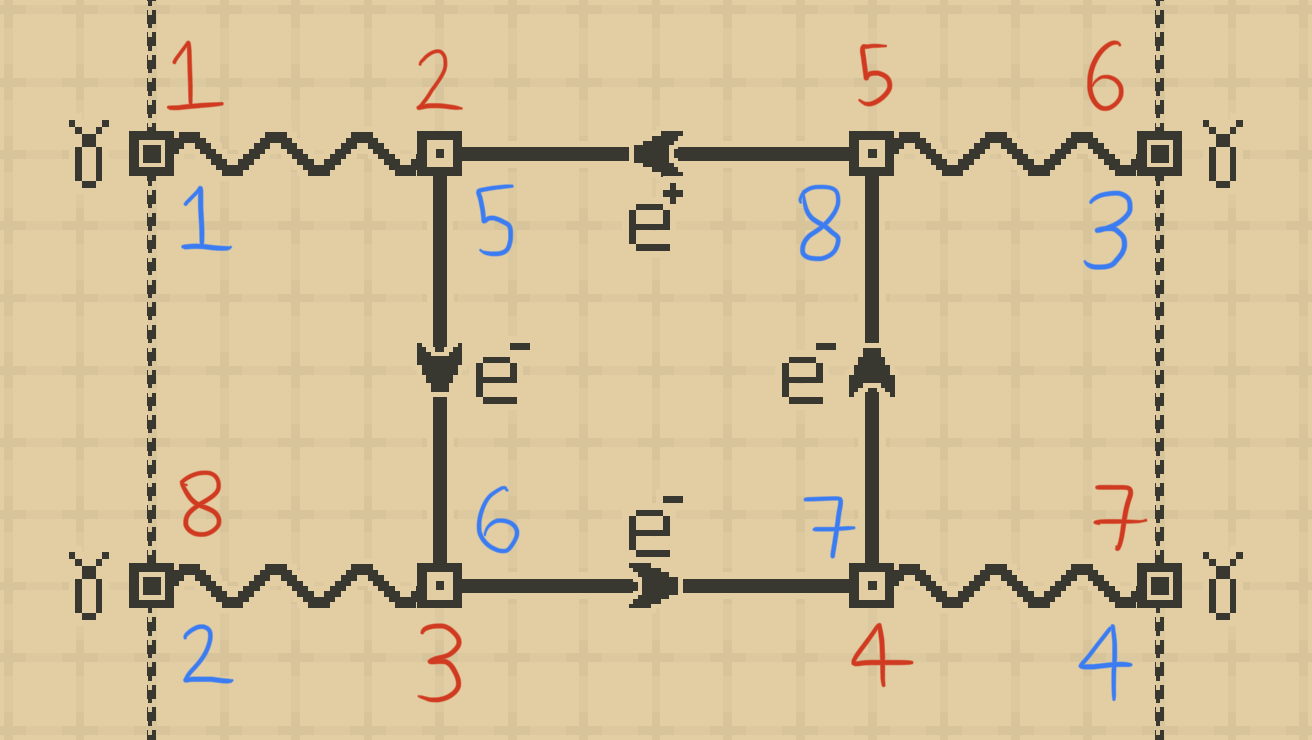}
    \includegraphics[width=0.45\linewidth]{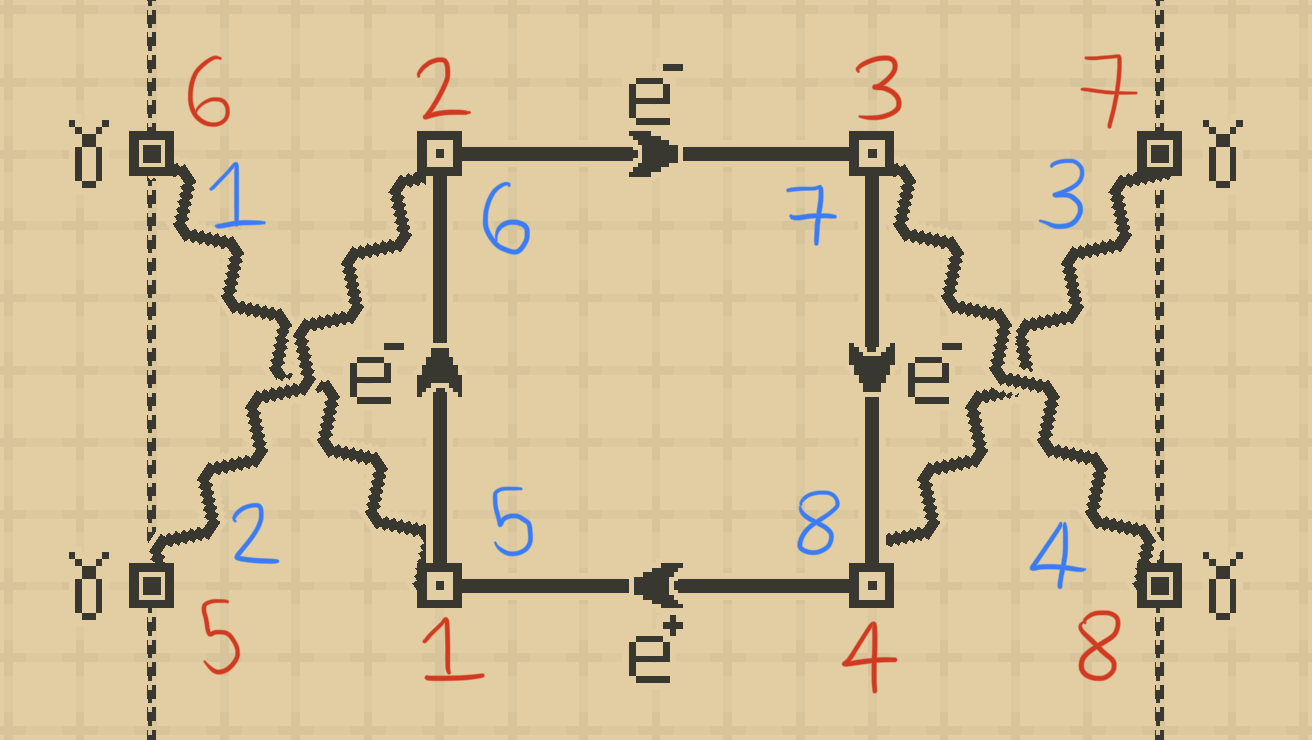}
    \caption{Two differently arranged duplicate diagrams. The red numbers indicate the order in which the interactions were drawn, the blue numbers indicate the order after sorting.}
    \label{fig:duplicate_diagram_example}
\end{figure}

Take the two diagrams shown in \figref{fig:duplicate_diagram_example}, ignoring the blue numbering for now. The red numbers indicate the order each interaction was drawn and added to the `connection matrix', which represents the diagram in code. If we `flip' the box on the right hand side of \figref{fig:duplicate_diagram_example} vertically, moving (red) interactions 2 and 3 to the bottom, and interactions 1 and 4 to the top, then we can indeed see these two diagrams are duplicates; however, as the interactions were drawn in a different order, and thus have a different numbering, we cannot immediately see that the diagrams are duplicates of each other in the code. For example, in the first diagram interaction-5 is connected to 6 via a photon, but in the second 5 is not connected to 6 at all. We have to sort these diagrams (re-index them) to be able to compare the diagrams directly. How should we do this? We can first sort the state interactions based on the state particle and their vertical position, starting with the initial state. The sorted order is shown by the blue numbering in the same figure, where we can see the states have now been ordered the same way in both diagrams. For the  interactions inside the diagram (`middle interactions'), however, we cannot sort based on position as their position on the diagram does not differentiate them. Instead, we follow a path, starting from the lowest numbered state interaction (1), and order middle interactions in the order we see them along the path. When this path splits, we choose which particle to follow next from a definite ordering of particles, which ensures the same path in two diagrams will be ordered and re-indexed the same way. We can follow this re-ordering in the example diagrams by starting at the re-ordered (blue) interaction-1. This photon is connected to an interaction which we then order as interaction-5 (as 4 has already been used by a final-state interaction). We follow the fermion path around the loop, re-ordering interactions as 6, 7, and 8. Once this process is completed, we can directly compare two diagrams, e.g. are all interactions connected to the same interactions and via the same particles? If so, they are duplicate. We can see from the blue numbering in the example we can now directly compare the connections in each diagram: 1 is connected to 5 via a photon, 5 is connected to 6 via an electron as so on. We find that all connections are identical as so declare these diagrams duplicate.\\

When following a path, we might have to choose between two or more interactions which are connected to our current interaction by the the same particle. How should we choose between them, and decide which to re-index first? If we find that these interactions are not re-indexed by another path, we decide which particle to follow first based on how each particle is connected further into the diagram. This includes if one interaction is connected to the other by a charged particle, which particles these interactions are connected via, and which re-indexed interactions they are connected to. If these checks fail again to distinguish the interactions, we look further into the diagram again.\\

\section{Conclusions}
\label{sec: conclusions}

In this manuscript we have introduced \FeynCraft{}, a browser-based game designed to aid players in learning how to draw Feynman diagrams in the Standard Model of particle physics. The game has a broad suite of features linked to this aim. Users can draw diagrams and, if the diagram is invalid, the code will show where and why this is the case. One can solve example Feynman diagram problems -- either randomly generated ones, or curated ones created by the developers or other users of \FeynCraft{}. Alternatively, one can specify a scattering process (as well as numerous other options, such as the forces/interactions one wants to be involved) and ask \FeynCraft{} to generate the Feynman diagrams for the process. Finally there is the possibility to add overlays to the diagram that convey additional information (interaction strength, QCD colour flow, and the analogous `weak flow'), and export drawn diagrams to \LaTeX\ code.

There are various potential avenues for improvement for \FeynCraft{} in future. One possibility is to create a new \LaTeX{} package that would more smoothly work alongside \FeynCraft{} exports, in order to get round certain limitations of the \lstinline{tikz} and \lstinline{tikz-feynman} packages. Another interesting possibility would be to expand \FeynCraft{} to allow the inclusion of interactions beyond the Standard Model -- for example, to allow the inclusion of operators in the Standard Model Effective Field Theory (SMEFT) (note that some of these operators, like the four-fermion interactions, can already be drawn in \FeynCraft{}, but are marked as invalid due to the fact that we restrict to SM interactions). SMEFT is now extensively used for searches for beyond the SM physics, and given the plethora of SMEFT operators and somewhat complex notation used for operators and operator coefficients, an implementation of SMEFT vertices within \FeynCraft{} could be useful for students and young researchers to build familiarity with these (we note that there are already some useful tools in these direction available on the website \cite{SMEFTsimwebsite} of the package SMEFTsim \cite{Brivio:2017btx, Brivio:2020onw}). Finally, one could add to \FeynCraft{} the feature to draw all of the QCD colour flows for a given graph, and compute from these the associated colour factor of the graph, following \cite{Kilian:2012pz}. This could be useful to researchers that wish to acquaint themselves with this colour-flow representation.

\section*{Acknowledgements}

The work of JRG, and part of the work of AO, has been supported by the Royal Society through Grant URF\textbackslash{}R1\textbackslash{}201500. JRG would like to thank Stephanie Gaunt for the daffodils.





\bibliographystyle{elsarticle-num}
\bibliography{feyncraft}

\begin{thebibliography}{10}
\expandafter\ifx\csname url\endcsname\relax
  \def\url#1{\texttt{#1}}\fi
\expandafter\ifx\csname urlprefix\endcsname\relax\def\urlprefix{URL }\fi
\expandafter\ifx\csname href\endcsname\relax
  \def\href#1#2{#2} \def\path#1{#1}\fi

\bibitem{Braid}
\url{https://store.steampowered.com/app/499180/Braid_Anniversary_Edition/}, Accessed on 10/10/2025.

\bibitem{Talos}
\url{https://store.steampowered.com/app/257510/The_Talos_Principle/}, Accessed on 10/10/2025.

\bibitem{Baba}
\url{https://store.steampowered.com/app/736260/Baba_Is_You/}, Accessed on 10/10/2025.

\bibitem{Portal1}
\url{https://store.steampowered.com/app/400/Portal/}, Accessed on 10/10/2025.

\bibitem{Portal2}
\url{https://store.steampowered.com/app/620/Portal_2/}, Accessed on 10/10/2025.

\bibitem{Thomson:2013zua}
M.~Thomson, {Modern particle physics}, Cambridge University Press, New York, 2013.
\newblock \href {https://doi.org/10.1017/CBO9781139525367} {\path{doi:10.1017/CBO9781139525367}}.

\bibitem{Kilian:2012pz}
W.~Kilian, T.~Ohl, J.~Reuter, C.~Speckner, {QCD in the Color-Flow Representation}, JHEP 10 (2012) 022.
\newblock \href {http://arxiv.org/abs/1206.3700} {\path{arXiv:1206.3700}}, \href {https://doi.org/10.1007/JHEP10(2012)022} {\path{doi:10.1007/JHEP10(2012)022}}.

\bibitem{EdExcel}
{A-level physics specification: Pearson Edexcel Level 3 Advanced GCE in Physics (9PH0). First teaching from September 2015. }, \url{https://qualifications.pearson.com/content/dam/pdf/A%20Level/Physics/2015/Specification%20and%20sample%20assessments/pearsonedexcel-alevel-physics-spec.pdf}, Accessed on 10/10/2025.

\bibitem{OCR}
{OCR A level specification Physics A H556. For first assessment in 2017}, \url{https://www.ocr.org.uk/images/171726-specification-accredited-a-level-gce-physics-a-h556.pdf}, Accessed on 10/10/2025.

\bibitem{AQA}
{AQA AS and A-level physics specification. For AS and A-level exams in 2016 onwards}, \url{https://www.aqa.org.uk/subjects/physics/a-level/physics-7408/specification/subject-content/particles-and-radiation}, Accessed on 10/10/2025.

\bibitem{Binosi:2008ig}
D.~Binosi, J.~Collins, C.~Kaufhold, L.~Theussl, {JaxoDraw: A Graphical user interface for drawing Feynman diagrams. Version 2.0 release notes}, Comput. Phys. Commun. 180 (2009) 1709--1715.
\newblock \href {http://arxiv.org/abs/0811.4113} {\path{arXiv:0811.4113}}, \href {https://doi.org/10.1016/j.cpc.2009.02.020} {\path{doi:10.1016/j.cpc.2009.02.020}}.

\bibitem{Harlander:2020cyh}
R.~V. Harlander, S.~Y. Klein, M.~Lipp, {FeynGame}, Comput. Phys. Commun. 256 (2020) 107465.
\newblock \href {http://arxiv.org/abs/2003.00896} {\path{arXiv:2003.00896}}, \href {https://doi.org/10.1016/j.cpc.2020.107465} {\path{doi:10.1016/j.cpc.2020.107465}}.

\bibitem{Harlander:2024qbn}
R.~Harlander, S.~Y. Klein, M.~C. Schaaf, {FeynGame-2.1 -- Feynman diagrams made easy}, PoS EPS-HEP2023 (2024) 657.
\newblock \href {http://arxiv.org/abs/2401.12778} {\path{arXiv:2401.12778}}, \href {https://doi.org/10.22323/1.449.0657} {\path{doi:10.22323/1.449.0657}}.

\bibitem{Bundgen:2025utt}
L.~B\"undgen, R.~V. Harlander, S.~Y. Klein, M.~C. Schaaf, {FeynGame 3.0} (1 2025).
\newblock \href {http://arxiv.org/abs/2501.04651} {\path{arXiv:2501.04651}}.

\bibitem{Feynman:1949hz}
R.~P. Feynman, {The Theory of positrons}, Phys. Rev. 76 (1949) 749--759.
\newblock \href {https://doi.org/10.1103/PhysRev.76.749} {\path{doi:10.1103/PhysRev.76.749}}.

\bibitem{TongQFT}
D.~Tong, {Quantum Field Theory, University of Cambridge Part III Mathematical Tripos}, \url{https://www.damtp.cam.ac.uk/user/tong/qft/qft.pdf}, Accessed on 10/10/2025.

\bibitem{Brivio:2019ius}
I.~Brivio, S.~Bruggisser, F.~Maltoni, R.~Moutafis, T.~Plehn, E.~Vryonidou, S.~Westhoff, C.~Zhang, {O new physics, where art thou? A global search in the top sector}, JHEP 02 (2020) 131.
\newblock \href {http://arxiv.org/abs/1910.03606} {\path{arXiv:1910.03606}}, \href {https://doi.org/10.1007/JHEP02(2020)131} {\path{doi:10.1007/JHEP02(2020)131}}.

\bibitem{Ellis:2020unq}
J.~Ellis, M.~Madigan, K.~Mimasu, V.~Sanz, T.~You, {Top, Higgs, Diboson and Electroweak Fit to the Standard Model Effective Field Theory}, JHEP 04 (2021) 279.
\newblock \href {http://arxiv.org/abs/2012.02779} {\path{arXiv:2012.02779}}, \href {https://doi.org/10.1007/JHEP04(2021)279} {\path{doi:10.1007/JHEP04(2021)279}}.

\bibitem{ATLAS:2022xyx}
A.~collaboration, \href{https://cds.cern.ch/record/2816369}{{Combined effective field theory interpretation of Higgs boson and weak boson production and decay with ATLAS data and electroweak precision observables}}, ATL-PHYS-PUB-2022-037 (2022).
\newline\urlprefix\url{https://cds.cern.ch/record/2816369}

\bibitem{Celada:2024mcf}
E.~Celada, T.~Giani, J.~ter Hoeve, L.~Mantani, J.~Rojo, A.~N. Rossia, M.~O.~A. Thomas, E.~Vryonidou, {Mapping the SMEFT at high-energy colliders: from LEP and the (HL-)LHC to the FCC-ee}, JHEP 09 (2024) 091.
\newblock \href {http://arxiv.org/abs/2404.12809} {\path{arXiv:2404.12809}}, \href {https://doi.org/10.1007/JHEP09(2024)091} {\path{doi:10.1007/JHEP09(2024)091}}.

\bibitem{CMS:2025ugn}
V.~Chekhovsky, et~al., {Combined effective field theory interpretation of Higgs boson, electroweak vector boson, top quark, and multi-jet measurements} (4 2025).
\newblock \href {http://arxiv.org/abs/2504.02958} {\path{arXiv:2504.02958}}.

\bibitem{BILENKY2003395}
S.~Bilenky, \href{https://www.sciencedirect.com/science/article/pii/B0122274105004774}{Neutrinos}, in: R.~A. Meyers (Ed.), Encyclopedia of Physical Science and Technology (Third Edition), third edition Edition, Academic Press, New York, 2003, pp. 395--417.
\newblock \href {https://doi.org/https://doi.org/10.1016/B0-12-227410-5/00477-4} {\path{doi:https://doi.org/10.1016/B0-12-227410-5/00477-4}}.
\newline\urlprefix\url{https://www.sciencedirect.com/science/article/pii/B0122274105004774}

\bibitem{Peskin:1995ev}
M.~E. Peskin, D.~V. Schroeder, {An Introduction to quantum field theory}, Addison-Wesley, Reading, USA, 1995.
\newblock \href {https://doi.org/10.1201/9780429503559} {\path{doi:10.1201/9780429503559}}.

\bibitem{Martin:2016xsp}
S.~P. Martin, {Top-quark pole mass in the tadpole-free $\overline {MS}$ scheme}, Phys. Rev. D 93~(9) (2016) 094017.
\newblock \href {http://arxiv.org/abs/1604.01134} {\path{arXiv:1604.01134}}, \href {https://doi.org/10.1103/PhysRevD.93.094017} {\path{doi:10.1103/PhysRevD.93.094017}}.

\bibitem{Martin:2019lqd}
S.~P. Martin, D.~G. Robertson, {Standard model parameters in the tadpole-free pure $\overline{\rm{MS}}$ scheme}, Phys. Rev. D 100~(7) (2019) 073004.
\newblock \href {http://arxiv.org/abs/1907.02500} {\path{arXiv:1907.02500}}, \href {https://doi.org/10.1103/PhysRevD.100.073004} {\path{doi:10.1103/PhysRevD.100.073004}}.

\bibitem{Ellis:2016jkw}
J.~Ellis, {TikZ-Feynman: Feynman diagrams with TikZ}, Comput. Phys. Commun. 210 (2017) 103--123.
\newblock \href {http://arxiv.org/abs/1601.05437} {\path{arXiv:1601.05437}}, \href {https://doi.org/10.1016/j.cpc.2016.08.019} {\path{doi:10.1016/j.cpc.2016.08.019}}.

\bibitem{kobourov2013forcedirected}
S.~G. Kobourov, Force-directed Drawing Algorithms, in Handbook of Graph Drawing and Visualization, CRC Press, 2013, Ch.~12, p. 383–408.

\bibitem{SMEFTsimwebsite}
{SMEFTsim website}, \url{https://smeftsim.github.io/}, Accessed on 10/10/2025.

\bibitem{Brivio:2017btx}
I.~Brivio, Y.~Jiang, M.~Trott, {The SMEFTsim package, theory and tools}, JHEP 12 (2017) 070.
\newblock \href {http://arxiv.org/abs/1709.06492} {\path{arXiv:1709.06492}}, \href {https://doi.org/10.1007/JHEP12(2017)070} {\path{doi:10.1007/JHEP12(2017)070}}.

\bibitem{Brivio:2020onw}
I.~Brivio, {SMEFTsim 3.0 \textemdash{} a practical guide}, JHEP 04 (2021) 073.
\newblock \href {http://arxiv.org/abs/2012.11343} {\path{arXiv:2012.11343}}, \href {https://doi.org/10.1007/JHEP04(2021)073} {\path{doi:10.1007/JHEP04(2021)073}}.

\end{thebibliography}







\end{document}